\documentclass[twocolumn]{aastex631}
\usepackage{amsmath}
\usepackage{comment}
\usepackage{cleveref}
\usepackage{longtable}

\crefformat{section}{\S#2#1#3} 
\crefformat{subsection}{\S#2#1#3}
\crefformat{subsubsection}{\S#2#1#3}

\newcommand{\approximately}{\raisebox{0.5ex}{\texttildelow}}

\begin{document}

\title{Short-period Heartbeat Binaries from TESS Full-Frame Images}

\author[0000-0001-6541-734X]{Siddhant Solanki}
\affiliation{Department of Astronomy, University of Maryland, 7901 Regents Drive, College Park, MD 20742, USA}
\author[0009-0006-9864-0517]{Agnieszka M. Cieplak}
\affiliation{Space Sciences Laboratory, University of California, Berkeley, 7 Gauss Way, Berkeley, CA 94720-7450, USA}
\author[0000-0002-2942-8399]{Jeremy Schnittman}
\affiliation{NASA Goddard Space Flight Center, 8800 Greenbelt Road, Greenbelt, MD 20771, USA}
\author{John G. Baker}
\affiliation{NASA Goddard Space Flight Center, 8800 Greenbelt Road, Greenbelt, MD 20771, USA}
\author{Thomas Barclay}
\affiliation{NASA Goddard Space Flight Center, 8800 Greenbelt Road, Greenbelt, MD 20771, USA}
\author{Richard K. Barry}
\affiliation{NASA Goddard Space Flight Center, 8800 Greenbelt Road, Greenbelt, MD 20771, USA}
\author{Veselin Kostov}
\affiliation{NASA Goddard Space Flight Center, 8800 Greenbelt Road, Greenbelt, MD 20771, USA}
\affiliation{SETI Institute, 189 Bernardo Ave, Suite 200, Mountain View, CA 94043, USA}
\author{Ethan Kruse}
\affiliation{NASA Goddard Space Flight Center, 8800 Greenbelt Road, Greenbelt, MD 20771, USA}
\author{Greg Olmschenk}
\affiliation{NASA Goddard Space Flight Center, 8800 Greenbelt Road, Greenbelt, MD 20771, USA}
\author[0000-0003-0501-2636]{Brian P. Powell}
\affiliation{NASA Goddard Space Flight Center, 8800 Greenbelt Road, Greenbelt, MD 20771, USA}
\author[0000-0003-2267-1246]{Stela {Ishitani Silva}}
\affiliation{NASA Goddard Space Flight Center, 8800 Greenbelt Road, Greenbelt, MD 20771, USA}
\affiliation{Oak Ridge Associated Universities, Oak Ridge, TN 37830, USA}
\author{Guillermo Torres}
\affiliation{Center for Astrophysis $\vert$ Harvard \& Smithsonian, 60 Garden St., Cambridge, MA 02148, USA}

\begin{abstract}
We identify $240$ short-period ($P \lesssim 10$ days) binary systems in the TESS data, $180$ of which are heartbeat binaries (HB). The sample is mostly a mix of A and B-type stars and primarily includes eclipsing systems, where over $30\%$ of the sources with primary and secondary eclipses show a secular change in their inter-eclipse timings and relative eclipse depths over a multi-year timescale, likely due to orbital precession. The orbital parameters of the population are estimated by fitting a heartbeat model to their phase curves and Gaia magnitudes, where the model accounts for ellipsoidal variability, Doppler beaming, reflection effects, and eclipses. We construct the sample's period-eccentricity distribution and find an eccentricity cutoff (where $e \rightarrow 0$) at a period $1.7$ days.  Additionally, we measure the periastron advance rate for the $12$ of the precessing sources and find that they all exhibit prograde apsidal precession, which is as high as $9^{\circ}$ yr$^{-1}$ for one of the systems. Using the inferred stellar parameters, we estimate the general relativistic precession rate of the argument of periastron for the population and expect over $30$ systems to show a precession in excess of $0.3^{\circ}$ yr$^{-1}$.
\end{abstract}

\section{Introduction} \label{sec:intro}
Heartbeat systems are a class of detached binary stars with relatively eccentric $(e\gtrsim0.2)$ orbits and typical periods of less than $100$ days \citep{susan_thompson_2012ApJ...753...86T,shporer_2016ApJ...829...34S,kirk_2016AJ....151...68K}. The brighter star can be a main-sequence (MS) star or a giant \citep{beck_2014AA...564A..36B}. The physical separation between the two stars at periastron is small, on the order of a few stellar radii, $R_*$. At these short distances, the tidal interactions give rise to detectable flux variations in the light curve of the system as the shapes and the temperatures of the stars get perturbed. These systems get their name from a characteristic cardiogram-like signature in their light curves when the stars pass the periastron \citep{susan_thompson_2012ApJ...753...86T}.  The fractional peak-to-peak flux variations in HBs are usually between $10^{-3}$ and $10^{-2}$ but can be as large as $40\%$ \citep{jayasinghe_2021MNRAS.506.4083J}. Some of the heartbeat systems also show clear tidally excited oscillations (TEOs) in their light curves. These are resonantly-excited g-modes in the stars from the tidal perturbation of the companion, with frequencies that are multiples of the orbital period \citep{burkart_elliot_2012MNRAS.421..983B,fuller_lai_2012MNRAS.420.3126F,fuller_2017MNRAS.472.1538F,shelley_cheng_2020ApJ...903..122C,teos_2020ApJ...888...95G}.

The light curves of HB systems can provide information about the orbits, even without radial velocity measurements, since the orbital parameters can be estimated by fitting a model for ellipsoidal variability to their light curves \citep{morris_1985ApJ...295..143M,Kumar_1995ApJ...449..294K,Engel_2020MNRAS.497.4884E}. Furthermore, the parameters can be constrained even more if the systems are eclipsing.

Heartbeat systems are more eccentric than other binaries with similar orbital periods and a fraction of them probe the upper envelope of the period-eccentricity (PE) distribution \citep{shporer_2016ApJ...829...34S}. Therefore, the PE distribution for relatively young, short-period HBs can help us understand tidal dissipation and orbital circularization in these systems as well as the physical mechanisms behind short-period binary formation, which still remain uncertain \citep{toonen_2020A&A...640A..16T,moe_kratter_2018ApJ...854...44M}.

Only a few HBs were known before the 2010s, prior to the Kepler mission \citep{high_ecc_B_2009A&A...508.1375M}. Since then, a large number of HBs have been detected in the Kepler \citep{welsh_koi_2011ApJS..197....4W,susan_thompson_2012ApJ...753...86T,hambleton_2013MNRAS.434..925H,shporer_2016ApJ...829...34S} and Optical Gravitational Lensing Experiment (OGLE) data \citep{ogle_2_2022ApJ...928..135W,ogle_2022ApJS..259...16W}. Many of these systems are old or have relatively long periods, with $P \gtrsim 10$ days.  Additionally, while Kepler's photometry is unrivaled, its limited sky coverage prevents the discovery of HBs in other parts of the sky. Discovering a class of bright and short-period HBs is possible using data from the \textit{Transiting Exoplanet Survey Satellite} (TESS) \citep{tess_2015JATIS...1a4003R,tess_ebs_2022ApJS..258...16P}. Primarily intended to study exoplanets around bright host stars, TESS is an all-star sky survey covering about $85\%$ of the sky. The TESS data contain full-frame images (FFIs) taken at a cadence of $30$ minutes, $10$ minutes, or $200$ seconds, depending on the TESS sector. Each sector covers $24^{\circ} \times 96^{\circ}$ of the sky and lasts about a month. Furthermore, the multi-sector data from TESS provides a multi-year observing baseline that can be used to look for long-term trends in the light curves. Several heartbeat systems have already been discovered in the TESS data, a majority of them containing massive stars \citep{jayasinghe_2021MNRAS.506.4083J,tess_heartbeats_2021AA...647A..12K}.

In this work, we report the discovery of $180$ heartbeat systems in the TESS data, identified using two convolutional neural networks (CNNs). Most systems are A and B-spectral class stars with periods less than $10$ days and de-reddened B-V colors between $-0.4$ and $0.5$. Out of these, $133$ systems also show eclipses. The light curves are extracted from TESS FFI using the \textsc{eleanor} pipeline \citep{eleanor_2019PASP..131i4502F}, and the orbital parameters of the systems are then estimated by fitting a heartbeat model to the light curves, accounting for ellipsoidal variability, Doppler beaming, reflection effects, and eclipses \citep{Engel_2020MNRAS.497.4884E}. Close to $30$ systems that have both primary and secondary eclipses per orbit show a secular change in inter-eclipse timing and the relative eclipse depths over a multi-year baseline in their phase-folded light curves, which is likely due to orbital precession. We pick and model $12$ such systems and find that all show prograde-apsidal precession.

Our method of sample selection and some of the representative light curves are given in Section \ref{sec:TESS_lightcurves}; the light curve modeling is described in Section \ref{sec:methods}; the parameter estimates for the entire population are presented in Section \ref{sec:lightcurve_fits}. Finally, we discuss our results in a broader context of binary evolution as in Section \ref{sec:discussion} and conclude in Section \ref{sec:conclusion}.

\section{TESS light curves} \label{sec:TESS_lightcurves}

\subsection{Light Curve Collection} \label{subsec:lightcurve_collection}
An initial list of $240$ heartbeat candidates was constructed using two CNNs that independently looked through the light curves generated from the TESS FFIs. $207$ of the $240$ sources were obtained from a neural network (NN) trained to look for eclipses in light curves with an apparent magnitude $m_{TESS} \leq 15$, discussed further in \citep{2021AJ....161..162P}. Light curves that resembled those of HBs were manually selected from the light curves identified by the NN. The effort to find HBs was not nearly rigorous, systematic or comprehensive, but rather a subset of a larger effort to identify binaries and ultimately larger hierarchical systems (see, e.g. \citealt{2022ApJS..259...66K,2024MNRAS.527.3995K}).  As such, the $207$ sources identified here are merely a sample.  Given the richness of the data retrieved from such a small fraction of TESS light curves -- 207 out of a few hundred -- our findings suggest that a dedicated effort to find HBs in TESS data would be richly rewarded. This search included the bulk of our candidates in the paper. 

Additionally, $33$ sources were taken from a different neural network and trained to look for HBs. This network was a 12-layer 1D CNN, identical to the one used by \citet{olmschenk2021transit} to identify transiting planet candidates. As this network was designed to be a generalized network for TESS data, a similar training setup was used as described in \citet{olmschenk2021transit}, except synthetic HBs were used as the positive training samples. The synthetic HBs were generated using the light curve model (refer to Sec. \ref{subsec:eBEER_model}) and TESS FFI light curves that were used for training and were inferred to find new candidates are identical to those used in \citet{olmschenk2021transit}.
These 33 sources are taken from a search of this neural network. During this search, the trained network was applied to a dataset of \approximately{}60M light curves and the top 100 results according to the network were manually inspected, resulting in these 33 identified HB sources. Preliminary results from a more complete analysis of this search suggests that this ratio of \approximately{}30\% true positives identified by the network holds far beyond the first 100 results, suggesting many more sources are available. However, as of this writing, analysis of this more complete network search is still ongoing.

After identifying the candidate HBs, light curves up through TESS sector $67$ were generated using the \textit{eleanor} pipeline \citep{eleanor_2019PASP..131i4502F}. The multi-sector light curves provided a long observing baseline, which was useful in constraining the HB periods. Light curves from sectors $1-26$ have a $30$ minute cadence, those from sectors $27-55$ have a $10$ minute cadence, and ones from sectors $56-67$ have a $200$ second cadence. The light curves were binned and phase-folded after their orbital periods were determined. The light curves had a mean photometric uncertainty between $10^{-4}$ and $10^{-3}$, which decreases somewhat from $m_{G}=11$ to $7$. The values reported by the TESS team \footnote{https://heasarc.gsfc.nasa.gov/docs/tess/observing-technical.html} suggest that the photometric uncertainty increases to $1\%$ for sources with $m_{TESS} = 15$, which is comparable to the strength of the heartbeat signal. Therefore, it is likely that we are biased towards finding brighter heartbeat systems.

\begin{figure}
    \centering
    \includegraphics[width=0.8\linewidth]{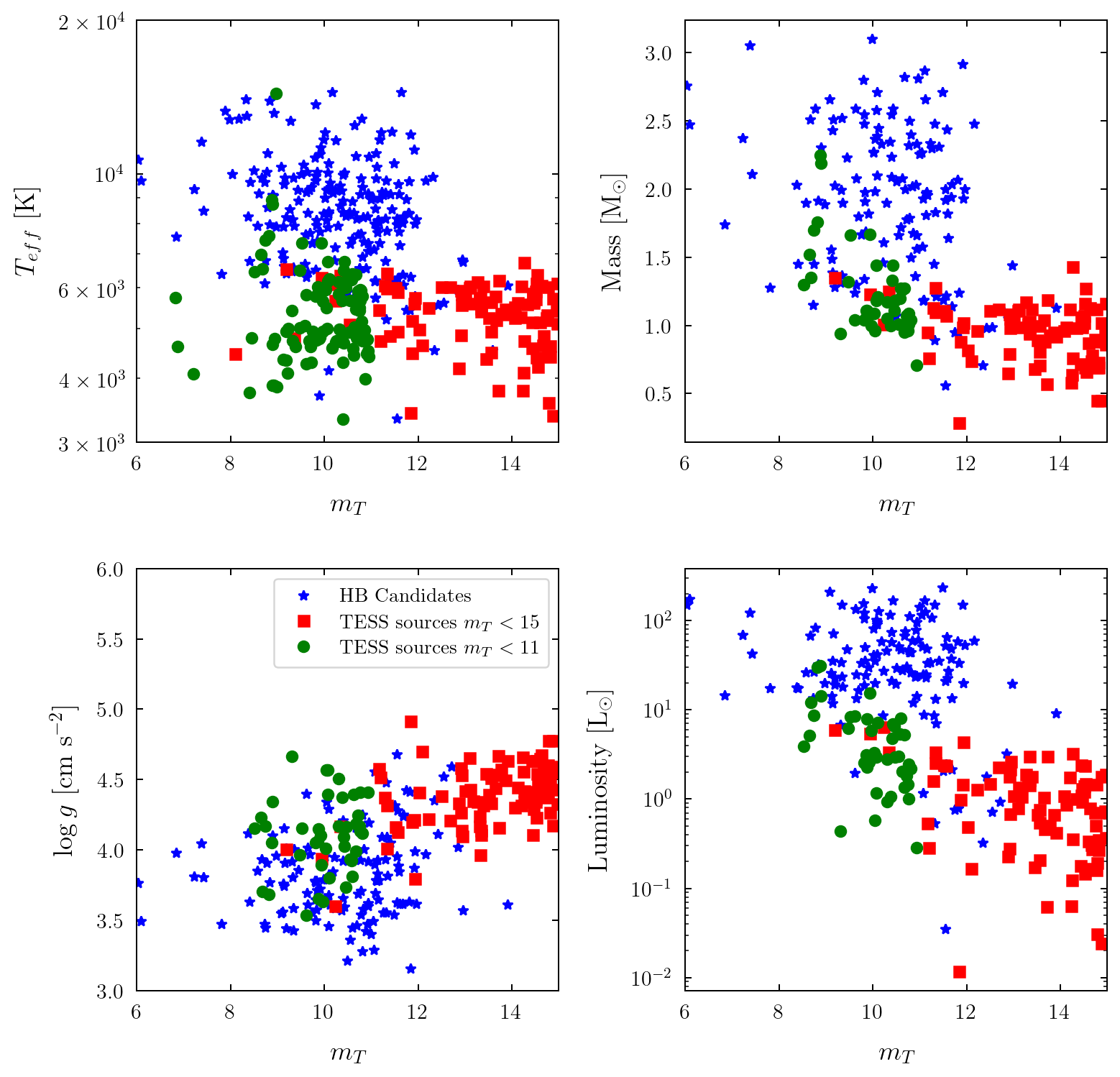}
    \caption{The figure shows the effective temperatures, stellar masses, surface gravities, and luminosities reported by the TESS Input Catalog for the candidate HB list and two magnitude-limited TESS samples. The first TESS sample has a maximum TESS magnitude of $15$, which is the same magnitude cut that applied to the TESS data to search for HBs, and the second sample has a maximum TESS magnitude of $11$, which is similar to the apparent magnitudes of the identified HBs in the data. Both these populations have smaller effective temperatures, masses, luminosities, and larger surface gravities than the stars on the candidate HB list. This points towards heartbeat stars generally being more massive stars as opposed to us selectively identifying more of them in a sample of bright stars.}
    \label{fig:population_comparison}
\end{figure}

To be more quantitative about how the candidate HBs differ from the typical TESS systems with similar magnitudes, we plot the effective temperatures $(T_{eff})$, inferred stellar masses $M$, surface gravities ($\log g$) and the luminosities of our candidate list and two magnitude-limited TESS samples. The first sample has a maximum TESS magnitude of $15$, and the second sample has a maximum magnitude equal to $11$. We plot 100 random sources from each of these two samples. The physical values for the three populations are all taken from the TESS Input Catalog and are generally not independent from each other. These are shown in Figure \ref{fig:population_comparison}. The HB candidate population is more massive and luminous than the randomly TESS sources with similar apparent magnitudes, which is qualitatively consistent with the masses that we infer from fitting the light curves with the heartbeat model. This points towards heartbeat stars generally being more massive stars as opposed to us selectively identifying more of them in a sample of bright stars, assuming that all the samples primarily contain main sequence stars.

After fitting the heartbeat model to the light curves and getting estimates for the orbital parameters, several cuts were made to the original list of $240$ sources, narrowing it down to $180$ HBs. First, $30$ systems that could not be fit by the model were removed. These primarily included sources with short periods ($P \lesssim 1.75 \text{ days}$), whose light curves resembled those of semi-detached binaries or contact binaries. These systems are discussed in Sec. \ref{subsec:short_period_sources}. Additionally, sources with eccentricities less than $0.2$ were removed, the cutoff point we use for defining HBs. Nevertheless, we keep these low eccentricity sources while presenting some population statistics, such as the period-eccentricity distribution. 

\subsection{Different Heartbeat Systems} \label{subsec:different_hb_systems}
The final light curve sample of $180$ HBs contains $133$ eclipsing systems, of which $35$ systems show a single eclipse and $97$ show two eclipses per orbit. Many systems also exhibit clear TEOs apparent in their light curves, although we do not estimate the exact percentage because of variable data quality between different TESS sectors and most systems not being observed in the same sectors. The light curves of some of those exhibiting large-amplitude TEOs are discussed in Sec. \ref{subsec:teos}. Out of the $97$ systems with two eclipses per orbit, over $30\%$ show changes in orbital parameters through variations in relative eclipse depths and, in some cases, changes in their inter-eclipse timings. We model the multi-sector data of $12$ such sources separately, and their light curves are discussed in Sec. \ref{subsec:eclipse_drift}. 

\begin{figure*}
    \centering
    \includegraphics[width=0.49\textwidth]{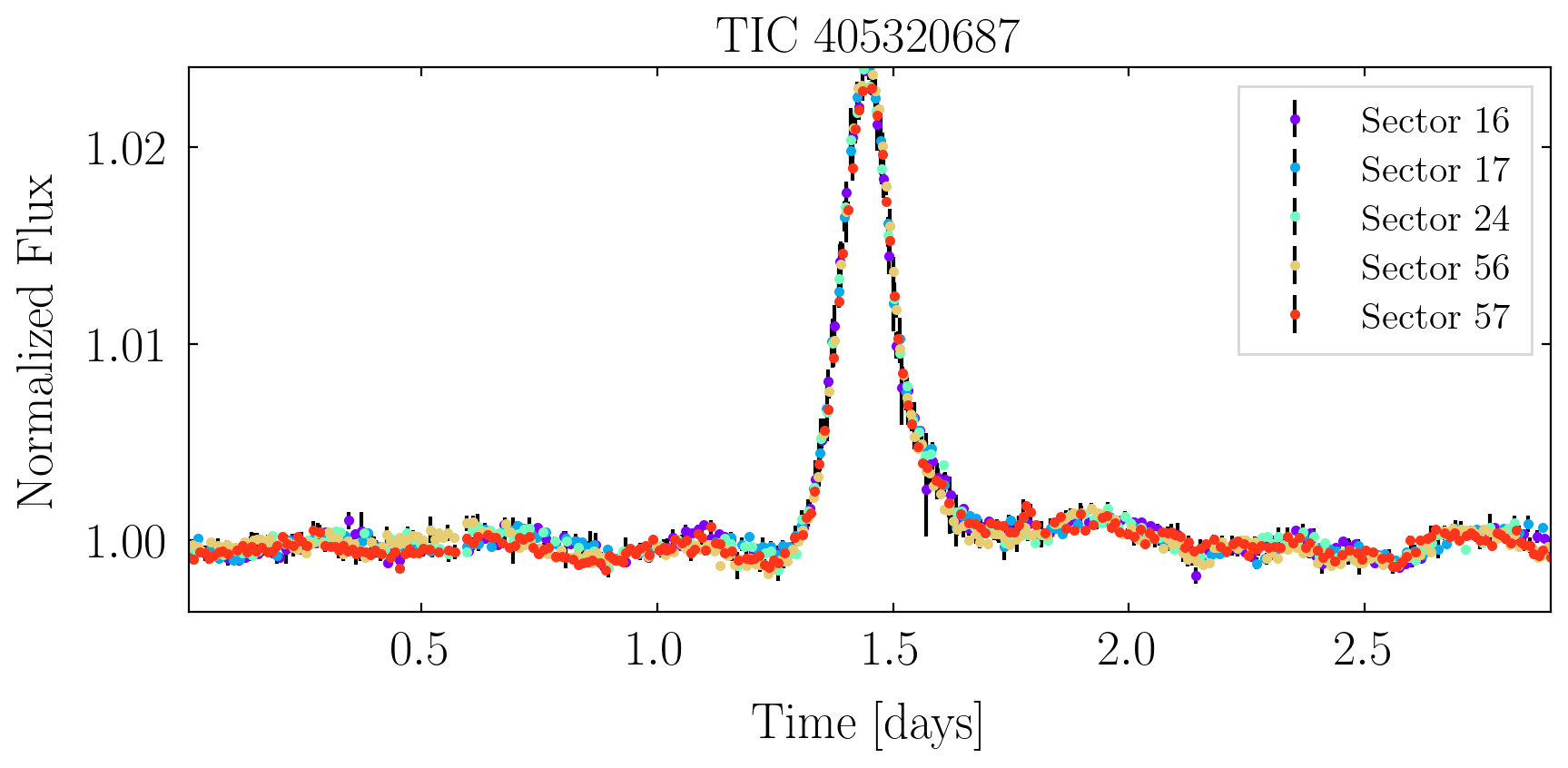}
    \includegraphics[width=0.49\textwidth]{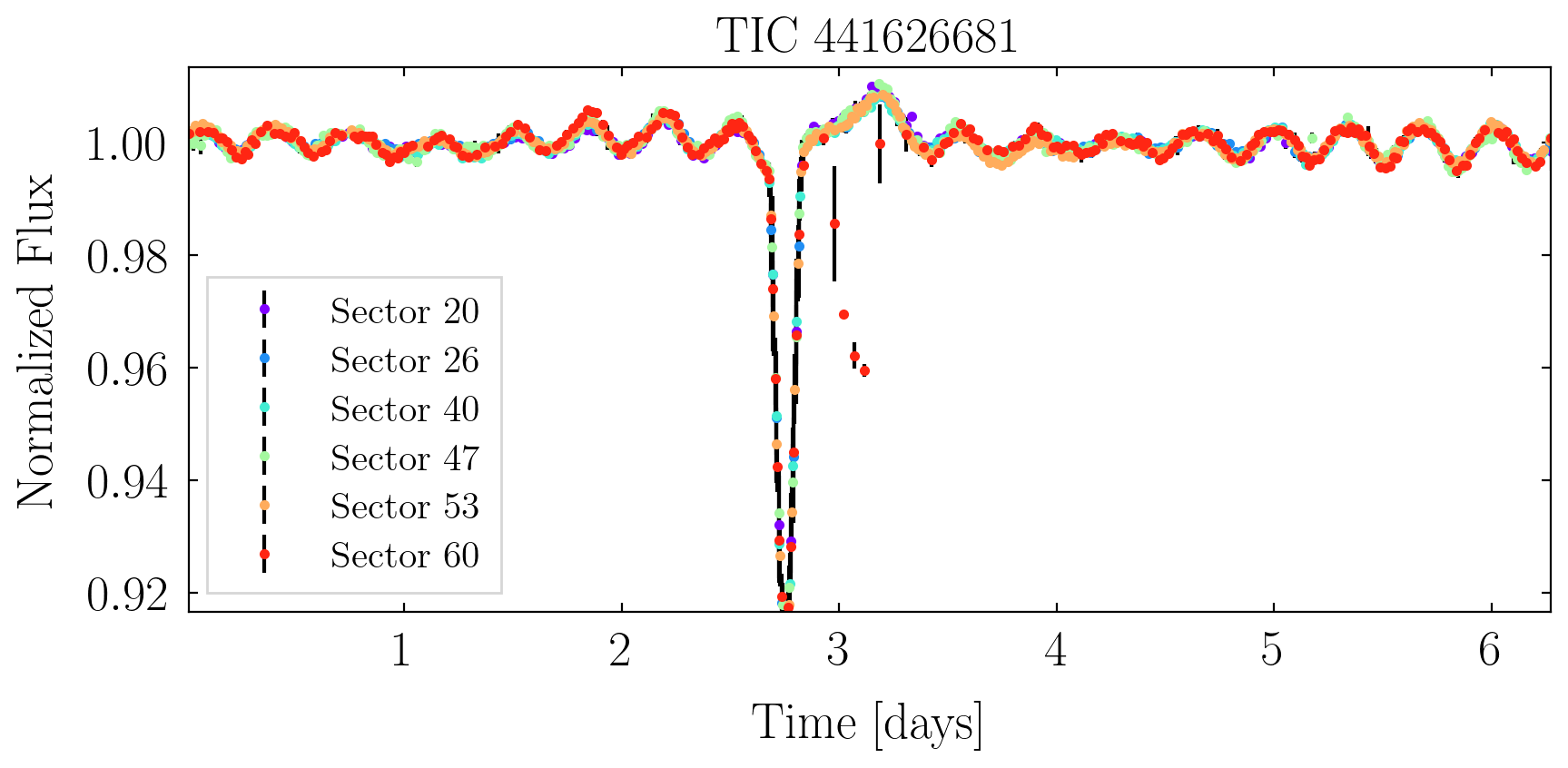}
    \includegraphics[width=0.49\textwidth]{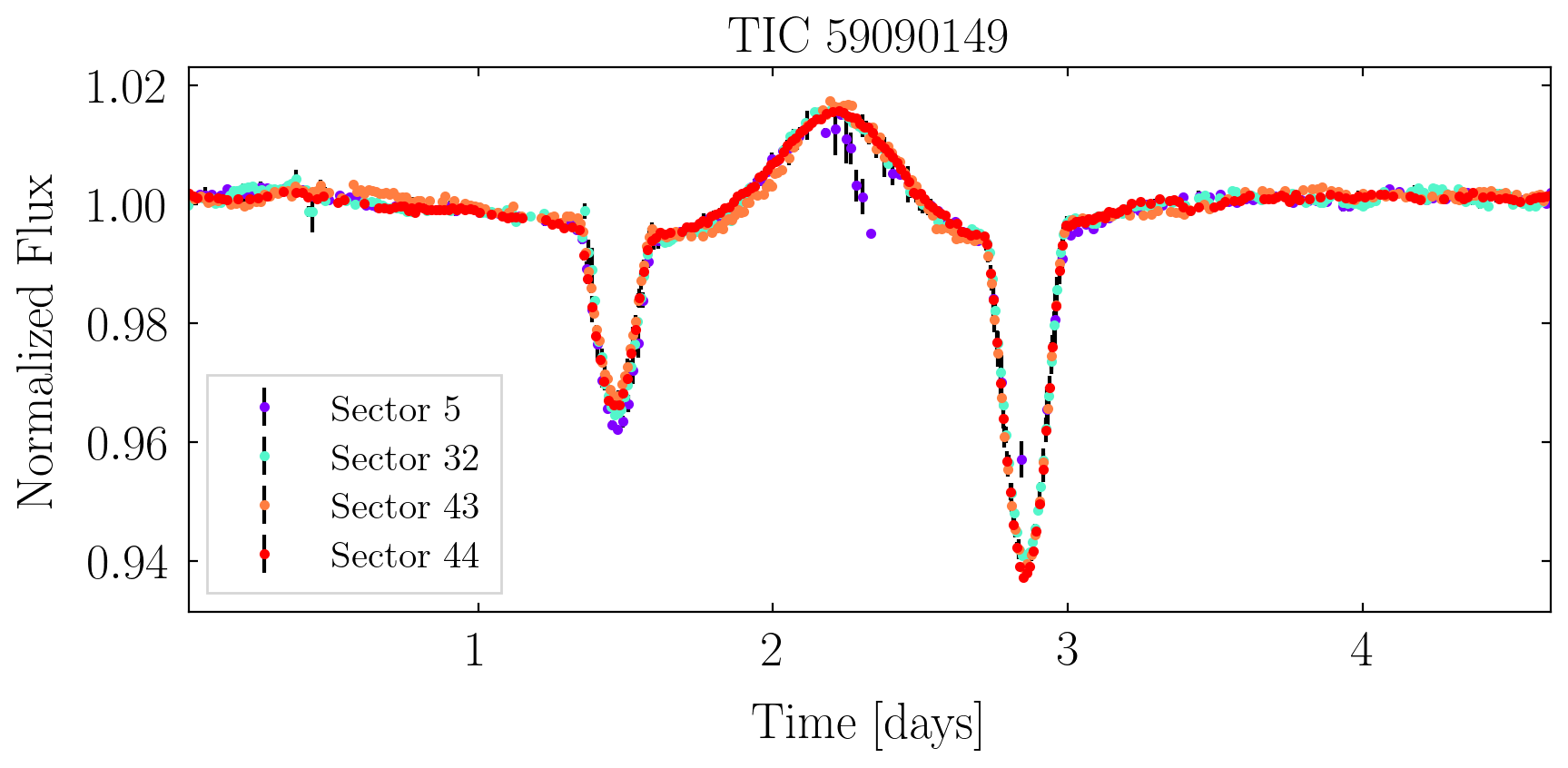}
    \includegraphics[width=0.49\textwidth]{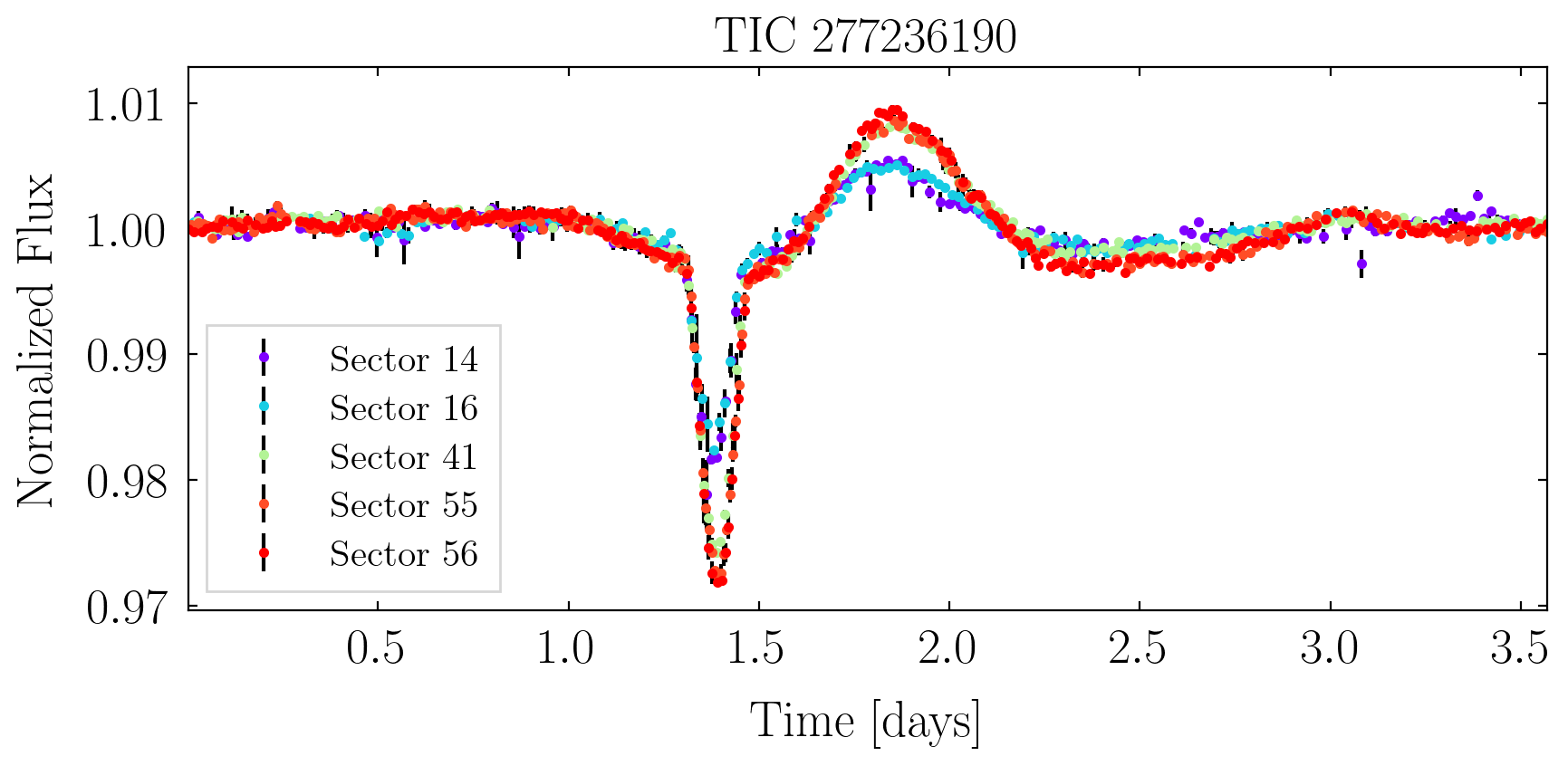}
    \caption{Multi-sector binned and phase folded light curves (phase curves) for some of the HBs in our sample. The error bars represent the Root Mean Square Deviation (RMSD) of the fluxes in each bin. The phase curves from every sector are normalized to the median flux of that sector. The zero phases of the curves are shifted in the plots above such that the periastron lies in the center of each panel, which is obtained after fitting the heartbeat model to the light curves; see Section\ref{sec:lightcurve_fits}. TIC 405320687 (top left) is one of the largest amplitude heartbeats in the sample. TIC 441626681 (top right) shows TEOs at $N=18,21$ times the orbital frequency. TICs 59090149 and 277236190 (bottom panel) are eclipsing sources, where the latter shows different amounts of blending in different TESS sectors.}
    \label{fig:heartbeat_classes}
\end{figure*}

The binned, phase-folded light curves (phase curves) of some representative sources from our sample are shown in Figure \ref{fig:heartbeat_classes}. The zero phases of the curves are shifted in the plots such that the periastron lies in the center of each panel, which is determined after fitting the heartbeat model to the phase curves; see Section \ref{sec:lightcurve_fits}. These sources include eclipsing and non-eclipsing sources, ones that show TEOs, and variable amounts of blending (flux contamination from other sources) in different TESS sectors. Each TESS sector is plotted in a different color.

TIC $405320687$ in the top-left panel is one of the largest amplitude heartbeats in our sample, with a fractional flux variation, $\delta F / F \gtrsim 2\%$. After modeling this system, we find it is also an outlier because of its very high eccentricity given its short orbital period. TIC $441626681$ in the top right panel shows TEOs with an amplitude of a few tenths of a percent at $N=18,21$ times the orbital frequency. In the bottom panel, TICs $59090149$ and $277236190$ are doubly and singly-eclipsing heartbeat systems, respectively. TIC $59090149$ shows modest changes in the relative eclipse depths between sectors $5$ and $44$. We do not model this system sector-by-sector, however, because there are several other systems with larger variations in their phase curves that we focus on in Section \ref{subsec:eclipse_drift}. Finally, the phase curve of TIC $277236190$ shows a higher amount of blending, or flux contamination,  in TESS sectors $14$ and $16$ than in sectors $41$, $55$, and $56$.

\begin{figure}
    \centering
    \includegraphics[width=0.43\textwidth]{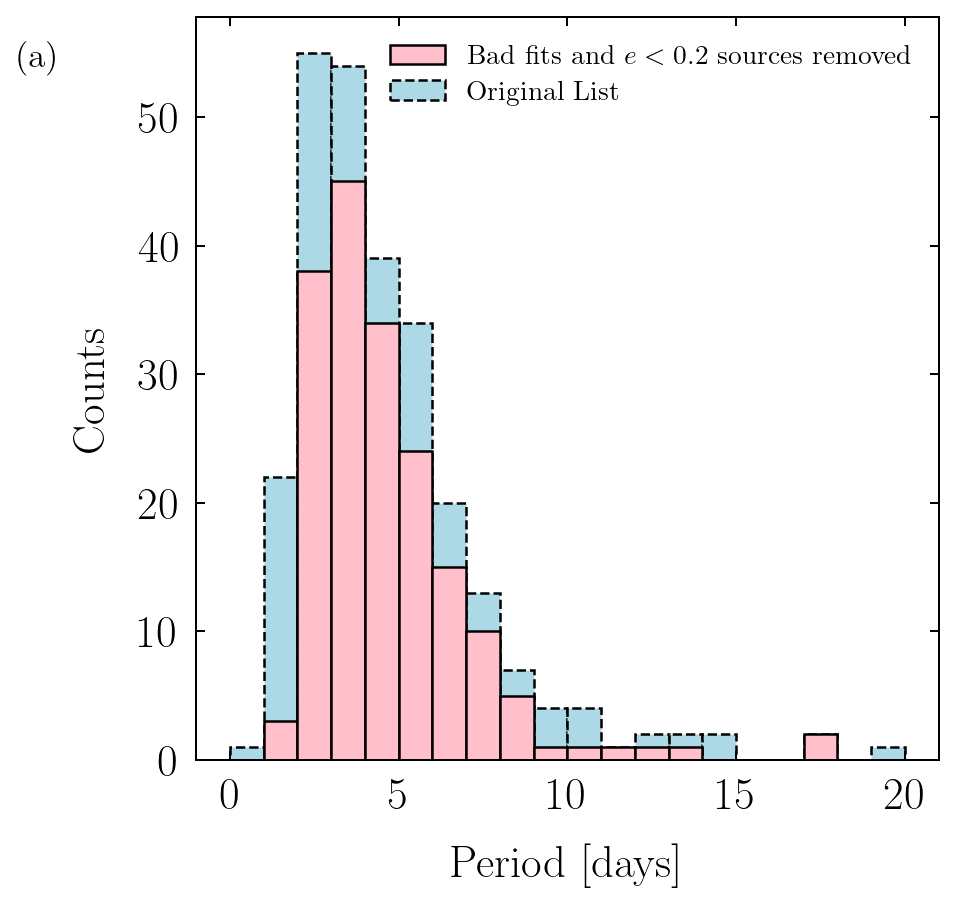}
    \includegraphics[width=0.47\textwidth]{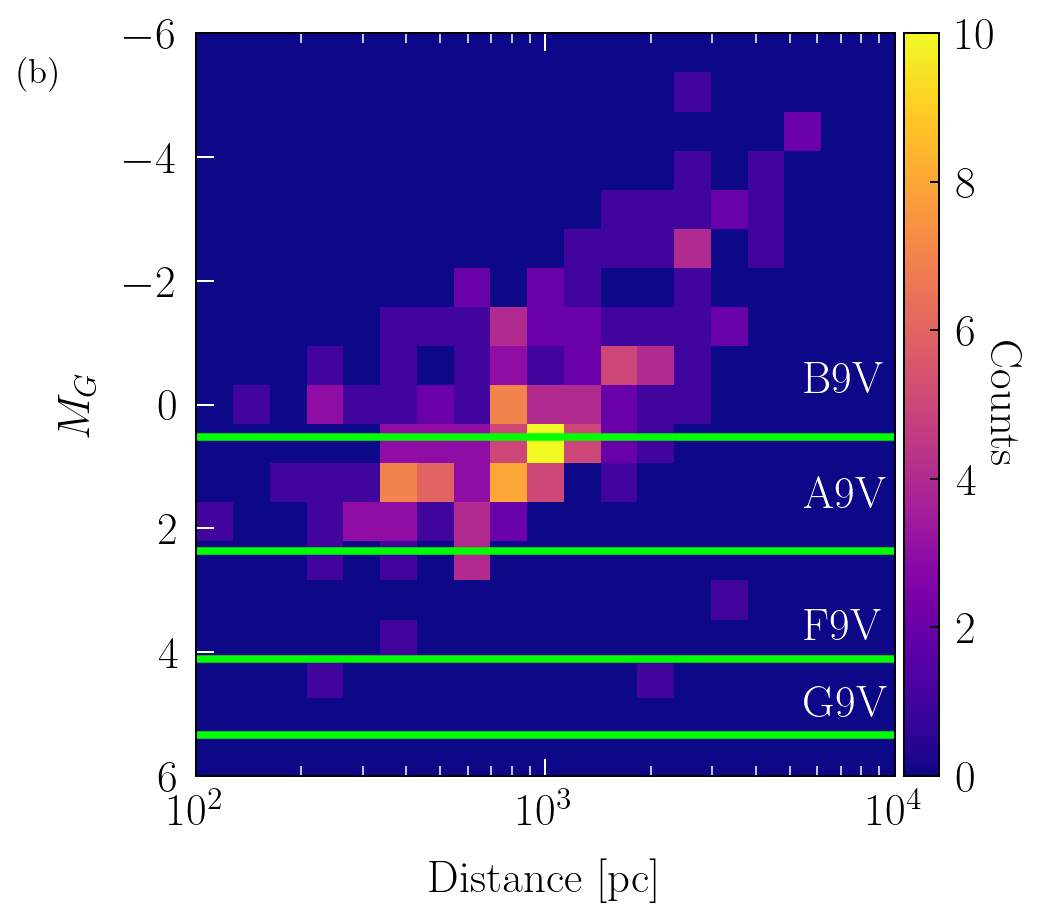}
    \caption{Histogram of the periods (panel a) and the 2D histogram of the absolute Gaia magnitude, $M_G$, and distance (panel b) for our heartbeat candidates. The period distributions of the original $240$ candidates and the final $180$ sources after applying the cuts are shown separately. Many short-period sources are excluded from the final HB list because they have either circularized or semi-detached or contact binaries without heartbeat-like light curves. The light curves in the final distribution have a mean period of $5.39$ days, with the shortest period being $P=1.47$ days and the longest being  $17.78$ days. We only show the magnitudes and distances for the $180$ HB systems. A and B-type stars dominate our sample, which may be a selection effect from the neural network classifier and a larger binary fraction for the massive stars.}
    \label{fig:periods_distances_magnitudes}
\end{figure}

Figure \ref{fig:periods_distances_magnitudes} contains the histogram of the periods of all the sources in panel (a). The period distributions of the original $240$ candidates and the final $180$ sources, obtained after applying the cuts, are shown separately in blue and pink, respectively. Many short-period sources are excluded from the final HB list because they have either circularized or have likely overfilled their Roche lobes, which is strongly suggestive by their phase curves. We are also unable to fit their phase curves using the heartbeat model. The light curves in the final distribution have a mean period of $5.39$ days, with the shortest period being $P=1.47$ days and the longest being $17.78$ days. 

In panel (b) of the same figure, we plot the 2D histograms of the extinction-corrected absolute magnitudes of the $180$ heartbeat systems and their distances, both taken from the TESS input catalog \citep{tic_catalog_2019AJ....158..138S}. The magnitudes are computed using the Gaia G-magnitudes ($m_G$) and distances available in the Gaia DR3 catalog. We apply a reddening correction to the Gmag using the catalog's $E(B-V)$ values, following \cite{tic_catalog_2019AJ....158..138S}. The different stellar types are separated by horizontal green lines as a function of the absolute magnitudes, where the $m_G$ thresholds are taken from \cite{pre-ms-colors_2013ApJS..208....9P}. Many sources are clustered around a distance of $1000$ pc from the Solar System, and have an average $m_G \sim 11$. The distribution of high stellar masses likely results from the neural network selectively identifying brighter TESS targets. Additionally, many of the TESS heartbeats that have been described in the literature are massive binaries, which may further highlight the preferential detection of these systems \citep{jayasinghe_2021MNRAS.506.4083J,tess_heartbeats_2021AA...647A..12K}.

\begin{figure}
    \includegraphics[width=0.46\textwidth]{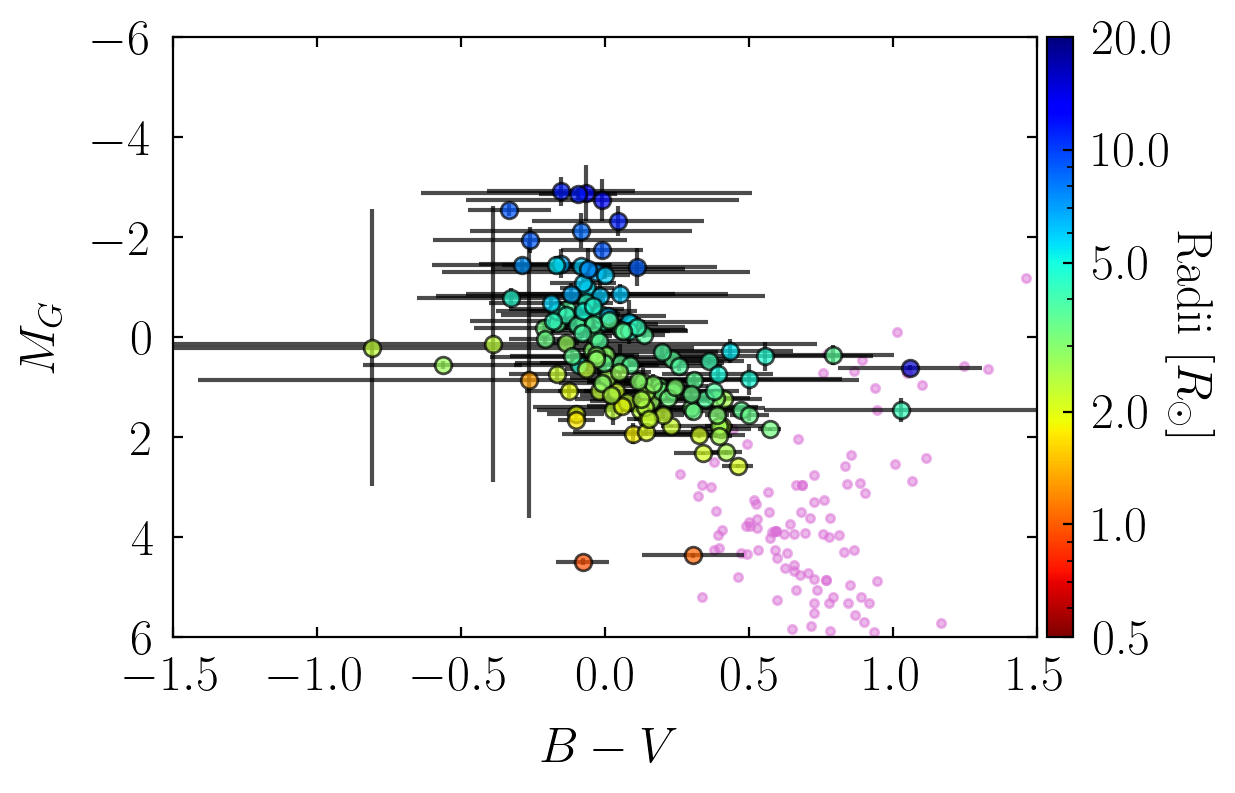}
    \caption{The color-magnitude values of the sources in the dataset, colored by the radius. Most of the systems have a small $B-V$ color, consistent with the population being primarily composed of A and B-type stars. The color is computed from the absolute magnitudes and distances taken from the TESS Input Catalog. We also plot the colors and magnitudes of roughly an equal number of randomly sampled TESS sources as pink points with $m_{TESS} <15$. These sources are, on average, have a larger $B-V$ than the systems in our sample. The radii are also taken from the TESS Input Catalog (TIC). These values can be very different from realistic ones since they are computed assuming that the sources are single stars.}
    \label{fig:color-mag-sources}
\end{figure}

In Figure \ref{fig:color-mag-sources}, we plot the $B-V$ color and magnitudes of the sources in our dataset. The points are colored by the values of stellar radii, also taken from the TESS Input Catalog. These radii should not be taken at face value since the systems are binaries, whereas the radii are calculated assuming the sources are single stars, derived from the Gaia parallax, magnitude, and inferred effective temperature \citep{tic_catalog_2019AJ....158..138S}. Nonetheless, we see a general trend that most of the heartbeat candidates in the data are bluer stars and have larger inferred radii, potentially increasing the ellipsoidal signal and making these systems easier to detect. About 48 sources do not have either $B$ or $V$ magnitudes or stellar radii information, and are therefore not plotted. For systems without the extinction information in the TIC, we set $E(B-V) = 1$ and use a conservative estimate of $\delta E(B-V) = 1$ for the error in the extinction. Therefore, the $4$ systems without $E(B-V)$ values have large error bars in their color and magnitudes.

Additionally, we plot $200$ randomly selected TESS sources as pink dots, with an apparent magnitude brighter than $m_{TESS}=15$, the same magnitude threshold used by the neural network to find HBs in the TESS data. The heartbeat candidates are bluer than the typical TESS sources, which is consistent with the population being primarily composed of massive A and B-type stars. 

\subsubsection{Sources with TEOs} \label{subsec:teos}
\begin{figure*}
    \centering
    \includegraphics[width=0.49\textwidth]{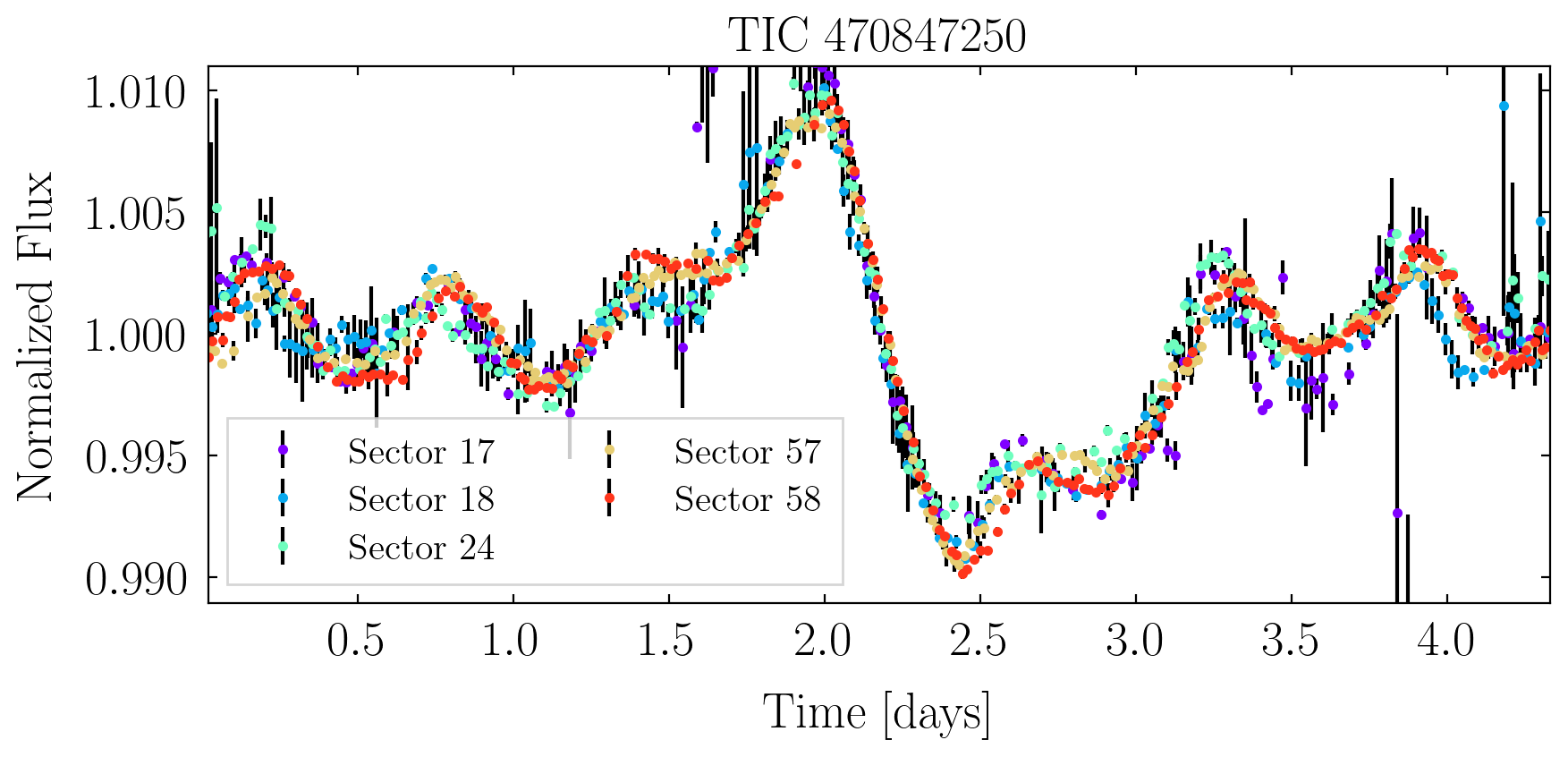}
    \includegraphics[width=0.49\textwidth]{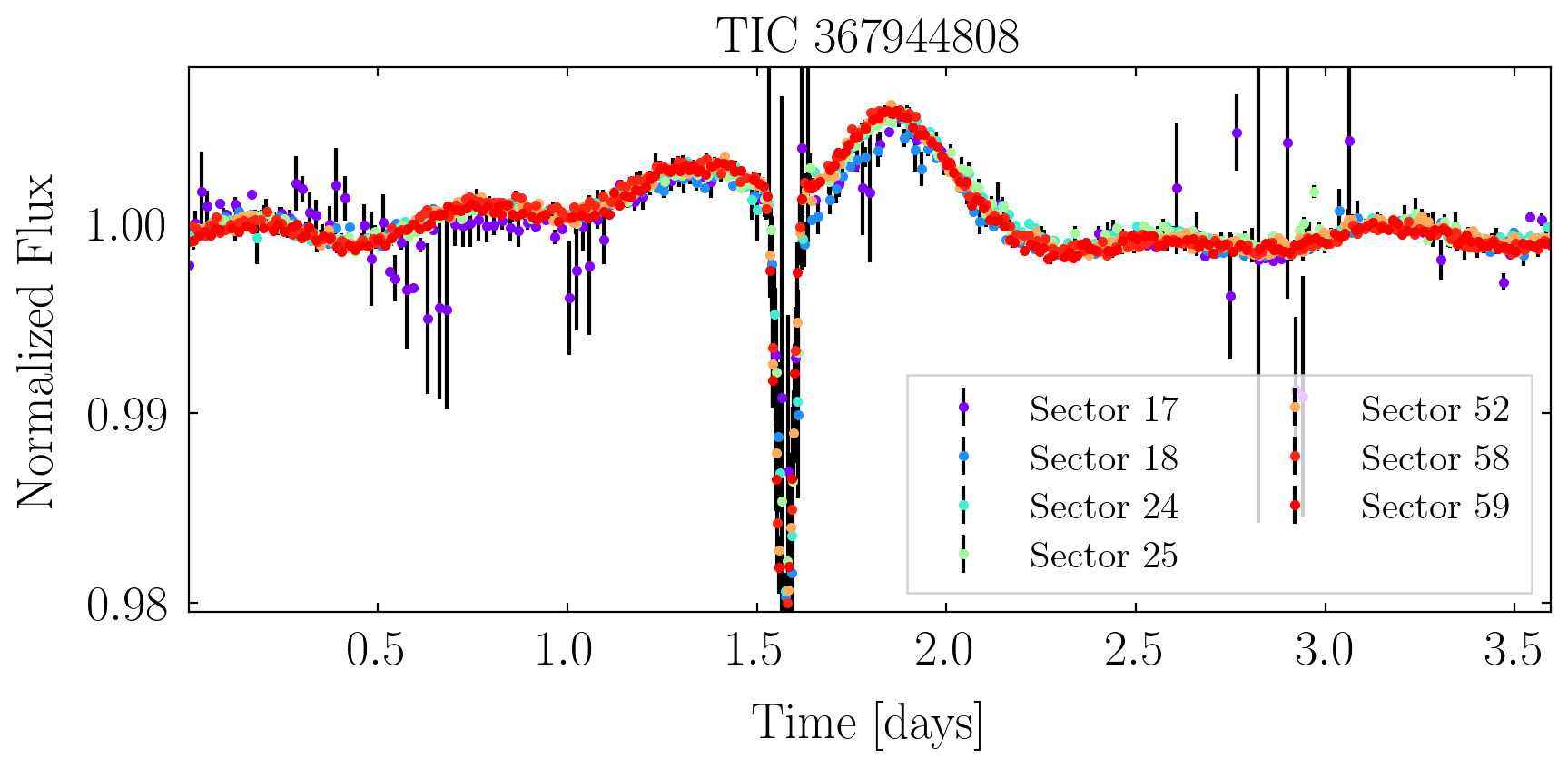}
    \includegraphics[width=0.49\textwidth]{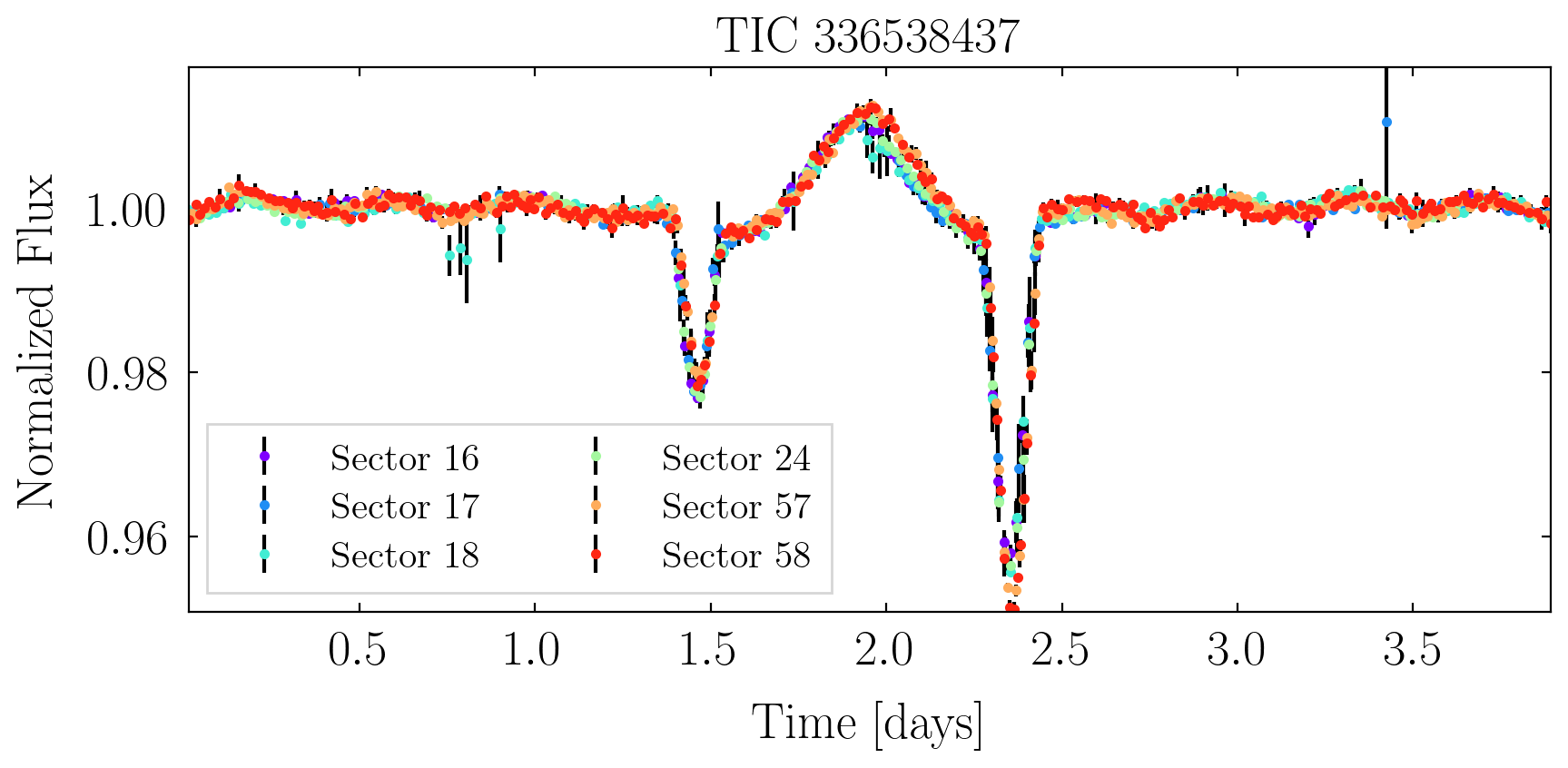}
    \includegraphics[width=0.49\textwidth]{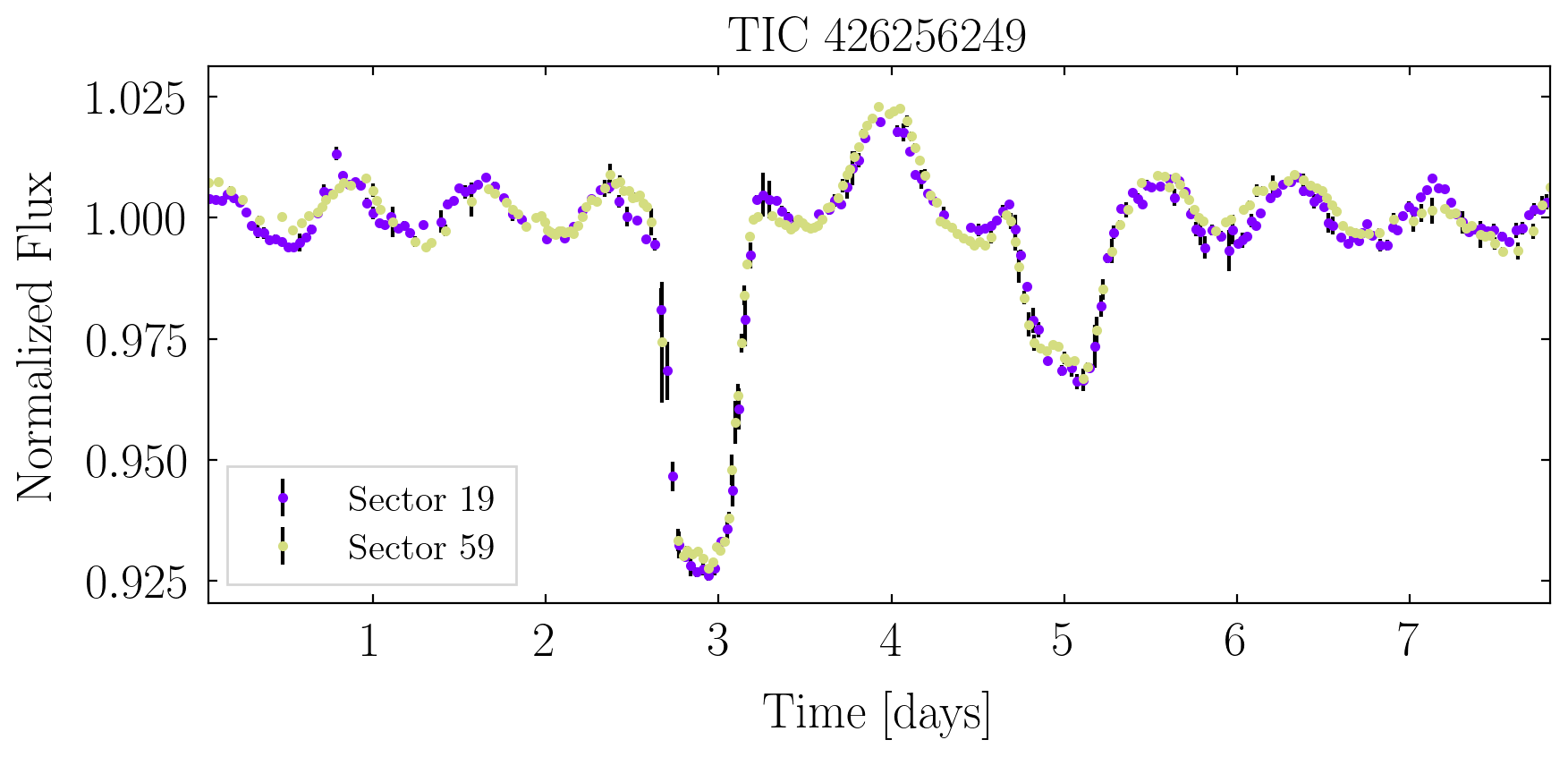}
    \caption{Some of the heartbeat systems have large-amplitude TEOs that are easily visible in their light curves, that occur on integer harmonics of the orbital period. The multi-sector data is useful in distinguishing TEOs from, e.g., starspots because, unlike the latter, TEOs should be persistent over a long observing window. We plot the phase-folded light curves of TICs 470847250, 367944808, 336538437 and 426256249. In addition to showing TEOs, TIC 336538437 shows variations in its phase curve profiles in different TESS sectors, possibly due to orbital precession from a tertiary companion.}
    \label{fig:HB_TEOs}
\end{figure*}

In general, all heartbeat systems should contain TEOs, although many are not directly visible in the light curves and need to be inferred from various periodogram techniques \citep{shelley_cheng_2020ApJ...903..122C}. With a large number of light curves and varying data quality in different sectors, making a quantitative statement about the fraction of systems with visible TEOs is difficult. Instead, we focus on a few interesting sources with large-amplitude TEOs and plot their light curves in Figure \ref{fig:HB_TEOs}. The flux variations from TEOs are visible through a `ringing' in the light curve, like in the case of TICs $470847250$, $367944808$, $336538437$, and $426256249$. In all these systems, the TEOs are seen at relatively lower harmonics of the orbital period ($n \lesssim 20$.). The TEO amplitudes can be comparable to the strength of the heartbeat signal, such as in the case of TIC $426256249$.

The multi-sector TESS data also helps us distinguish sources with TEOs from other features, such as starspots, the latter of which is not expected to last for multiple TESS sectors with a stable amplitude and phase.

\subsubsection{Short-Period Sources} \label{subsec:short_period_sources}
\begin{figure*}[htbp]
        \includegraphics[width=0.49\textwidth]{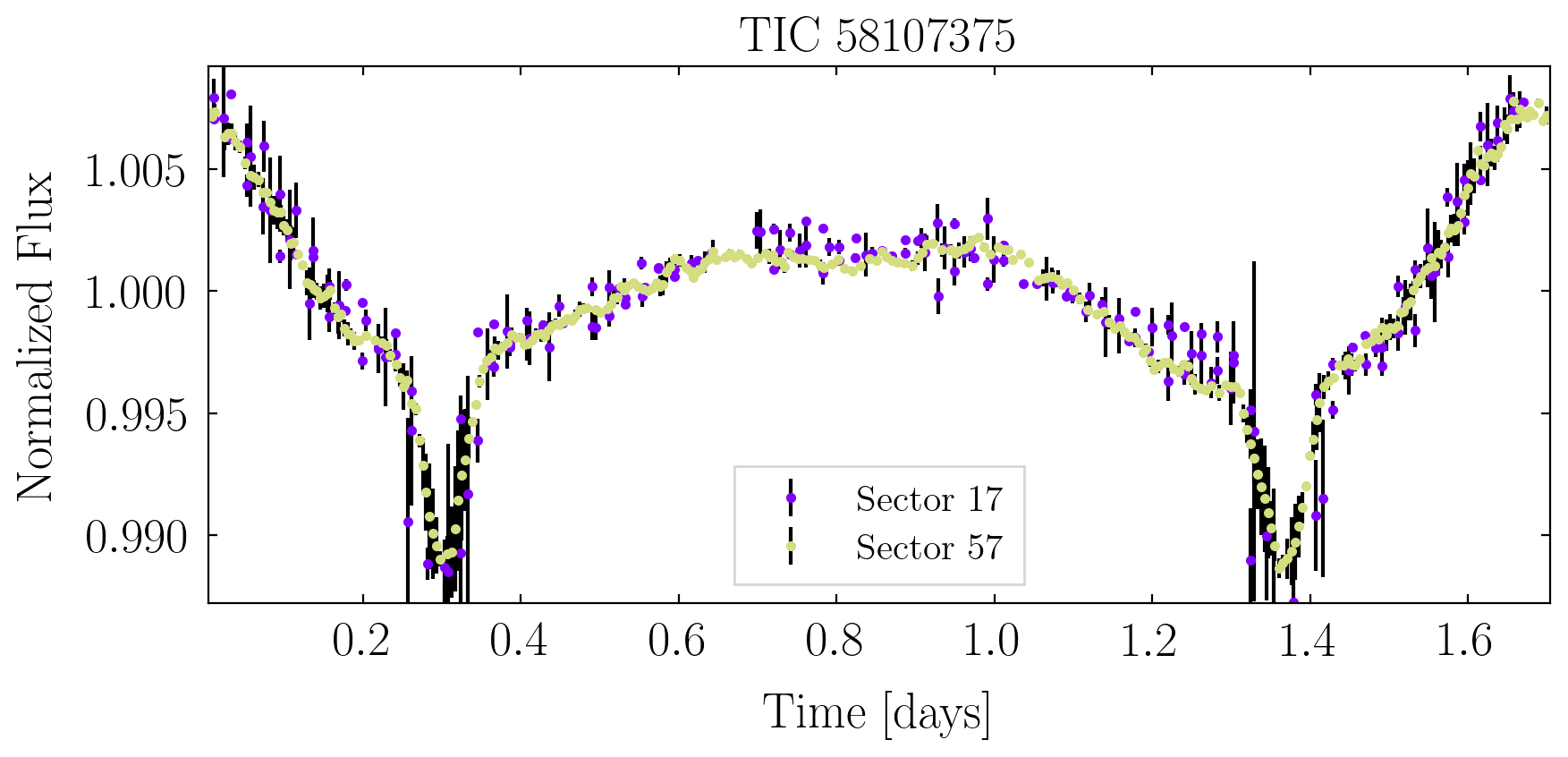}
        \includegraphics[width=0.49\textwidth]{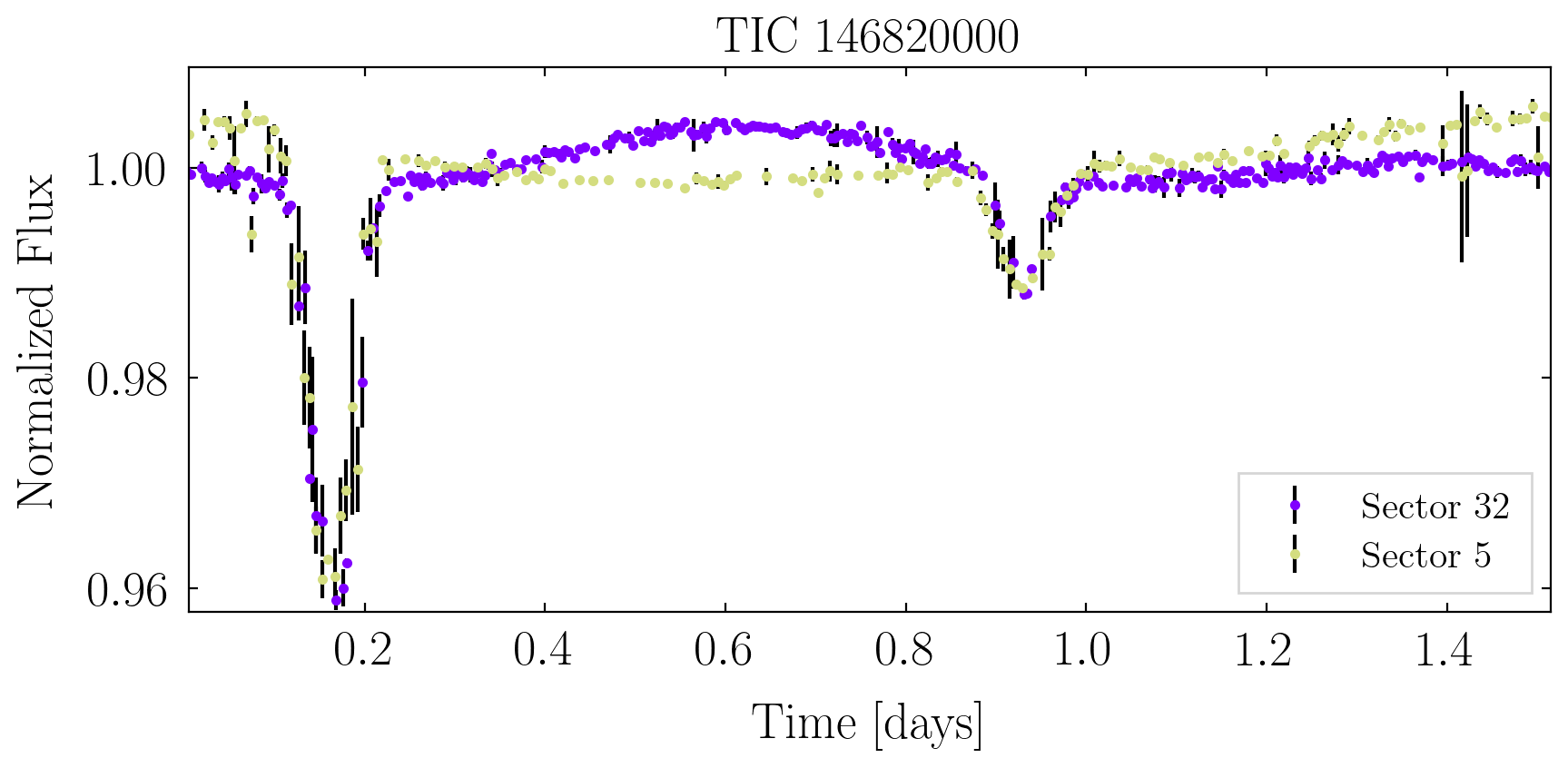}
        \includegraphics[width=0.49\textwidth]{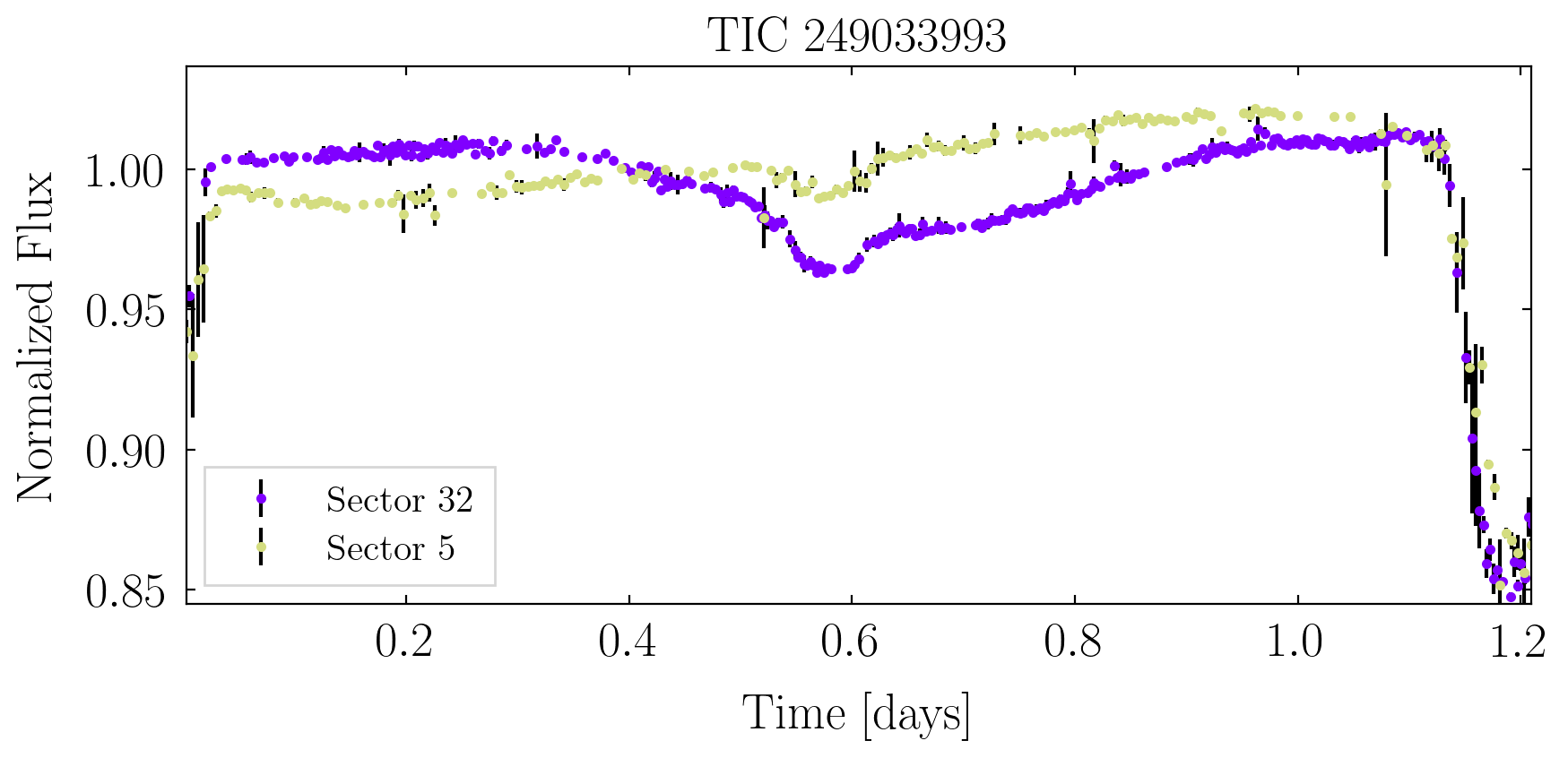}
        \includegraphics[width=0.49\textwidth]{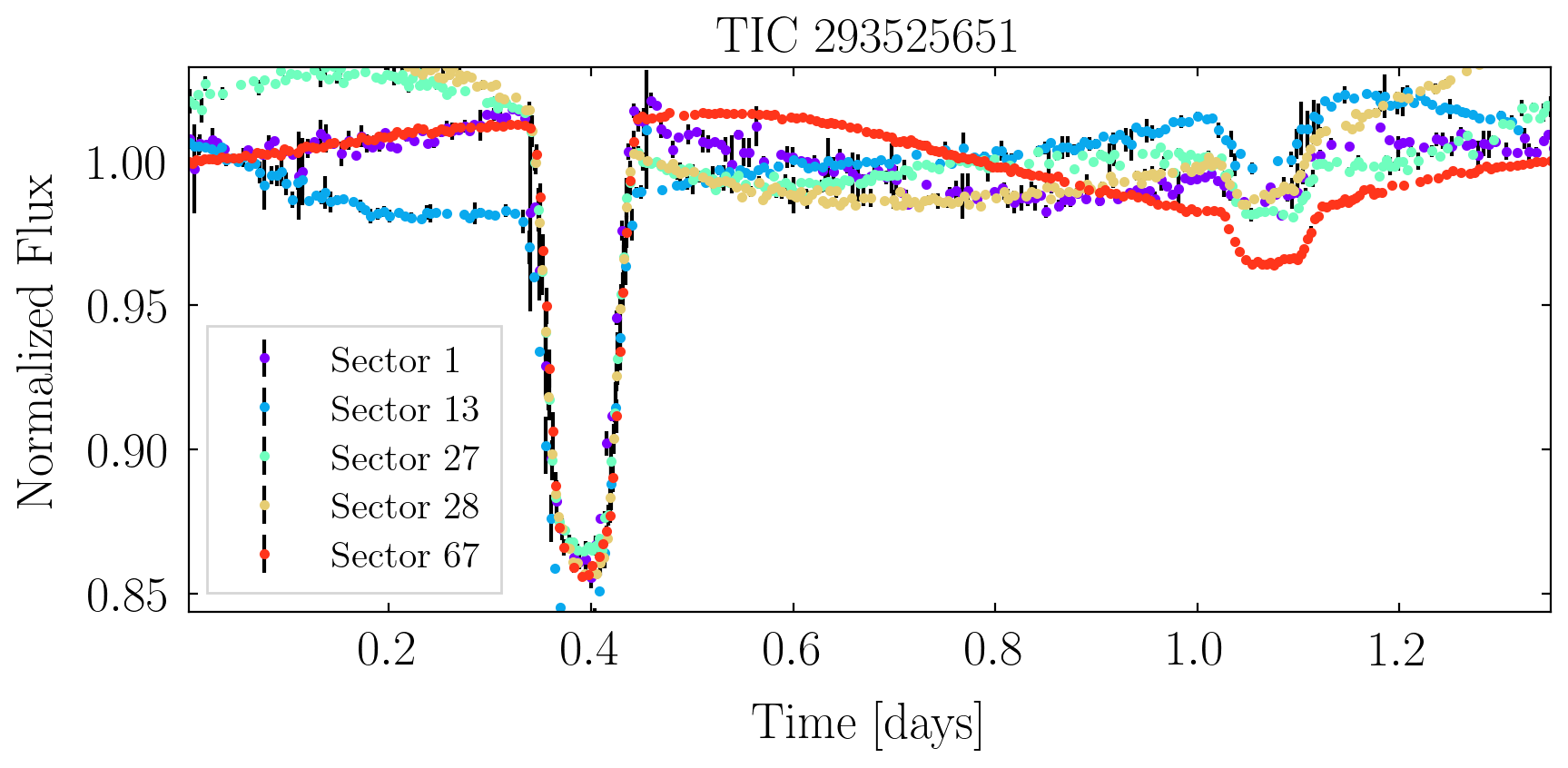}
        \caption{Some of the short period $(P < 1.75 \text{ days})$ systems that the network classified as heartbeat candidates. Many of these are likely semi-detached and contact binaries. Consequently, we could not fit the heartbeat model to these light curves.}
        \label{fig:short_period_lightcurves}
\end{figure*}

The neural networks classified several short-period $(P < 1.75 \text{ days})$ sources as potential heartbeat candidates. While these systems are eclipsing binaries, their lightcurves resemble those of semi-detached and contact binaries. Additionally, the heartbeat model, described in Appendix \ref{Appendix:ebeer_equations}, failed to give good fits for the light curves of these systems, as it had large discrepancies between the best-fit model and the light curves. We plot the light curves of several of these systems in Figure \ref{fig:short_period_lightcurves}. Many systems show significant variability throughout the light curve and sector-to-sector variations. For example, TIC $293525651$ is a doubly-eclipsing system that shows large sector-to-sector variations on top of the eclipses, which can result from e.g., orbital precession or possibly star spots \citep{star_spots_2017MNRAS.467.1830B}. These systems do not exhibit the typical heartbeat signal during their orbit. 

\section{Methods}\label{sec:methods}
\subsection{Light Curve Processing}\label{subsec:lightcurve_processing}

The light curves of the $240$ candidate HBs up to sector $67$ were generated using the \textsc{Eleanor} package. The fluxes, $F_i$, for every sector were normalized by dividing them by their median value in that sector, and points with $F_i < 0$ were removed. The TESS data for many systems also contained un-physical trails with large values of $F_i$ at the beginning and middle of many sectors. These points were removed by adopting a constant threshold of $F_{i, max} = 1.2$, where points with flux values exceeding $F_{i, max}$ were removed. None of the systems in our sample contained $F_i > 1.2$ in the heartbeat signal after they were normalized, and therefore, the threshold did not cut out any physically meaningful data.

For each source, the period was first estimated using the Lomb-Scargle periodogram from \textsc{astropy} \citep{astropy:2022}. It was then refined by folding the combined data of all sectors on a few trial periods centered around the original estimate. The period that minimized the dispersion in the phase curve of the combined data was then taken to be the period of the source. From this method, the typical fractional error in the period scales as $\Delta P / P \sim \Delta \phi P / \Delta t$, $\Delta \phi$ is the uncertainty in phase, of the order $10^{-2}$ (typically $1/$\# bins) and $\Delta t$ is the duration between the first and the last observation, which is around $10^3$ days for most systems. The systems that showed a variable inter-eclipse timing were folded on a period such that one of the eclipses lined up for all sectors. While this choice of period isn't at the exact radial period of the system, the difference between the two was less than $1$ part in $10^4$ for all such systems.

After estimating the period using the combined multi-sector data, the light curves from each sector were binned separately for the MCMC analysis. This was done to avoid combining the data from different sectors, potentially having different blending, i.e., flux contamination from other sources. Each bin contained $\text{max}(5, \text{int}\left[N/400 \right])$ points where $N$ is the number of data points in that sector. The latter constraint was employed to minimize the number of points per phase curve for, e.g., the high cadence data from the new TESS sectors. The error bar for each bin was computed by subtracting a linear fit from all of the data-points in the bin and computing the root mean squared deviation. The higher cadence data naturally had smaller error bars following the binning method. The binned phase curve for a given sector typically contained $\sim 200-400$ points. 

\subsection{Heartbeat Model} \label{subsec:eBEER_model}
The light curves of heartbeat systems are usually fit using the model in \cite{Kumar_1995ApJ...449..294K} under the assumption that the flux variations primarily result from tides raised on one of the stars. This works well if one of the stars is much more luminous than its companion, for example, a red giant. If the two stars are very similar, then contributions from both must be considered. Additionally, the effects of irradiation from the companion can be important for bright HB stars in their light curves \citep{faigler_mazeh_2011MNRAS.415.3921F,tess_heartbeats_2021AA...647A..12K,ogle_2_2022ApJ...928..135W}. These effects are taken into account by the `eBEER' (eccentric BEaming Ellipsoidal Reflection) model in \cite{Engel_2020MNRAS.497.4884E}, which includes the flux contributions from beaming, reflection, and ellipsoidal variations. The reflection contribution is computed by modeling the stars as ideal Lambertian surfaces \citep{lambert_2016arXiv161208846F,Engel_2020MNRAS.497.4884E}. 

We follow this model to estimate the orbital and other physical parameters of the HBs. Since most sources are eclipsing, we also add an eclipse contribution to the model, including limb-darkening effects. While adding the eclipsing terms makes the model different from the one presented in \cite{Engel_2020MNRAS.497.4884E}, we still refer to the model as the eBEER model. The equations for the flux contributions from each effect are provided in the Appendix \ref{Appendix:ebeer_equations}. We make the additional assumption that the spin frequencies of the two stars are equal to their angular velocities at the periastron. This prescription marginally differs from the pseudo-synchronous rotation rate given in \citep{hut_1981A&A....99..126H} by $\lesssim 10\%$ for $0 < e < 0.8$. The light curves depend on the eccentricity ($e$), inclination ($i$), argument of periastron ($\omega$) and the zero point phase ($\phi_0$). However, they are independent of the longitude of ascending node, ($\Omega$), and we fix it to be $0^{\circ}$ while fitting the light curves of the systems.

The stellar masses, radii, and temperatures are required to fit the light curves. We constrain the radius and temperature for each star using data from the TESS catalog. We fit a piece-wise polynomial to $\approx 3 \times 10^5$ TESS sources to construct mass-radius and mass-temperature relations. To allow for additional flexibility, we introduce radius and temperature re-scaling parameters, $\beta_{\text{R},j}, \beta_{\text{T},j}$ for each star $j$. 
These parameters are normalized such that for a given stellar mass, $68\%$ of radii (temperatures) of the $3 \times 10^5$ sources from the TIC are contained within the model computed radius (temperature) with $\beta_{\text{R}} (\beta_{\text{T}}) = \pm 1$. 
Additional details on mass-radius and mass-temperature relationships are provided in Appendix \ref{Appendix:Temperature_radius_estimatation}. The light curves further depend on the limb and gravity-darkening parameters. The flux contributions from beaming and reflection effects are further parameterized by dimensionless parameters (see e.g., \cite{faigler_mazeh_2011MNRAS.415.3921F}), which we keep as free model parameters.

For some of the systems, different TESS sectors contain different amounts of blending in the data. Additionally, they have a different number of data points per bin because of potentially different numbers of data points in the different sectors resulting from different sampling cadences. To address that, we add three additional parameters to every TESS sector: blending, noise re-scaling, and flux normalization.

To summarize, the model is parameterized by the stellar masses $(M_1, M_2)$, the orbital parameters $(e, i, \omega$ and $\phi_0)$, the temperature and radius re-scaling parameters for the two stars $(\beta_{\text{T,1}}, \beta_{\text{T,2}}, \beta_{\text{R,1}}$ and $\beta_{\text{R,2}})$, the gravity and limb-darkening parameters of the stars $(\mu_1, \tau_1, \mu_2$ and $\tau_2)$, the beaming and reflection coefficients for each star $(\alpha_{\text{beam},1}, \alpha_{\text{beam},2}, \alpha_{\text{ref},1}$ and $\alpha_{\text{ref},2})$,  and the three sector blending, flux normalization, and noise re-scaling parameters $(\delta_s, \Sigma_s \text{ and } \sigma_s)$. The semi-major axis is computed from the stellar masses and the orbital period, which is fixed during MCMC fitting. As stated previously, the model is independent of the longitude of the ascending node, $\Omega$. 

The light curve model is then fit to the binned, multi-sector phase curves simultaneously using MCMC, where the model uses the same physical parameters for all of the sectors but contains the three sector-specific parameters per TESS sector. The initial fits to the phase curves revealed that there is a potential degeneracy in constraining the stellar masses, radii, and temperatures using photometry alone. To obtain better estimates of the masses, we use the Gaia magnitudes and distances for all sources to estimate the absolute magnitude of the system, which we then use to constrain the stellar properties by assuming that the stars are blackbodies with temperatures $T_j$ and radii $R_j$. Better constraints on the stellar masses also help us constrain other orbital parameters. We expect modeling the stars as blackbodies to be a good assumption given that most of the stars in our sample are A and B-type stars, which do not contain many absorption and emission lines. Rather, we expect the dominant source of uncertainty to come from flux contamination from other sources, i.e., blending, which we account for using the blending parameter. This should not have a major impact on other orbital parameters such as inclination, eccentricity, and the argument of periastron, all of which depend sensitively on the shape of the light curve and not its normalization. 

\begin{figure*}
    \centering
    \includegraphics[width=0.99\textwidth]{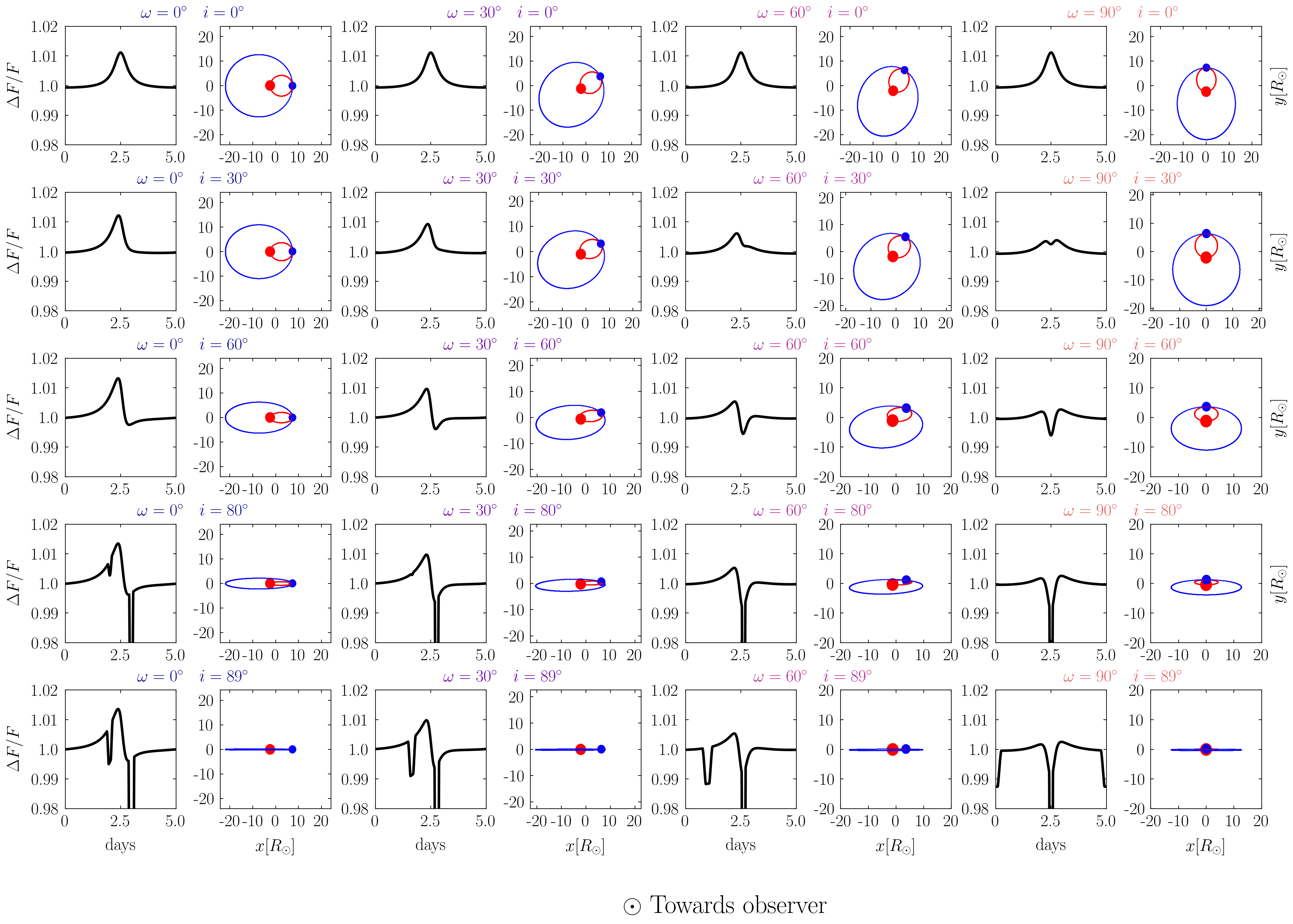}
    \caption{The light curve and orbits of $3M_{\odot}$ (red) and $1M_{\odot}$ (blue) stars that are constructed using the eBEER model, plotted as functions of the argument of periastron and inclination. The observer is located out of the plane of the page. All orbits have an eccentricity, period, and longitude of ascending node of $0.5$, $5$ days, and $90^{\circ}$ respectively. The shape of the light curves can change depending on the observer's position without any changes in the orbits themselves. At lower inclinations, the light curves are unaffected by the changing argument of periastron, making it difficult to estimate the orbital parameters. Likewise, the orbital parameters of the more inclined systems can be more strongly constrained.}
    \label{fig:orbits_ebeer_model}
\end{figure*}

Figure \ref{fig:orbits_ebeer_model} shows synthetic light curves and the corresponding orbits of a $3M_{\odot}$ and $1M_{\odot}$ star constructed using the eBEER model. They are plotted as functions of the argument of periastron and inclination, and the observer is located out of the plane of the page. The eccentricity ($e$), orbital period ($P$), and longitude of the ascending node ($\Omega$) are taken to be $0.5$, $5$ days, and $90^{\circ}$, respectively, and the light curves with the same angle of periastron are plotted with the same color. Note that while the light curves are independent of $\Omega$, it is required to plot the orbits. The orbits are plotted on the right of the light curves, where the red and blue colors correspond to the orbits of the $3$ and $1 M_{\odot}$ stars. The stars are depicted by circles at the periastron of the orbit, which occurs at $t=2.5$ days for all of the panels. While the physical orbits are the same in all panels, the light curves change depending on the observer's viewing angle, which is what enables the orbital parameters of the HB systems to be constrained even if they are not eclipsing. There is a degeneracy in determining the argument of periastron from the light curves at low inclinations, which get lifted for more inclined orbits. Additionally, at higher inclinations, when the argument of periastron is not close to $0^{\circ} (\text{ or } 180^{\circ})$ the heartbeat signal can add constructively to one of the eclipses and destructively to the other one. This can make what would be otherwise a doubly-eclipsing system look like a singly-eclipsing one. A similar trend is discussed in \cite{ogle_2022ApJS..259...16W}.

The fitting is done with a custom MCMC wrapper. The wrapper is OpenMP-parallelized and written in C for faster likelihood evaluations. Every run is initialized with $50$ walkers with different temperatures. We use a combination of uniform and differential-evolution sampling. The MCMC wrapper is described in more detail in Appendix \ref{Appendix:MCMC}.


\section{Light Curve Fits} \label{sec:lightcurve_fits}

\subsection{eBEER Fits on Systems with Pre-constrained Orbits} \label{subsec:comparison with observations}
We first test the eBEER model by fitting it on the light curves of systems that already have their orbits relatively well-constrained through radial velocity measurements. To that end, we take $19$ heartbeat systems from \cite{shporer_2016ApJ...829...34S} and perform the MCMC fitting on their publicly available Kepler lightcurves via the Mikulski Archive for Space Telescopes (MAST) Portal \footnote{https://archive.stsci.edu/missions-and-data/kepler}. The analysis could not be performed on the TESS light curves for the same systems because they had an extremely low signal-to-noise ratio and almost all of them could not be identified as heartbeats from a visual inspection.

\begin{figure*}
    \includegraphics[width=0.49\textwidth]{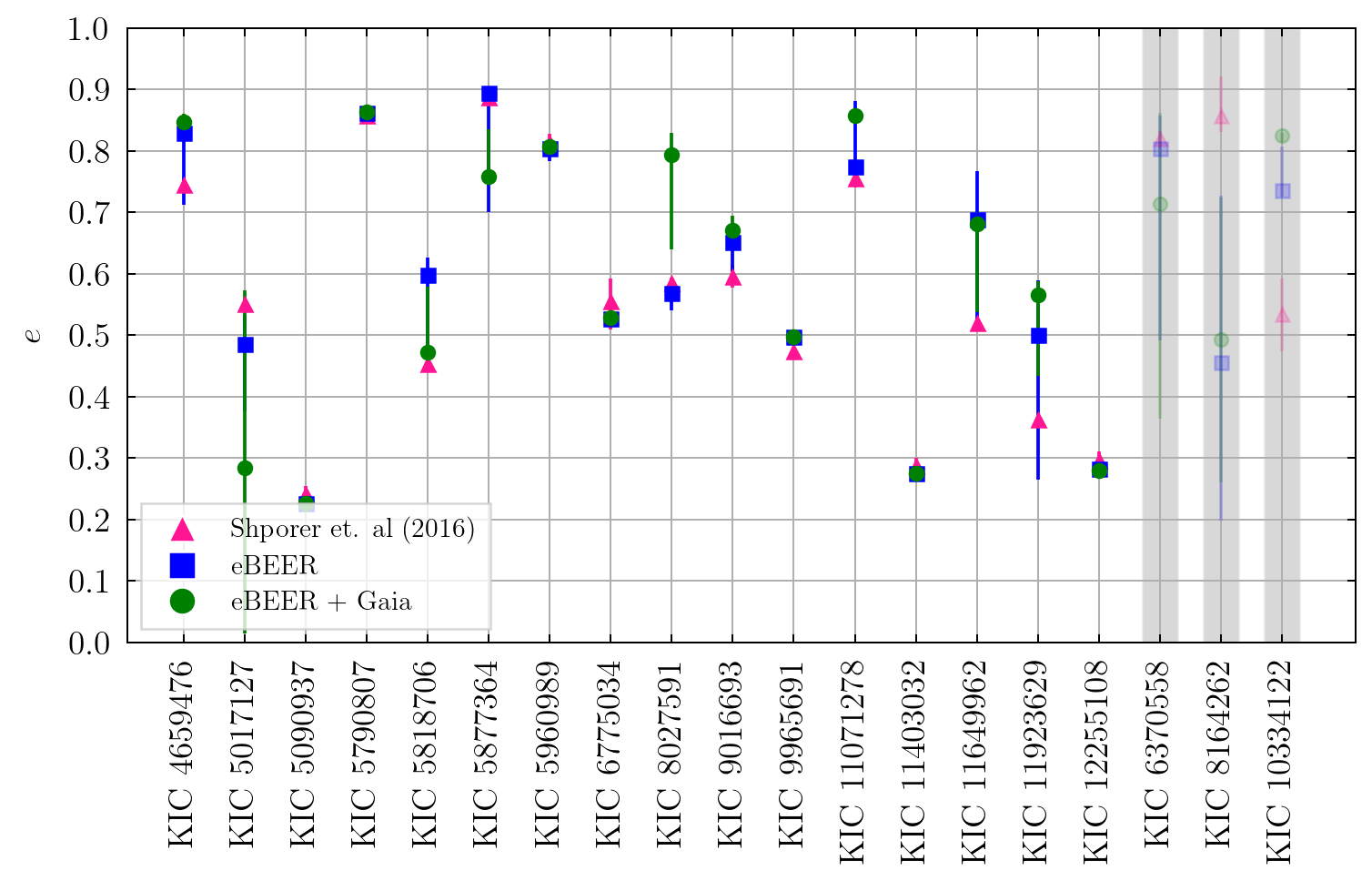}
    \includegraphics[width=0.49\textwidth]{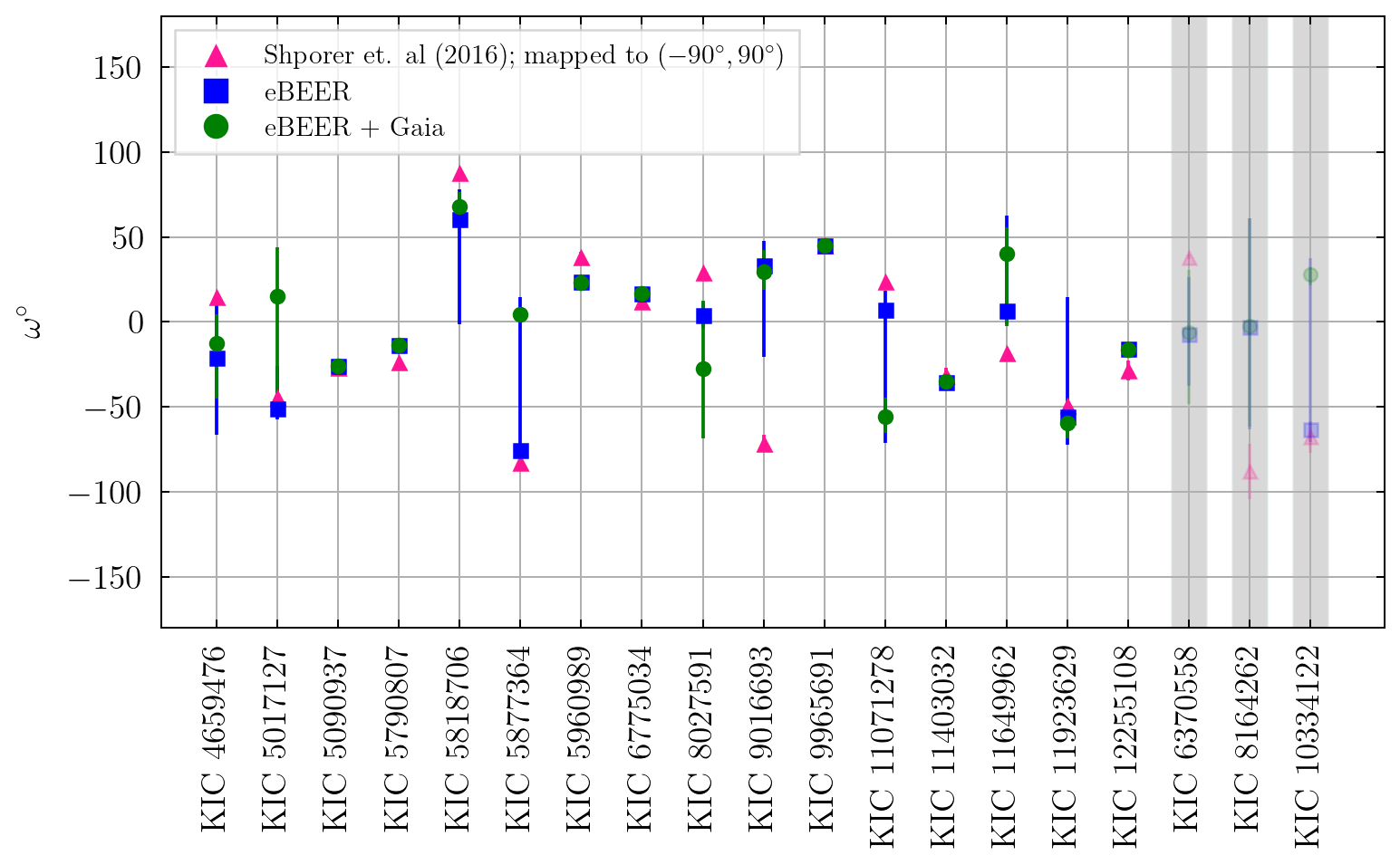}
    \caption{The estimates of eccentricities (left) and arguments of periastron (right) for the $19$ heartbeat binaries taken from \cite{shporer_2016ApJ...829...34S} using the eBEER model. We perform two sets of fits for each of the systems. In the first case (blue squares), we fit the eBEER model to the light curves only, and in the second case (green circles), we also fit the model to the corresponding Gaia absolute magnitudes of the systems, assuming that the stars are blackbodies. 
    The error bars for the eBEER fits represent the $16^{\text{th}}$ to $84^{\text{th}}$ percentile values in the posteriors, and the error bars of the pink triangles represent the $\pm 1 \sigma$ uncertainties taken from \cite{shporer_2016ApJ...829...34S}. There is a degeneracy between $\omega$ and $\omega + 180^{\circ}$ in the eBEER model posteriors that is removed by plotting $\arctan(\tan(\omega))$ for the model posteriors and the \cite{shporer_2016ApJ...829...34S} measurements.}
    \label{fig:shporer_comparison}
\end{figure*}

Two series of fits are performed on all the light curves. In the first case, we fit the model to only the systems' light curves, and in the second case, we also fit it to the Gaia absolute magnitude. We first compare the eccentricities ($e$) and the arguments of periastron ($\omega$) between the posteriors from our model and that taken from \cite{shporer_2016ApJ...829...34S} in Figure \ref{fig:shporer_comparison}. The eBEER model has a degeneracy in $\omega$, where $\omega \longleftrightarrow \omega + 180^{\circ}$ gives the same light curves under the exchange of the stellar masses $M_1 \longleftrightarrow M_2$. Additionally, when the stellar masses, radii, and temperatures are close to equal, the degeneracy persists under the exchange of the reflection parameters $\alpha_{\text{ref},1} \longleftrightarrow \alpha_{\text{ref},2}$, i.e., by making either of the stars brighter. We, therefore, limit the range of $\omega$ to $[-90^{\circ}, 90^{\circ}]$ by performing the transformation $\omega \longrightarrow \arctan(\tan(\omega))$ on the posteriors and the \cite{shporer_2016ApJ...829...34S} data.

The $e$ and $\omega$ values from the fits to only the light curves are shown as blue squares, and from the fits that also include the Gaia magnitude information are plotted as green circles. The results from \citep{shporer_2016ApJ...829...34S} are plotted as pink triangles, where the $e$ and $\omega$ are determined from the combined photometric and radial velocity measurements of the Kepler systems. We could not obtain good fits for the light curves of KICs $6370558$, $8164262$, and $10334122$ using the eBEER model. This was partially because these light curves had a relatively lower signal to noise ratio compared to the other Kepler sources. These three sources are highlighted with gray bands on the right of both panels in Figure \ref{fig:shporer_comparison}. The error bars for the eBEER fits represent the $1 \sigma$ confidence range from the model posteriors. Similarly, the error bars for the pink triangles are the $1 \sigma$ deviations taken from \cite{shporer_2016ApJ...829...34S}. The eccentricities and the remapped arguments of periastron from the eBEER fits are very close to the published values, where most measurements are within $1 \sigma$.
Adding Gaia magnitude and distance information to the data gives consistent estimates $e$ and $\omega$ and results in better constraints on the stellar masses and temperatures as we show next.

\begin{figure}
    \centering
    \includegraphics[width=0.49\textwidth]{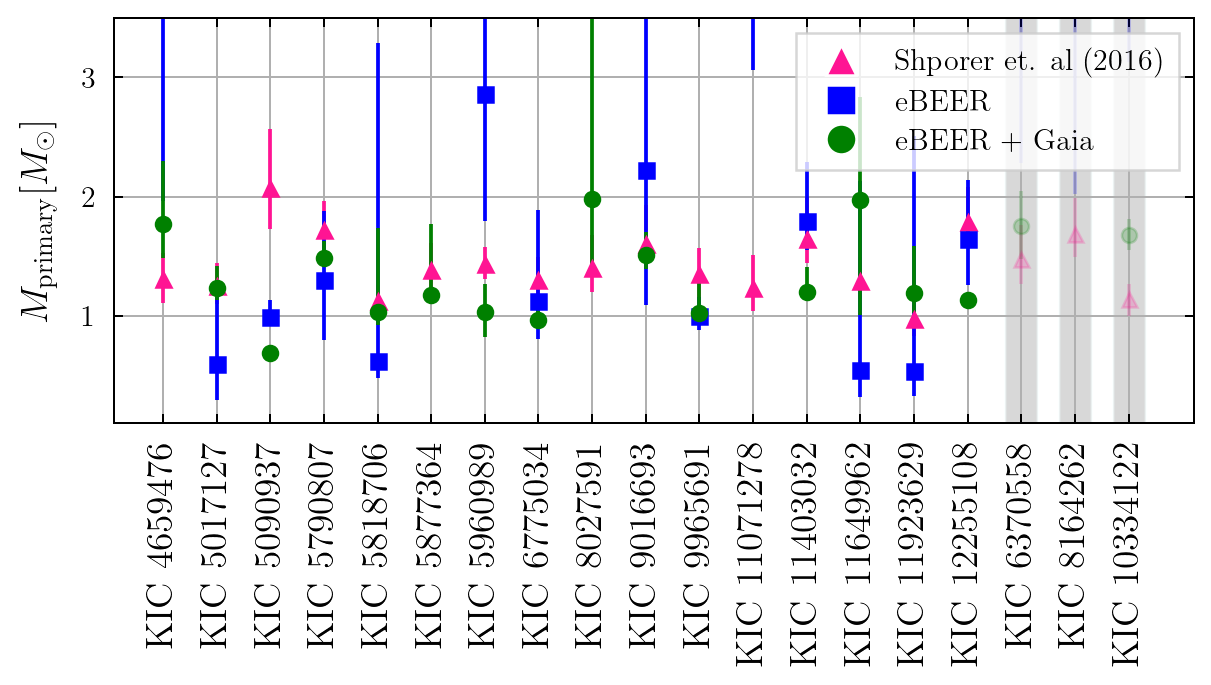}
    \includegraphics[width=0.49\textwidth]{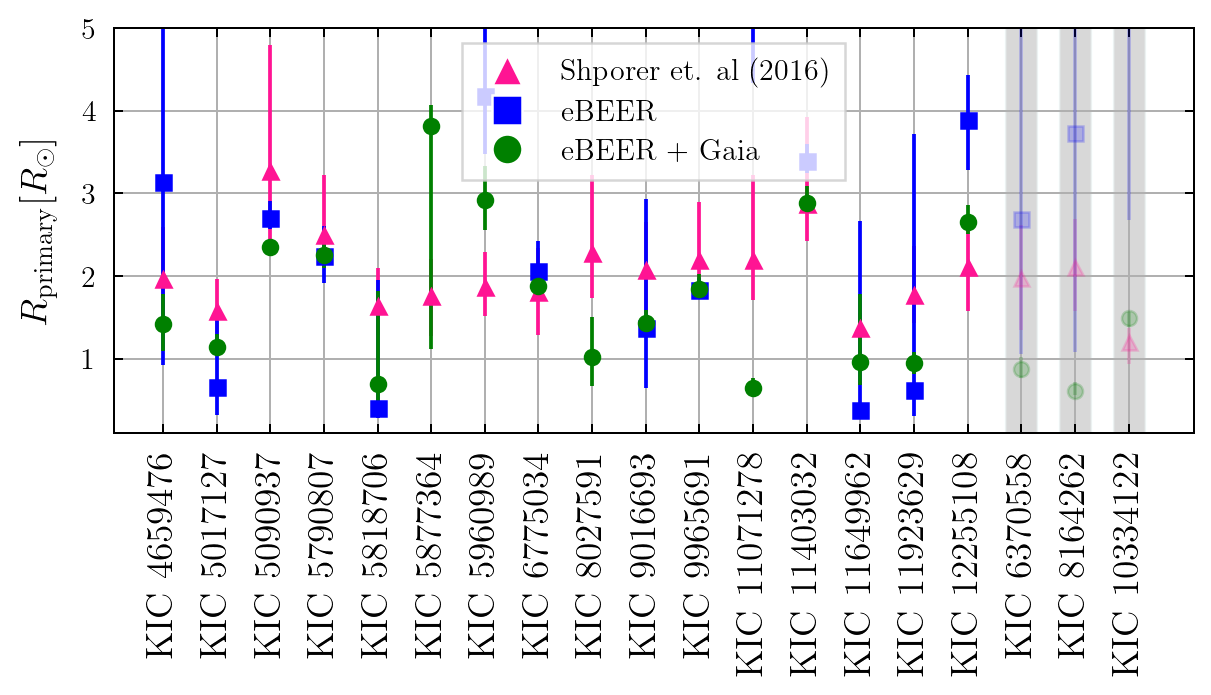}
    \caption{We compare the primary masses and radii estimated from the eBEER and eBEER + Gaia magnitude fits with the measurements from \cite{shporer_2016ApJ...829...34S}. The figure has the same format as Figure \ref{fig:shporer_comparison}. 
    }
    \label{fig:shporer_comarison_2}
\end{figure}

We now compare the eBEER estimates of the primary masses and radii with measurements from \cite{shporer_2016ApJ...829...34S} in Figure \ref{fig:shporer_comarison_2}. Jointly fitting the Gaia magnitudes and the phase curves gives much better estimates of the stellar masses and radii. The measurements for most of the sources are within $\pm 1\sigma$ uncertainties. Additionally, even when the eBEER mass estimates deviate significantly from the observed values, such as for KIC 11071278, the predicted eccentricities are close to the observed values. The eccentricity measurements are robust despite the large uncertainties in the stellar parameters.

\subsection{Population Demographics} \label{subsec:fits_to_multi_sector_tess_data}
We fit the binned, multi-sector TESS phase curves for all $187$ heartbeat candidates with the eBEER model. Since adding Gaia magnitudes helps constrain the orbital parameters better, we include it for all fits. The multi-sector data are then jointly fit so that the phase curves from all the sectors share the same physical parameters but are fit with a different blending, flux normalization, and noise re-scaling parameter for every sector. Therefore, there are a total of $19 + 3N_{s}$ parameters, where $N_s$ is the number of TESS sectors in the combined data for that system (see Table \ref{tab:orbital_parameters}). 

\begin{figure*}
    \centering
    \includegraphics[width=0.75\textwidth]{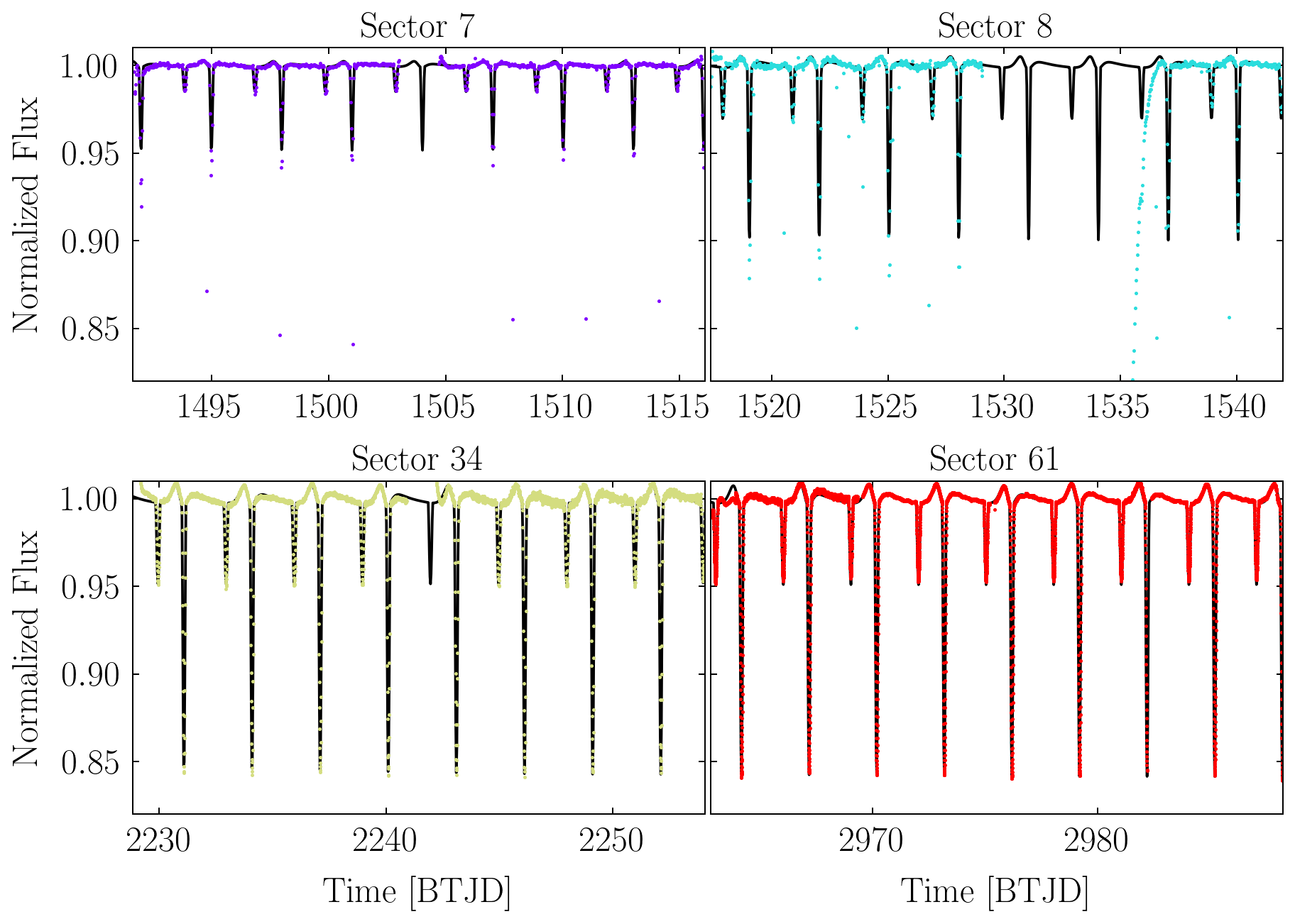}
    \caption{The TESS data for TIC $150284425$ in different sectors, plotted as colored dots on the corresponding best-fit model, shown as the solid black curve. The light curves in the earlier sectors have a lower sampling cadence of one datapoint every $30$ minutes in sectors $7-8$, which goes up to one datapoint every $200$ seconds for sector $61$. Additionally, the earlier sectors have a larger blending, resulting in shallower minima for the eclipses and a smaller amplitude heartbeat signal.  }
    \label{fig:fits_to_multi_sector_data}
\end{figure*}

\begin{figure*}
     \includegraphics[width=0.99\textwidth]{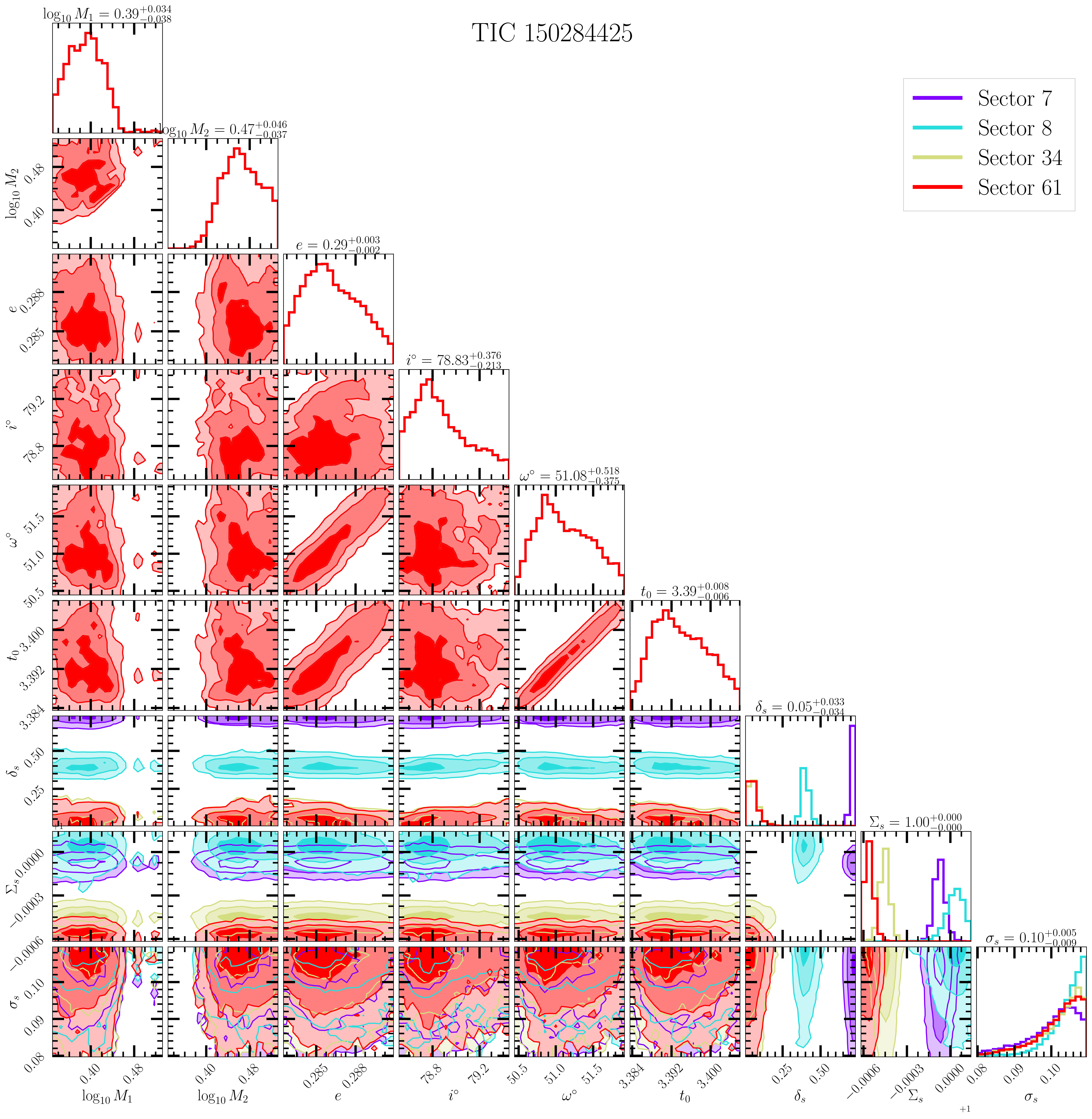}
    \caption{The corner plot for the fit to the multi-sector data of TIC 150284425. All sectors are fit simultaneously with the same physical parameters but a different blending ($\delta_s$), flux normalization ($\Sigma_s$), and noise-rescaling ($\sigma_s$). Therefore, the posteriors for $\log M, e, i, \omega$ and $t_0$ overlap for all 4 sectors. The blending is much higher in sector $7$ than in sectors $34$ and $61$, which is expected from the light curves in Figure \ref{subsec:fits_to_multi_sector_tess_data}.}
    \label{fig:corner_multi_sector_data}
\end{figure*}

To illustrate the importance of sector-specific blending, we plot the light curve model using the best-fit parameters for one of the sources, TIC $150284425$, in Figure \ref{fig:fits_to_multi_sector_data}.  Every panel contains data from a different TESS sector, plotted as colored dots, and the corresponding best-fit model is represented by a solid black curve. The break in the TESS data in the middle of every sector results from downlinking \footnote{https://heasarc.gsfc.nasa.gov/docs/tess/the-tess-space-telescope.html} the data back to Earth. Sectors $7$ and $8$ show a larger amount of blending, as evidenced by the relatively lower eclipse depths and a weaker heartbeat signal at the periastron than in sectors $34$ and $61$. Additionally, the $30$ minute cadence data in the earlier sectors have very few points capturing the primary eclipse, resulting in smaller minima and larger error bars during the primary eclipse once the data are phase-folded and binned. Despite the seemingly different light curves in different sectors, the eBEER model is able to fit the data from all sectors using the same physical parameters but different blending, flux normalization, and noise re-scaling parameters for every sector, providing additional confidence in the validity of the model. The corresponding posteriors of some model parameters are shown in the corner plot in Figure \ref{fig:corner_multi_sector_data}. The predicted blending ($\delta_s$) is much higher in sector $7$ than in sectors $34$ and $61$, which we expect from the light curves. All the parameters without the subscript `s' are identical between all sectors and hence plotted with the same color. 

Next, we present the population statistics of all the systems whose phase curves could be fit with the eBEER model. These include $180$ HBs and $29$ other binaries with estimated eccentricities less than $0.2$, which the NN flagged as HBs. The best-fit parameters determined from the posteriors are presented in Appendix \ref{appendix:list_of_hb_candidates}.

\begin{figure}
    \centering
    \includegraphics[width=0.49\textwidth]{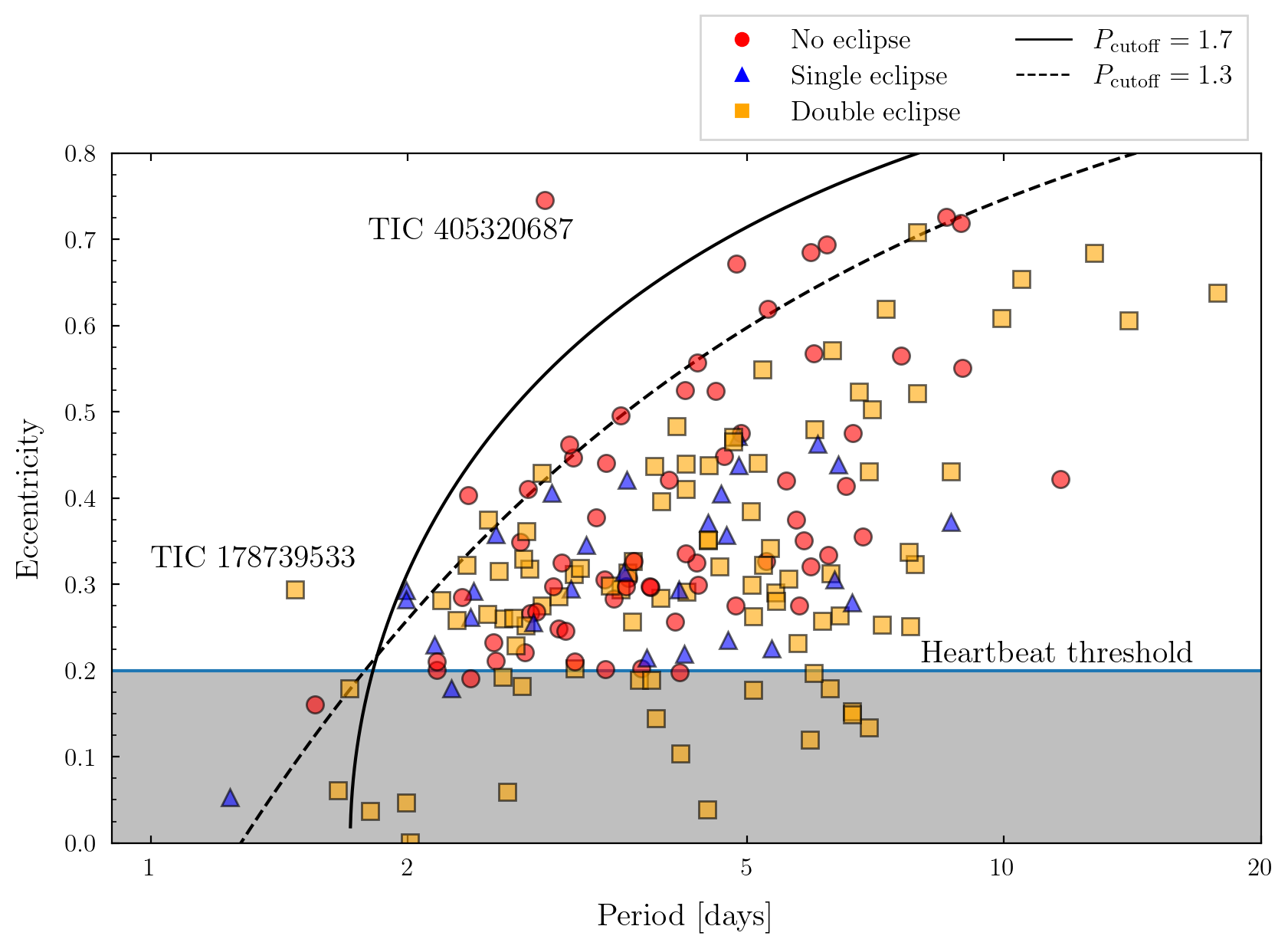}
    \caption{The period-eccentricity (PE) distribution for the HB candidates, where the eccentricities are computed using the posteriors from the eBEER model. The sources are further differentiated based on the number of eclipses that they show per orbit. The heartbeat signal gets progressively weaker at low eccentricities; therefore, all of the non-eclipsing candidates have eccentricities $\gtrsim 0.2$. The solid and dashed black curves represent curves of constant specific angular momentum and periastron distance, which are fit to the most eccentric sources in the period range of $1-10$ days, excluding the two outliers, TICs $405320687$ and $178739533$. These two systems stand out from the rest of the population because of their high eccentricity, given their orbital periods. TIC $405320687$ is a non-eclipsing heartbeat system whose phase curve is shown in Figure \ref{fig:heartbeat_classes}. TIC $178739533$ is a doubly-eclipsing system and shows a large change in its eclipse depths in different TESS sectors. We discuss this source with the $11$ additional sources showing orbital precession in Section \ref{subsec:eclipse_drift}. }
    \label{fig:period_ecc}
\end{figure}

Since HBs typically have very high eccentricities for their orbital periods, the period-eccentricity (PE) relation for such systems can offer insight into their evolution. Therefore, in Figure \ref{fig:period_ecc}, we plot the periods and eccentricities of the sample, where the latter are computed using the $50^{\text{th}}$ percentile MCMC posteriors. We further differentiate the sources based on the number of eclipses that they show per orbit. Systems with two eclipses are plotted using yellow squares, ones with a single eclipse are plotted as blue triangles, and non-eclipsing systems are plotted using red circles. The eccentricities of all non-eclipsing sources are greater than $\sim 0.2$ because the heartbeat signal near the periastron gets progressively weaker at lower eccentricities, making such systems harder to detect. Note that many doubly-eclipsing systems in the figure are below this minimum threshold; the neural networks can still easily identify these systems from their eclipses even though they do not show a strong HB signal. The break at $\sim 1.7$ days in the PE diagram reflects the lack of good fits to the light curves at shorter periods. Systems with shorter orbital periods typically overflow their Roche lobes and are semi-detached or contact binaries; see Section \ref{subsec:short_period_sources}. Two sources, TICs $405320687$ and $178739533$, stand out from the rest of the population because of their high eccentricity given their short orbital periods. TIC $405320687$ is a non-eclipsing heartbeat system whose phase curve is shown in Figure \ref{fig:heartbeat_classes}. TIC $178739533$ is a doubly-eclipsing system and shows a large change in its eclipse depths in different TESS sectors. We discuss this source and the $11$ additional sources showing orbital precession in Section \ref{subsec:eclipse_drift}.

Next, we measure the upper envelope of the period-eccentricity distribution by fitting two curves to the most eccentric systems for a given orbital period but excluding the outliers TICs $405320687$ and $178739533$.  These curves conserve angular momentum and periastron distance and are plotted as solid and dashed black curves, respectively. The eccentricity is related to the period as $e^2|_{\text{const }J} = 1 - (P/P_{\text{cutoff}})^{-2/3}$ and $e|_{\text{const }r_P} = 1 - (P/P_{\text{cutoff}})^{-2/3}$ for the constant angular momentum and constant periastron distance cases respectively. The curves are normalized using the cutoff period, $P_{\text{cutoff}}$, at which the eccentricity goes to $0$. We find the respective normalization parameters by fitting the curves to the upper envelope, i.e., the most eccentric sources in $20$ logarithmically spaced period bins between $1$ and $10$ days. The fits are performed using \textsc{SCIPY}'s \textsc{curve\_fit} and give cutoff periods of $1.7$ and $1.3$ days for the dashed and solid curves, respectively. Moreover, the $P =1.7$ day cutoff value corresponds to the typical minimum period for which we find good fits with the eBEER model; see Section \ref{subsec:short_period_sources}.

\begin{figure}
    \centering
    \includegraphics[width=0.49\textwidth]{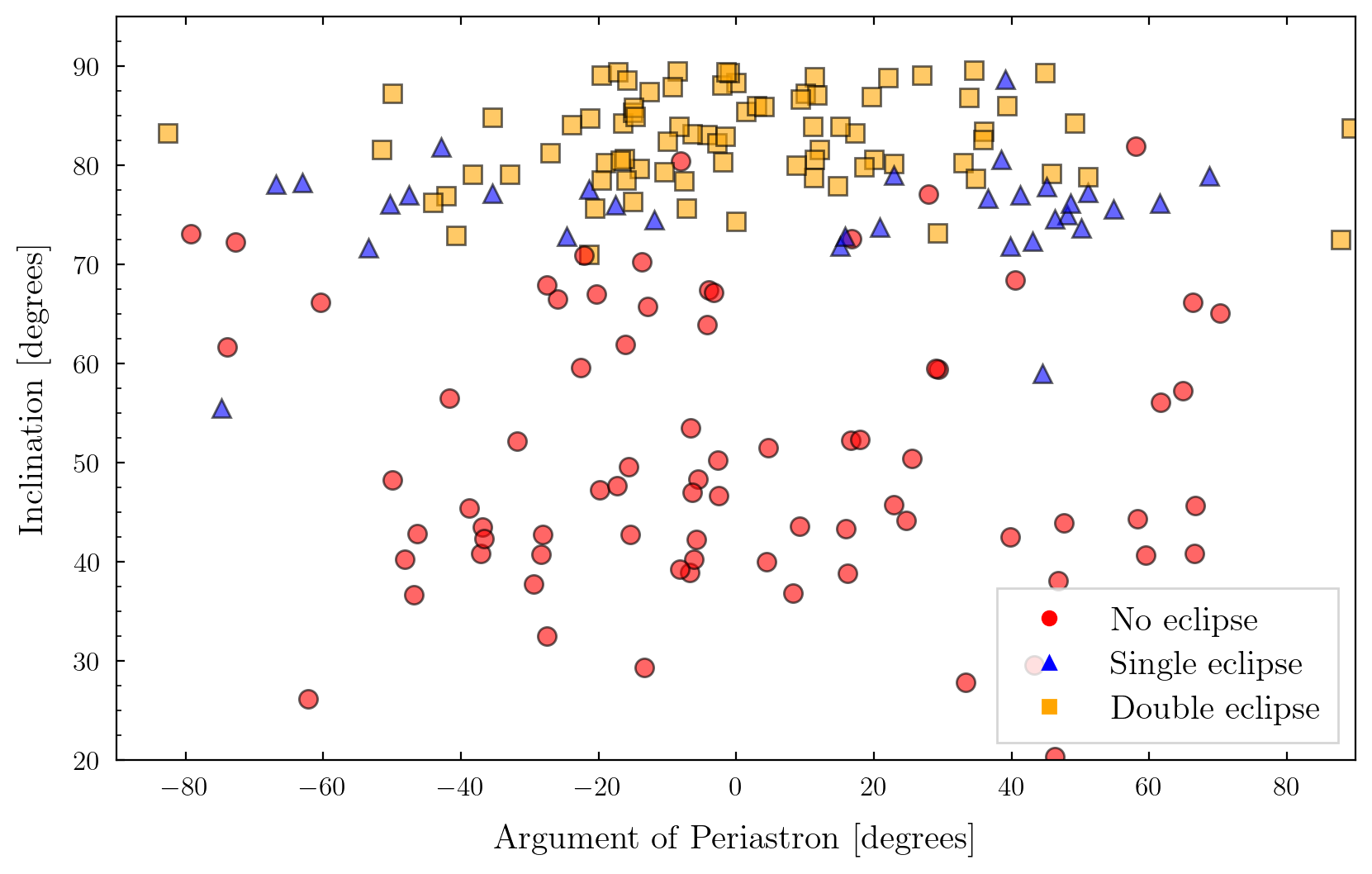}
    \caption{The distribution of the inclinations and arguments of periastron from the fits to the light curves. Systems with small inclinations do not naturally do not show eclipses. Additionally, the doubly-eclipsing sources tend to have their arguments of periastron biased towards smaller values of $|\omega|$. Meanwhile, singly-eclipsing systems have $|\omega| \gtrsim 20^{\circ}$. This trend can be understood from Figure \ref{fig:orbits_ebeer_model}. Double and single eclipses are favored by the orientation of the orbits for small and large values of $|\omega|$, respectively.}
    \label{fig:inc_omega}
\end{figure}

The distribution of the inclinations ($i$) and arguments of periastron $(\omega)$ for the population is shown in Figure \ref{fig:inc_omega}. As stated previously, we limit $\omega$ to $[-90^{\circ} - 90^{\circ}]$ by performing the transformation $\omega' = \arctan(\tan(\omega)$ in the posteriors because of a degeneracy in $\omega$ and $\omega + 180^{\circ}$ in the model. The eclipsing systems naturally have inclinations closer to $90^{\circ}$. Additionally, while the distribution of $\omega$ is roughly flat for the non-eclipsing systems, it is bi-modal for the singly-eclipsing systems. They have $|\omega| \gtrsim 20^{\circ}$ while the doubly-eclipsing sources generally have $|\omega| \lesssim 20^{\circ}$. This happens for two reasons. At high values of $|\omega|$, that is, when the line of the sight is along the semi-major axis of the ellipse, the position of the orbit favors single eclipses; see Fig. \ref{fig:orbits_ebeer_model}. Conversely, when the line of sight is along the semi-minor axis, two eclipses are favored by the geometry. Another reason single eclipses are favored at large $|\omega|$ is that the heartbeat signal can add destructively to one of the eclipses at higher values of $|\omega|$, potentially eliminating it from the light curve. Therefore, sources that would otherwise be doubly eclipsing without the heartbeat signal show just one eclipse because of this effect. A similar trend is seen in the synthetic light curves constructed using the eBEER model in Figure \ref{fig:orbits_ebeer_model}, and in the OGLE data \citep{ogle_2022ApJS..259...16W}. 

\begin{figure}
    \centering
    \includegraphics[width=0.49\textwidth]{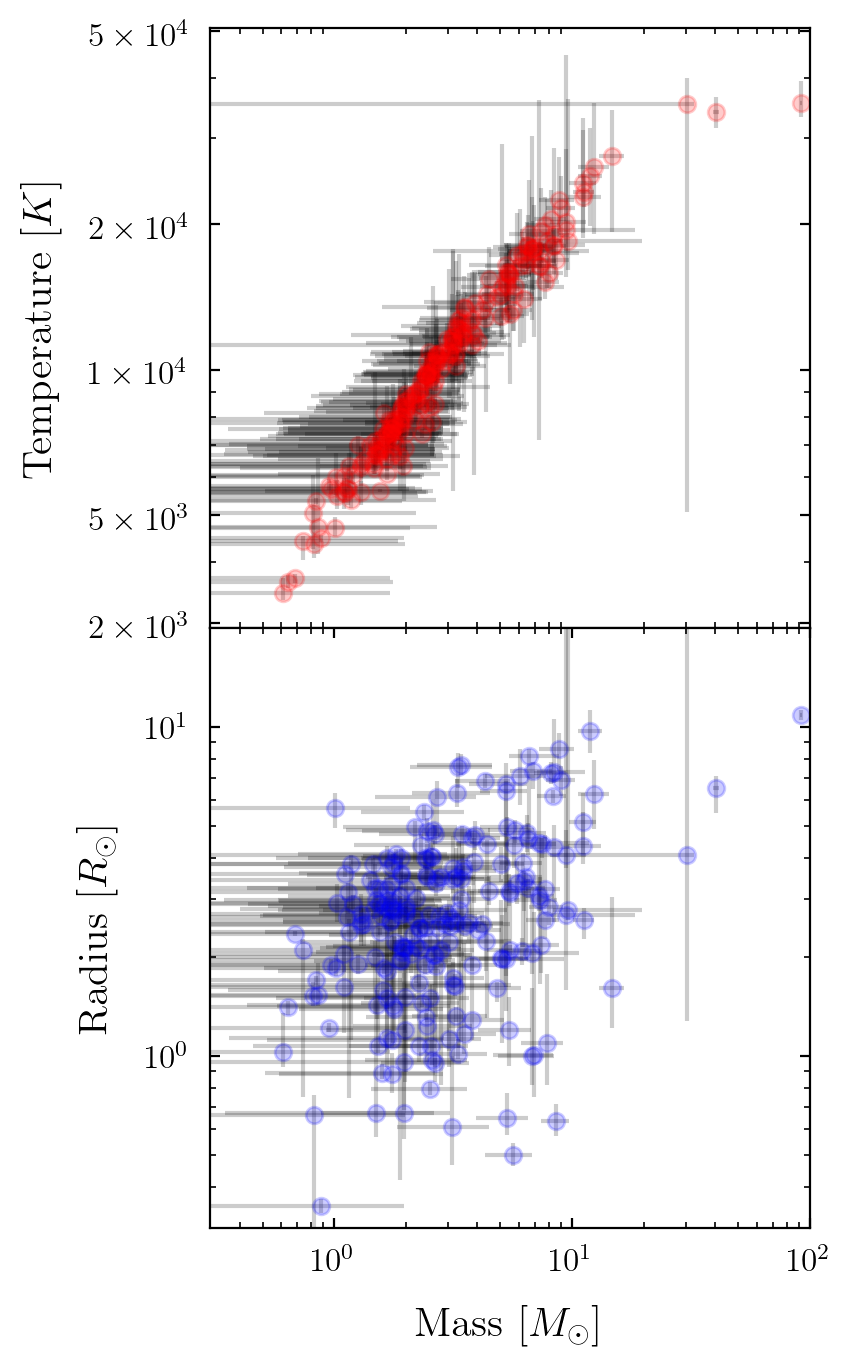}
    \caption{The distribution of the predicted temperatures and radii of the primary stars, plotted as functions of their masses. The error bars bracket the $16-84^{\text{th}}$ percentile values from the posteriors. While all the fits use the Gaia magnitude data, there is still significant uncertainty in determining stellar masses, temperature, and radii. This is because the Gaia magnitudes can only provide upper limits on the masses of the stars. Colder and less massive stars with larger radii can produce a similar phase curve with the same eccentricity and argument of periastron if they have a higher blending.}
    \label{fig:mass-temp-radius}
\end{figure}

While adding Gaia magnitudes helps better constrain some of the orbital parameters, there is still significant uncertainty in the mass, radius, and temperature estimates of the stars, which we also show in Fig. \ref{fig:shporer_comarison_2}. Figure \ref{fig:mass-temp-radius} shows the mass-radius and mass-temperature distributions of the primaries for all systems, where the error bars contain $68\%$ of the values from the posterior. The scatter arises from the blending and the radius and temperature re-scaling parameters. The Gaia magnitudes strongly constrain the upper limits of the masses, radii, and temperatures. However, less luminous stars can produce the same heartbeat signal with increased blending. Most systems have masses between $1.5$ and $8 M_{\odot}$, radii between $1$ and $10 R_{\odot}$, and temperatures between $6000$ and $20000$ K, consistent with the expectation of them being massive stars. The radii and temperatures are not strongly constrained from the data because their distribution follows the prior distributions; see Appendix \ref{Appendix:Temperature_radius_estimatation}. In order to perform a consistency check on the stellar masses, we select 10 systems with $q < 0.5$ as inferred from the eBEER model and compare the model masses with the masses inferred from the effective temperatures listed in the GAIA DR3 catalog. We find that the masses of all 10 systems are within 3-43\% within the main-sequence masses predicted from the GAIA temperatures, which is reasonable given the uncertainties on the effective temperatures in GAIA.

\begin{figure*}
    \centering
    \includegraphics[scale=0.55]{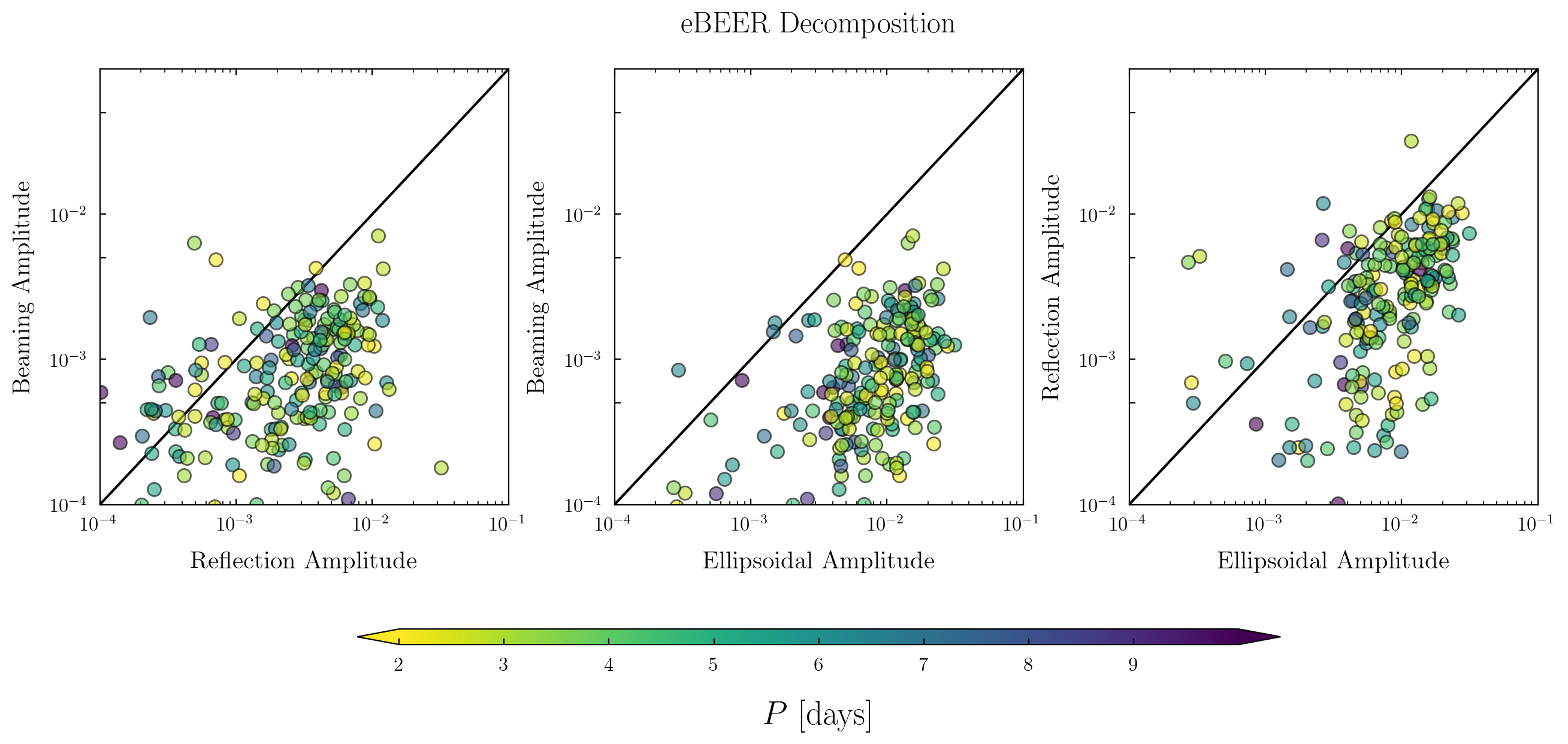}
    \caption{The peak-to-peak beaming, ellipsoidal, and reflection amplitudes from the fits to the heartbeat sample. The three panels show the relative flux contributions between beaming and reflection, beaming and ellipsoidal, and reflection and ellipsoidal effects, respectively. Ellipsoidal variations dominate their effect on the light curves, consistent with \cite{faigler_mazeh_2011MNRAS.415.3921F}. However, flux variations from reflection effects can be comparable to those from ellipsoidal variations.}
    \label{fig:lc_components}
\end{figure*}

Next, we measure the strength of beaming and the effects of reflection on the computed light curves. The flux is decomposed into beaming  ($\Delta F_{\text{beam}}$), ellipsoidal ($\Delta F_{\text{ellip}}$) and reflection ($\Delta F_{\text{ref}}$) contributions, and the peak-to-peak amplitudes of the 
three quantities are plotted against each other in Figure \ref{fig:lc_components}. Each source is represented by a circle, colored by the respective orbital period. We generally see $\Delta F_{\text{ellip}} \gtrsim \Delta F_{\text{ref}} \gtrsim \Delta F_{\text{beam}}$, however, the flux contribution from reflection can be a significant fraction of that from ellipsoidal variations. The flux variations from Doppler beaming are typically not more than $2 \times 10^{-3}$ in the population. Somewhat surprisingly, we see no strong correlation between any of the three amplitudes and the periods of the sources, perhaps because of the additional flux dependence on other orbital parameters. 

\subsection{Evolving Systems} \label{subsec:eclipse_drift}
 We model the $12$ systems that show the largest secular changes in their phase curves in the different sectors. For each of these sources, we fit the eBEER model to every sector separately without constraining the fits to have the same parameters for all the sectors. All sources show changes in their argument of periastron, where all have prograde apsidal precession, and some also show changes in eccentricity and inclination. A detailed analysis of one of these systems, TIC $378275980$, is presented below.

Figure \ref{fig:TIC_378275980} contains the different sector binned phase curves of TIC $378275980$ in the top left panel and the corresponding best-fit models in the top right panel. The light curves are folded on a period so that the primary eclipses align for all the sectors. This period can be different than the radial period of the system because of orbital precession; however, the relative difference is less than $10^{-4}$, seen from the phase shift of the secondary eclipse, which is of the order $\sim 1.75^{\circ}$ in $3$ years. Additionally, there is a secular change in the depth and timing of the secondary eclipse from sector $14$ to $55$, plotted in purple and red, respectively. In the bottom panel, we overlay the corner plots of the posteriors for the different sector fits. The posterior distributions for the masses overlap despite the completely independent fits to the different sector data. The eccentricity and inclination distributions also have a significant overlap, whereas the argument of periastron increases secularly in time.

\begin{figure*}
    \centering
    \includegraphics[width=0.49\textwidth]{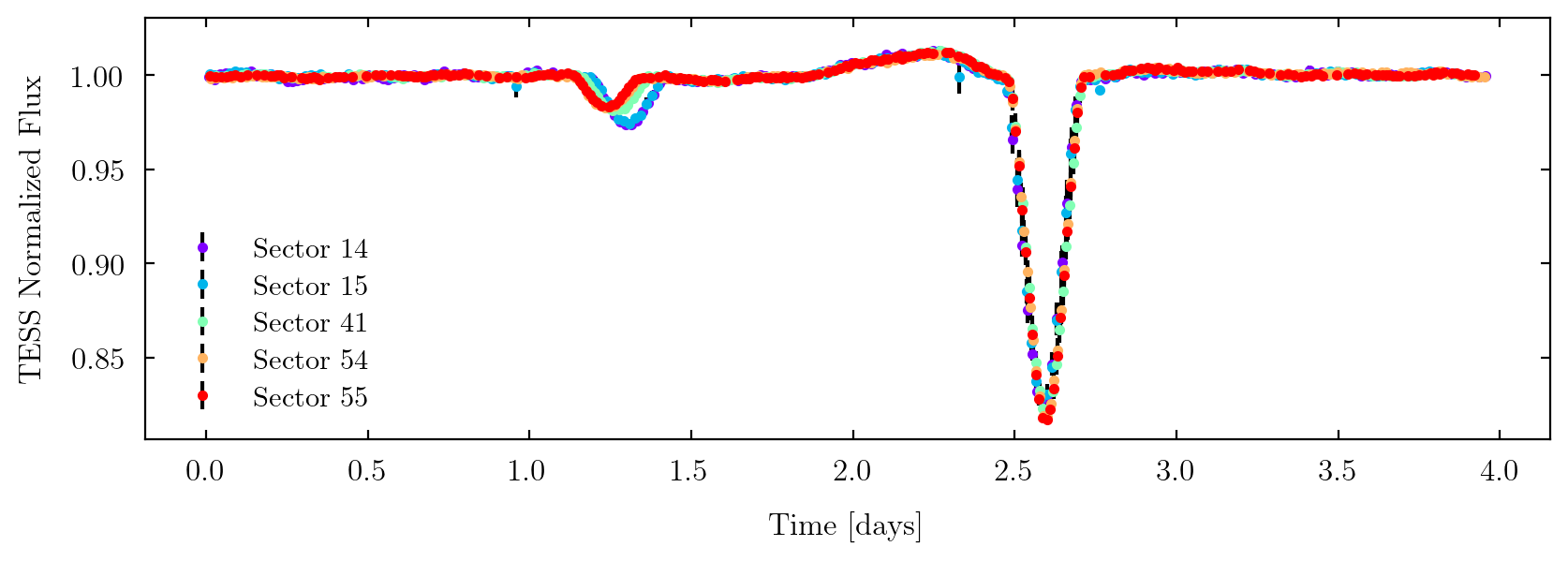}
    \includegraphics[width=0.49\textwidth]{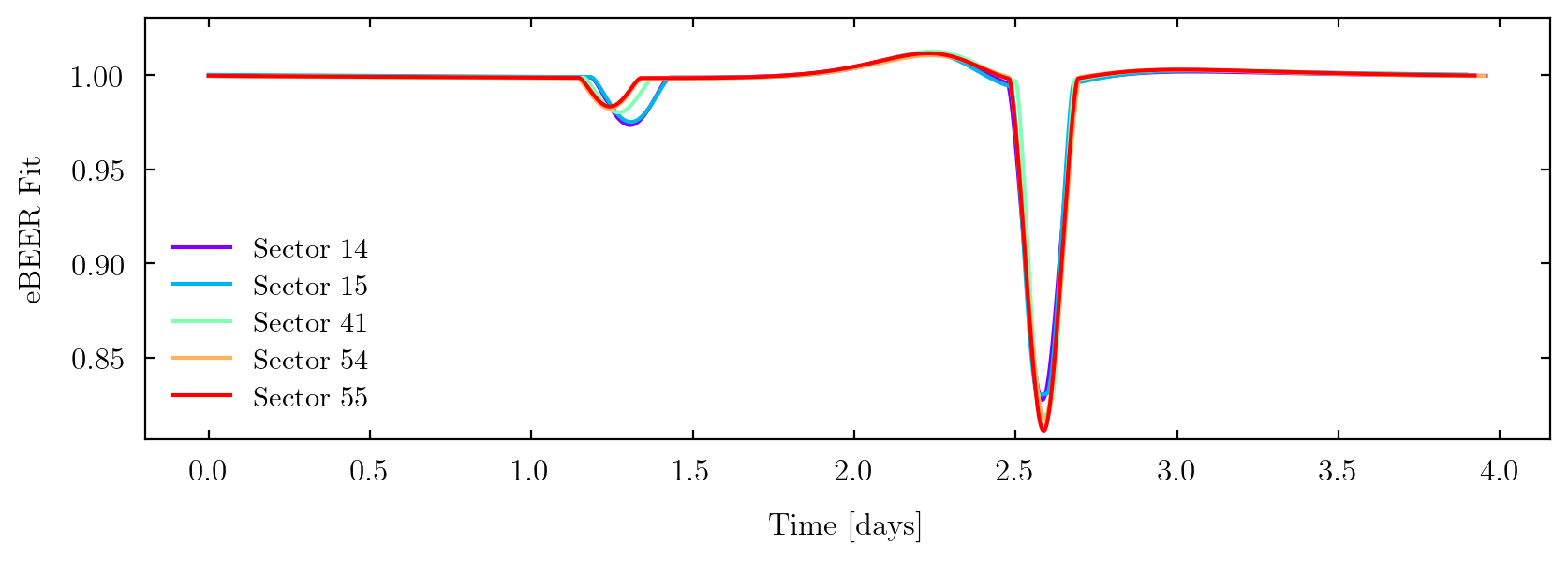}
    \includegraphics[scale=0.35]{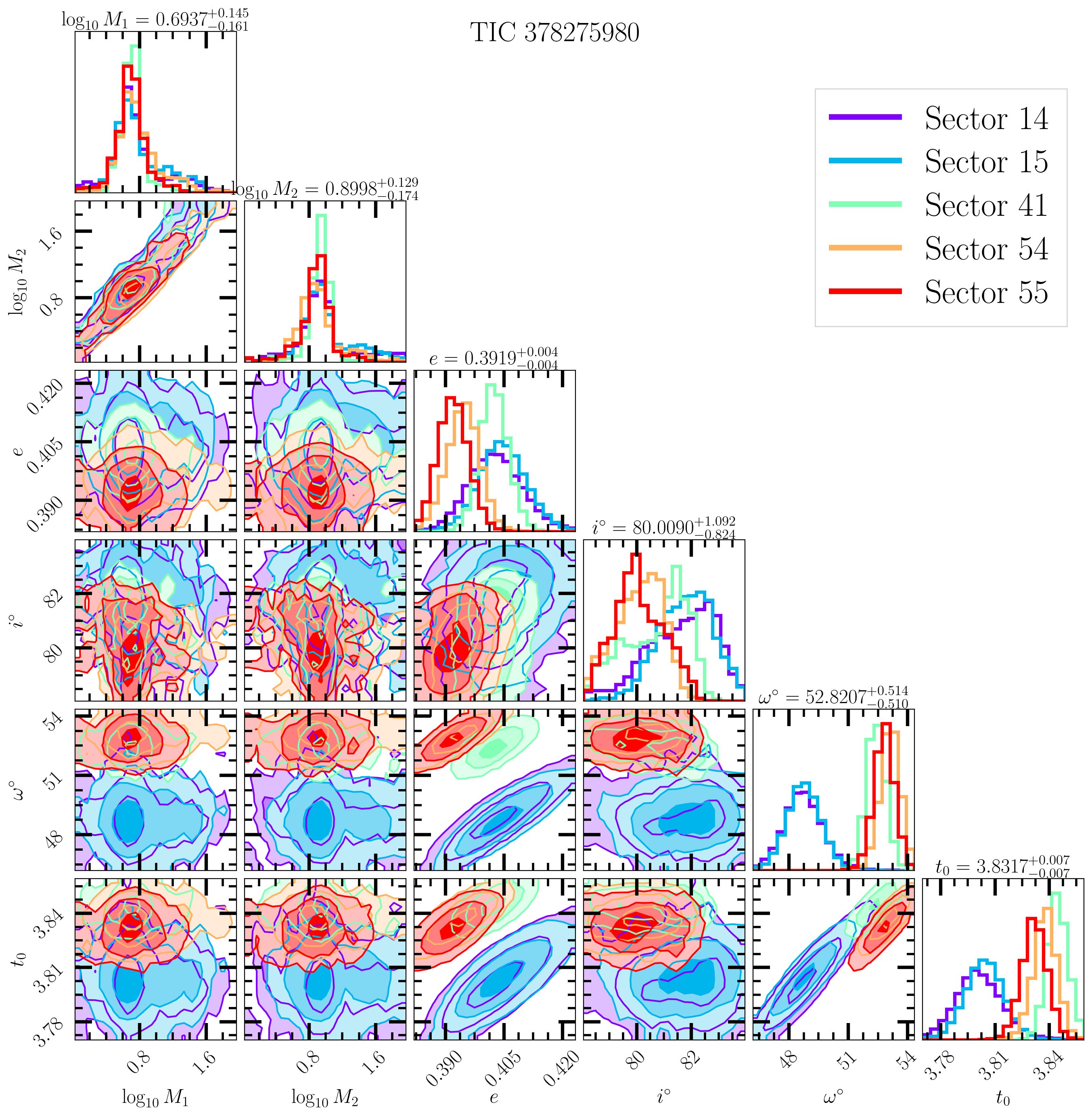}
    \caption{The multi-sector phase curves, eBEER fits, and the corresponding corner plots of TIC $378275980$. The data from sectors 14-55, which span over 4 years, show a long-term drift between the timing of the two eclipses, potentially from orbital precession. We fit the model to every sector separately. The posteriors yield consistent masses for the two stars but show a secular trend in the source's eccentricity for the data from different sectors. However, the eccentricity and inclination distributions are broadly consistent for all the sectors within $\pm 1 \sigma$.  }
    \label{fig:TIC_378275980}
\end{figure*}

\begin{figure*}
    \includegraphics[width=0.99\textwidth]{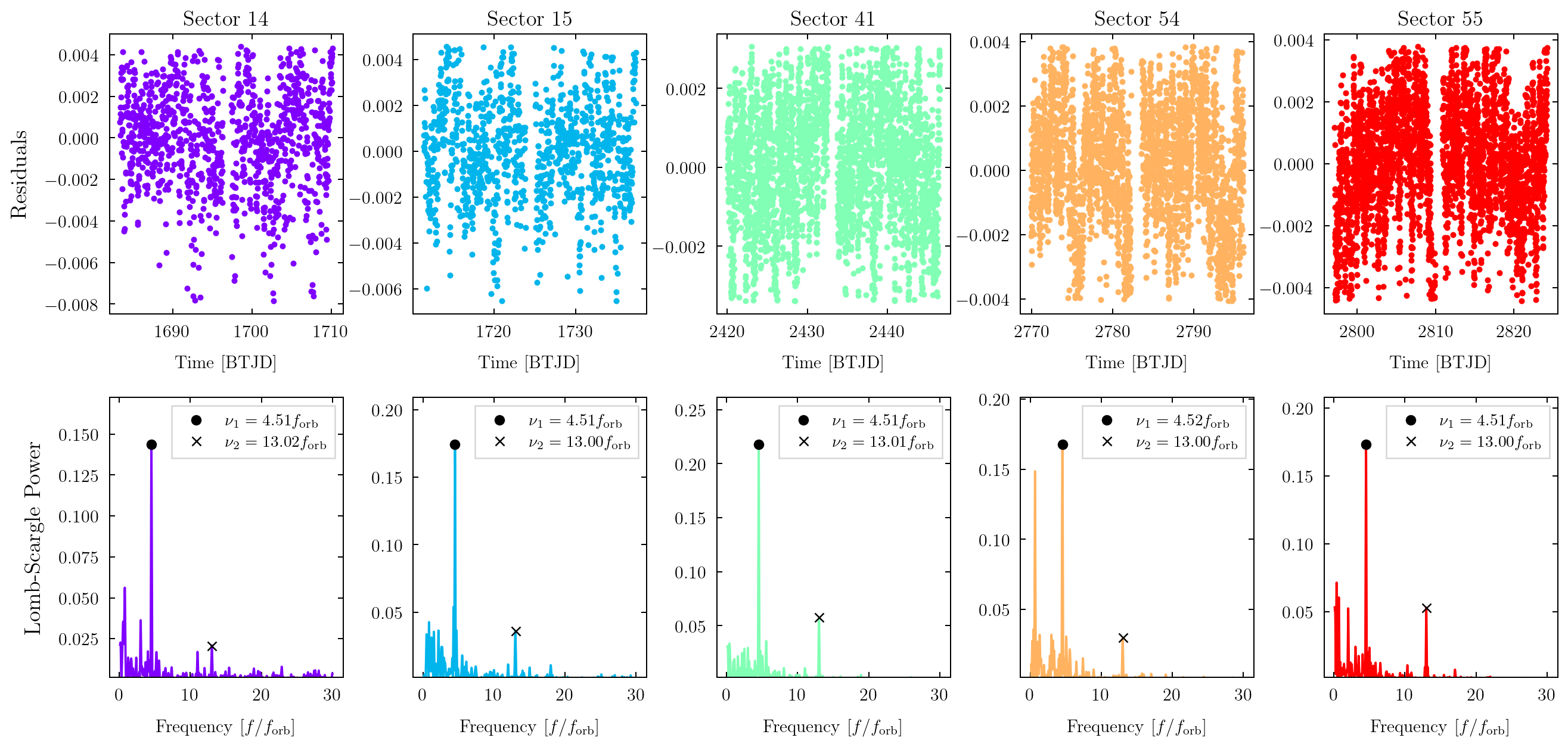}
    \caption{The residuals and their Lomb-Scargle periodogram of the TESS data of TIC $378275980$, where the residuals are obtained by subtracting the best-fit eBEER model from every sector. The frequencies on the $x-$axis are normalized by the orbital frequency of the system. We consistently find $2$ other periodicities in the system at $\sim 4.5$ and $13$ times the orbital frequency. If the first periodicity is interpreted as either of the spin frequencies of the stars, then it is higher than the expected pseudo synchronous frequency using an eccentricity of $0.4$. The latter frequency is TEO since it falls on an integer multiple of the orbital period.}
    \label{fig:periodogram_TIC_378275980}
\end{figure*}

We search for additional periodicities in the light curves by subtracting the best-fit eBEER model from the unfolded TESS data for every sector and removing the top and bottom $5^{\text{th}}$ percentile of the residuals. The latter eliminates the `trails' at the start and the middle of every TESS sector (see Figure \ref{fig:fits_to_multi_sector_data}), which can produce spurious peaks in the periodogram. We then take the Lomb-Scargle periodogram for the residuals and identify the periodicities in the signal.  The results for TIC $378275980$ are shown in Figure \ref{fig:periodogram_TIC_378275980}, where the top rows represent the residuals from every sector and the bottom rows contain the corresponding periodograms. We find two periodicities that consistently show up in all of the residuals, which have frequencies of $4.51 f_{\text{orb}} (0.88 \text{ days})$ and $13 f_{\text{orb}}  (0.31 \text{ days})$ respectively, where $f_{\text{orb}}$ is the system's orbital frequency. We check whether this frequency is close to the expected pseudo-synchronous frequency of the stars, where the latter is given by\citep{hut_1981A&A....99..126H}:

\begin{equation}
    f_{rot} = \frac{1 + \frac{15}{2} e^2 + \frac{45}{8}e^4 + \frac{5}{16}e^6}{(1 + 3e^2 + \frac{3}{8}e^4)(1-e^2)^{3/2}} f_{orb}\, .
    \label{eq:pseudo_syn_spin}
\end{equation}
Using the eccentricity of $0.29$ from the posteriors, the pseudo-synchronous spin frequency is $1.98 f_{\text{orb}}$, over $2\times$ slower than the observed frequency. Even if we consider a $m=2$ mode, the resulting frequency would still be closer to $4 f_{\text{orb}}$, which is still less than the observed frequency of $4.5 f_{\text{orb}}$. The frequency at $13 f_{\text{orb}}  (0.31 \text{ days})$ is almost certainly a TEO that remains in the light curve residuals. While the former frequency is not a tidally forced mode, it is in a $9:2$ resonance with the orbital frequency. Therefore, this mode can be an internal frequency of one of the stars, which happens to lie in close resonance with the orbital frequency \citep{fuller_2017MNRAS.472.1538F,burkart_elliot_2012MNRAS.421..983B}.

We perform the same procedure for the other $11$ HBs in the sample and find one additional source, TIC $240918551$, that shows a strong periodicity that is not an integer multiple of the orbital frequency. While TIC $240918551$ has data in sectors $17, 18$ and $58$, we could not obtain a good fit for the light curve of sector $17$ because the source displays very large amplitude pulsations, with  $\Delta F/F \sim 1.5 \%$, that possibly affected the fits. In this case, the dominant periodicity in the residuals is at $7.23$ times the orbital frequency, corresponding to a period of $\approx 0.53$ days. The predicted eccentricity of the system is $e = 0.51$, yielding a pseudo-synchronous frequency of $2.8 f_{\text{orb}} $, which is once again different from the dominant frequency in the residuals. These frequencies can be internal modes of the stars, completely independent of the orbital period of the binary.

\begin{figure*}
    \includegraphics[width=0.99\textwidth]{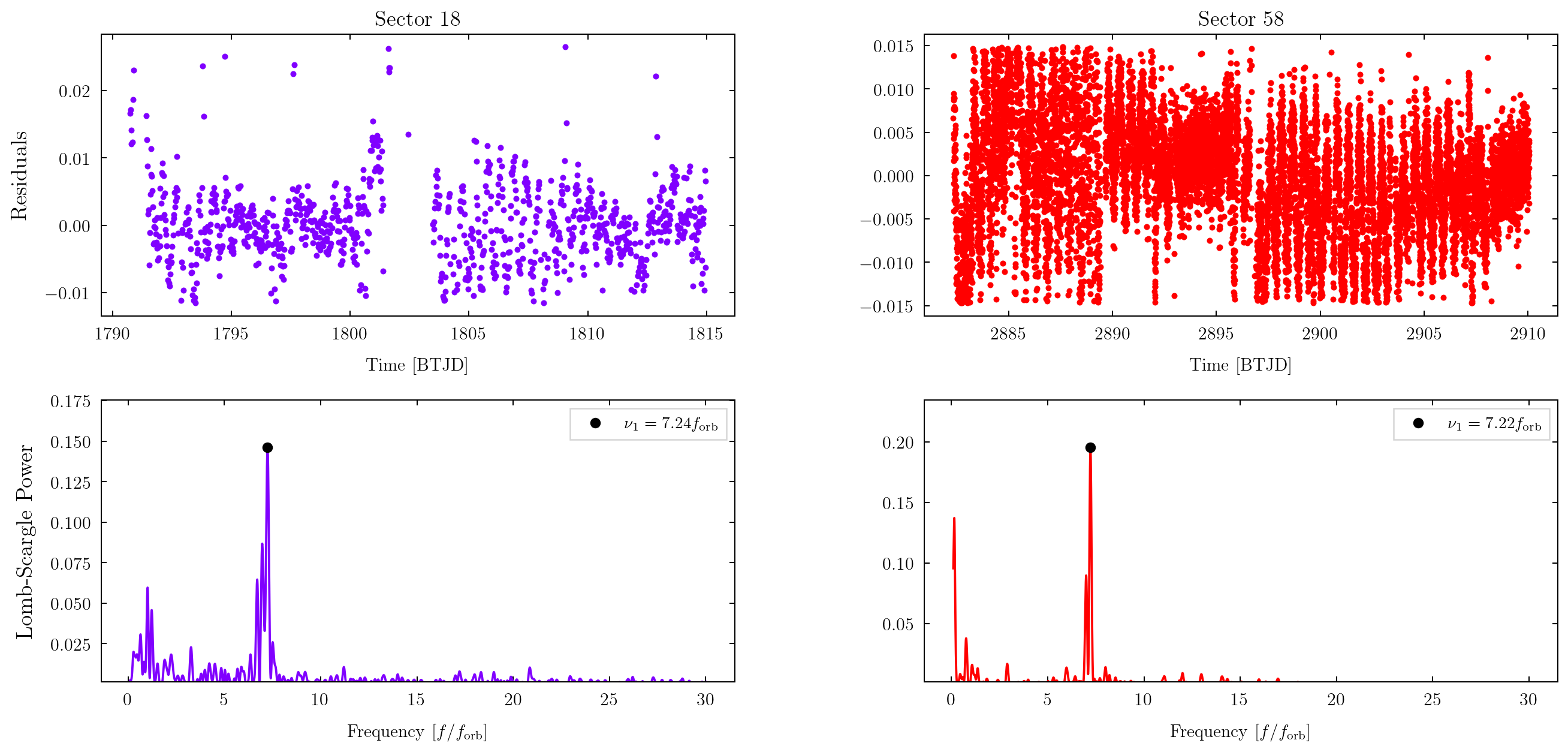}
    \caption{Same as Figure \ref{fig:periodogram_TIC_378275980} but for TIC $240918551$. The source displays large amplitude pulsations at $\approx 7.2$ times the orbital frequency, corresponding to a period of $\approx 0.53$ days. We could not obtain good fits on the sector $17$ light curve and, therefore, do not include the residuals and the periodogram for that sector. Using the $e = 0.51$ from the posteriors, the pseudo-synchronous rotation frequency of  TIC $240918551$ is $2.8 P^{-1}$, which is lower than the frequency in the residuals.}
    \label{fig:periodogram_TIC_240918551}
\end{figure*}

\begin{figure*}
    \centering
    \includegraphics[width=0.7\textwidth]{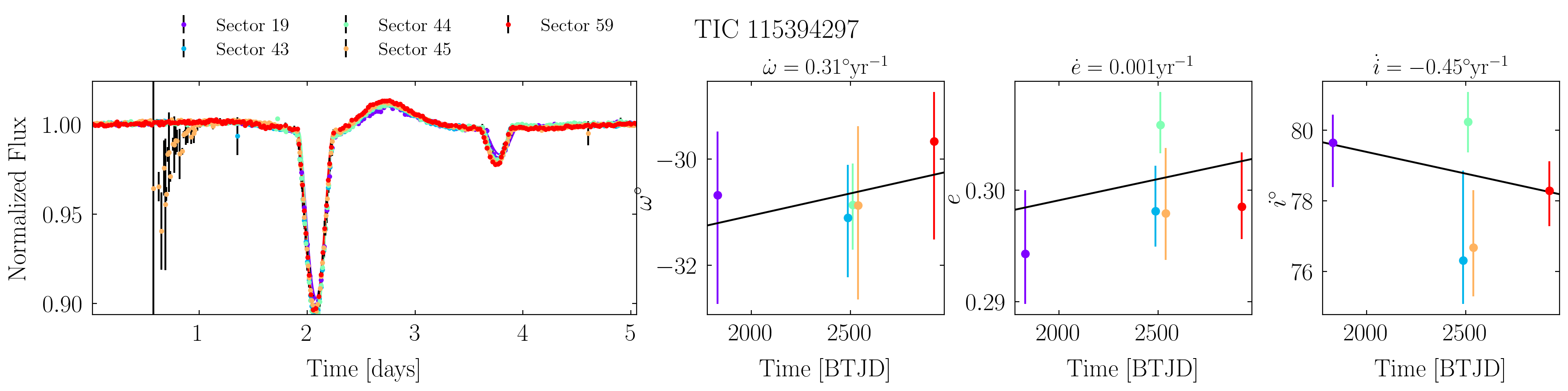}
    \includegraphics[width=0.7\textwidth]{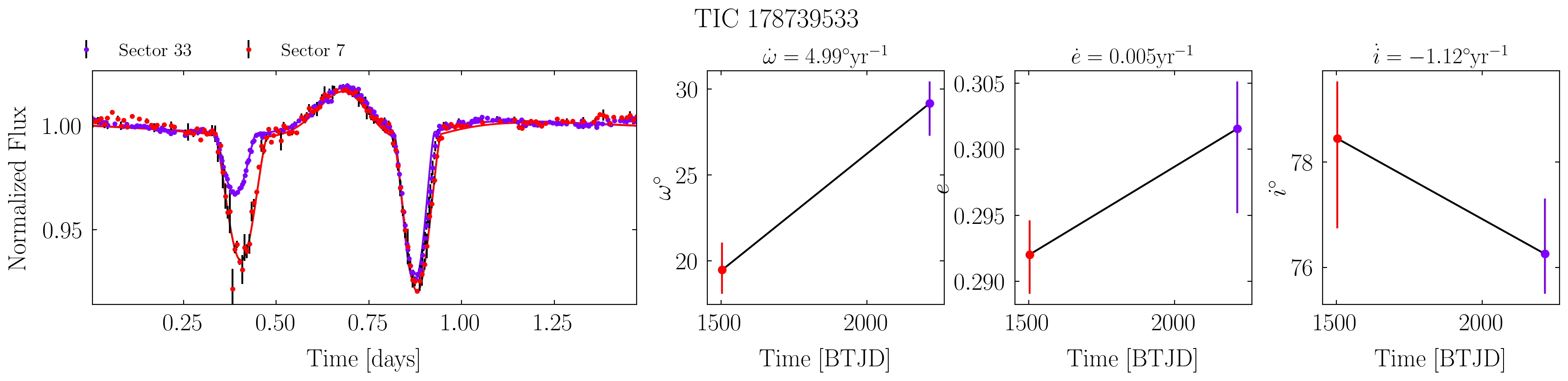}
    \includegraphics[width=0.7\textwidth]{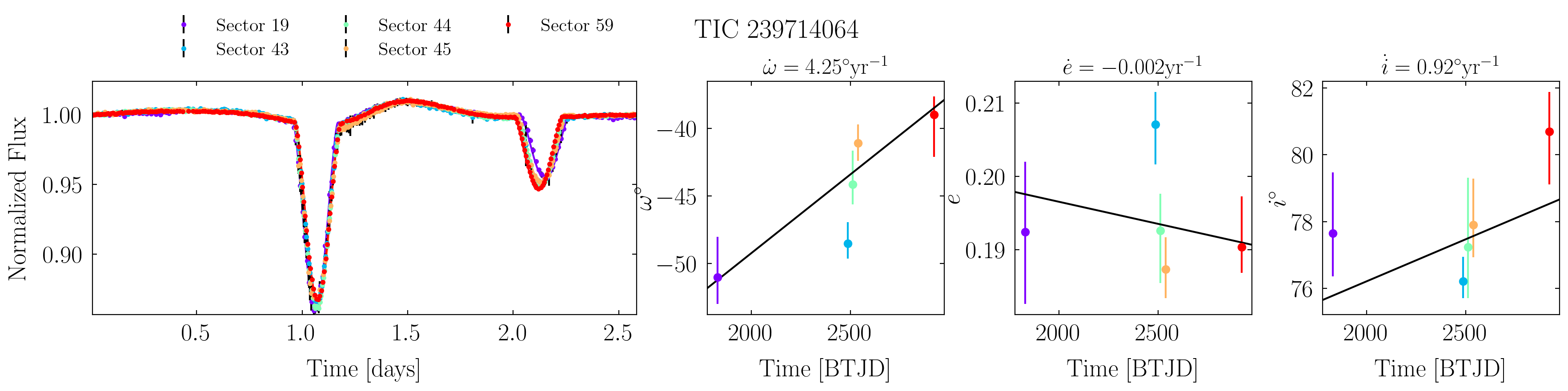}
    \includegraphics[width=0.7\textwidth]{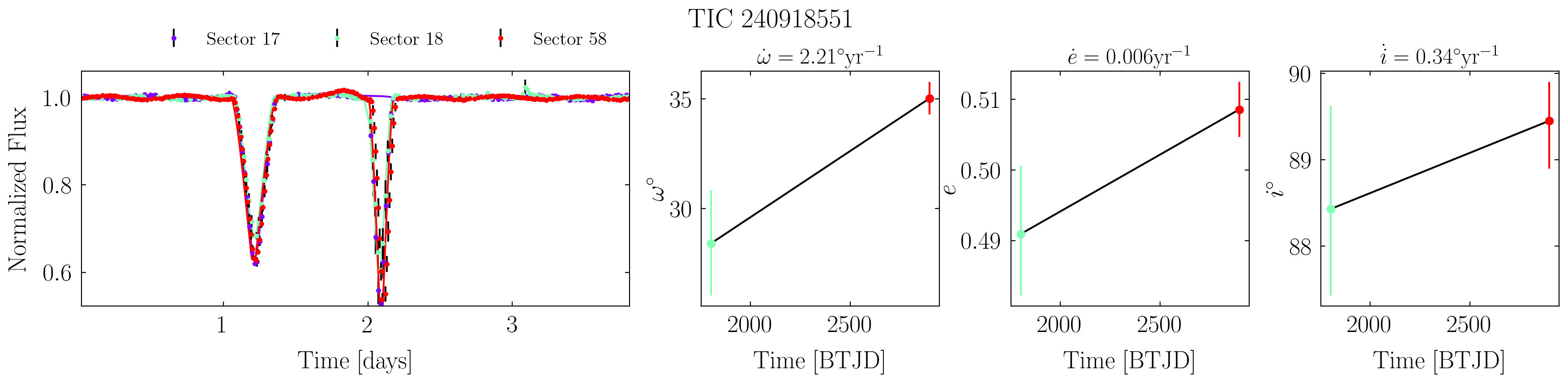}
    \includegraphics[width=0.7\textwidth]{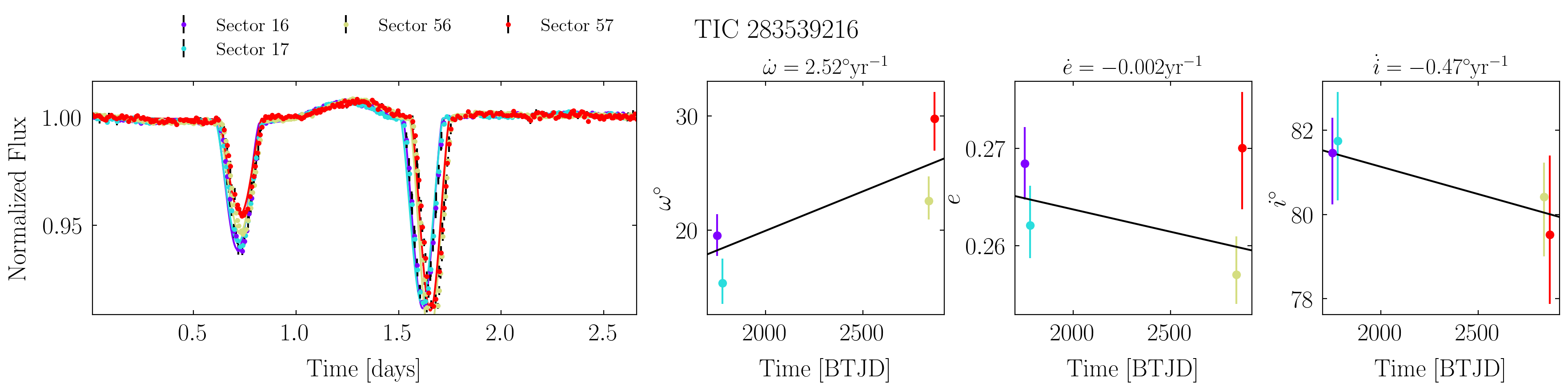}
    \includegraphics[width=0.7\textwidth]{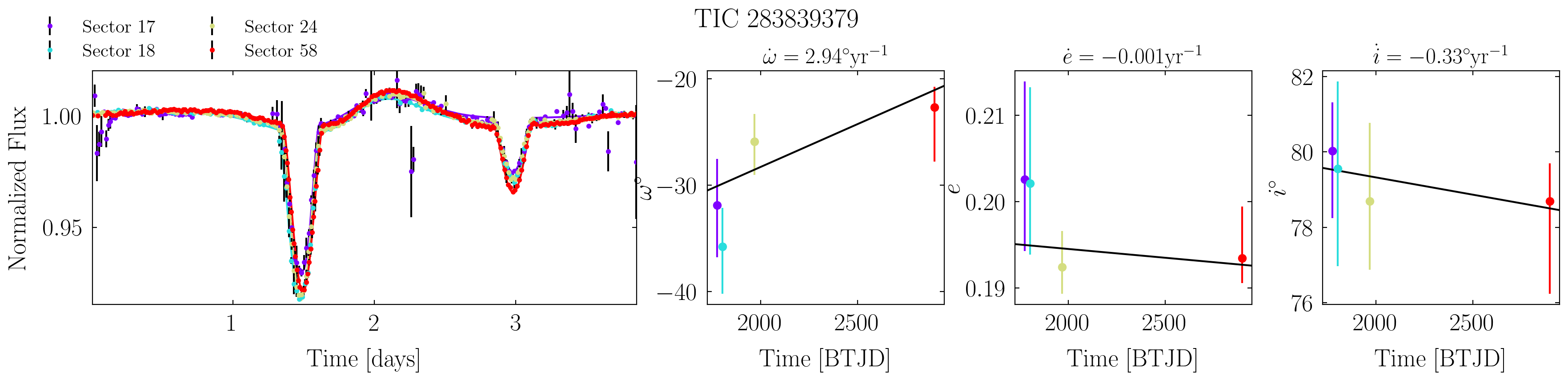}
    \caption{The phase curves with the best-fit models (left panels) and median posterior values of $\omega, e$ and $i$ as functions of time in TESS Barycentric Julian Day (BTJD) (right panels) for $6$ of the $12$ of the doubly-eclipsing systems with timing variations in their eclipses. The error bars in the orbital parameters indicate the $95\%$ confidence interval from the posteriors.  All systems show prograde apsidal precession. We perform a linear fit to the inferred orbital parameters for the different sectors shown as a solid black line.}
    \label{fig:periastron_advance}
\end{figure*}

Finally, we estimate the rate of periastron advance as well as the eccentricity and inclination changes for $12$ doubly-eclipsing systems that show a secular change in their light curves. We show the phase-folded data and the corresponding best-fit model for every sector on the left panels of every row in Figure \ref{fig:periastron_advance}. The model and data are plotted as solid curves and dots, respectively, with a unique color for every sector. As shown in the left panel, we could not obtain good fits for the light curves for a few sources. Namely, we could not fit the sector $11$ data for TICs $451708707$ and $305454334$, and the sector $17$ data for TIC $240918551$. The posteriors from these sectors were excluded while estimating the rate of periastron advance ($\dot{\omega}$), eccentricity change ($\dot{e}$) and inclination change ($\dot{i}$). The right panels show the inferred $\omega, e$ and $i$ as functions of time, where the error bars contain the $5-95^{\text{th}}$ percentile values in the posteriors. We perform a linear fit to the values taken from the posteriors to obtain the time rate of change of the orbital parameters. All systems show a prograde apsidal precession, the highest being $9.07^{\circ}$ degrees per year for TIC $451708707$. As we mentioned previously in Sec. \ref{subsec:eBEER_model}, all of the fits set $\Omega = 0^{\circ}$ because the light curves are independent of the longitude of the ascending node. Therefore, we cannot estimate nodal precession from the eBEER fits alone. If the sources show nodal precession, it will appear as apsidal prograde precession in our fits because the inclination of the systems is limited to $(0, 90^{\circ})$. $e$ and $i$ do not follow a secular trend for many of the systems, unlike the arguments of periastron.

\begin{figure*}
    \centering
    \includegraphics[width=0.7\textwidth]{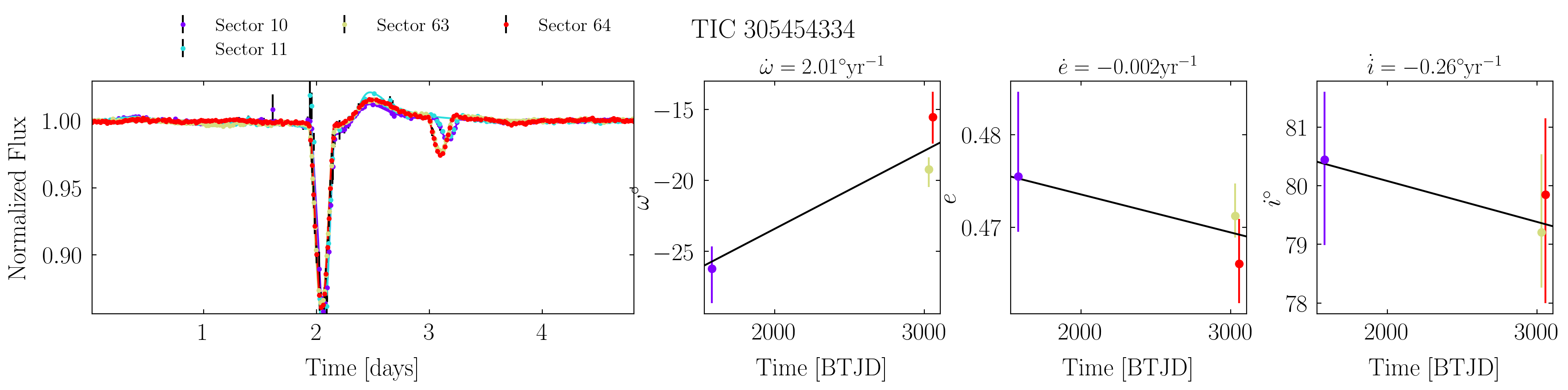}
    \includegraphics[width=0.7\textwidth]{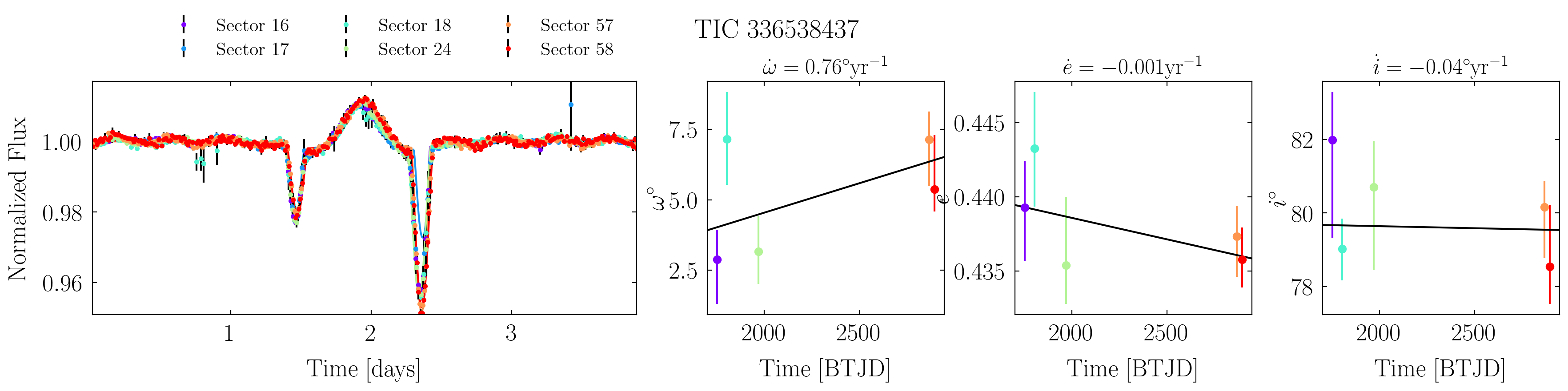}
    \includegraphics[width=0.7\textwidth]{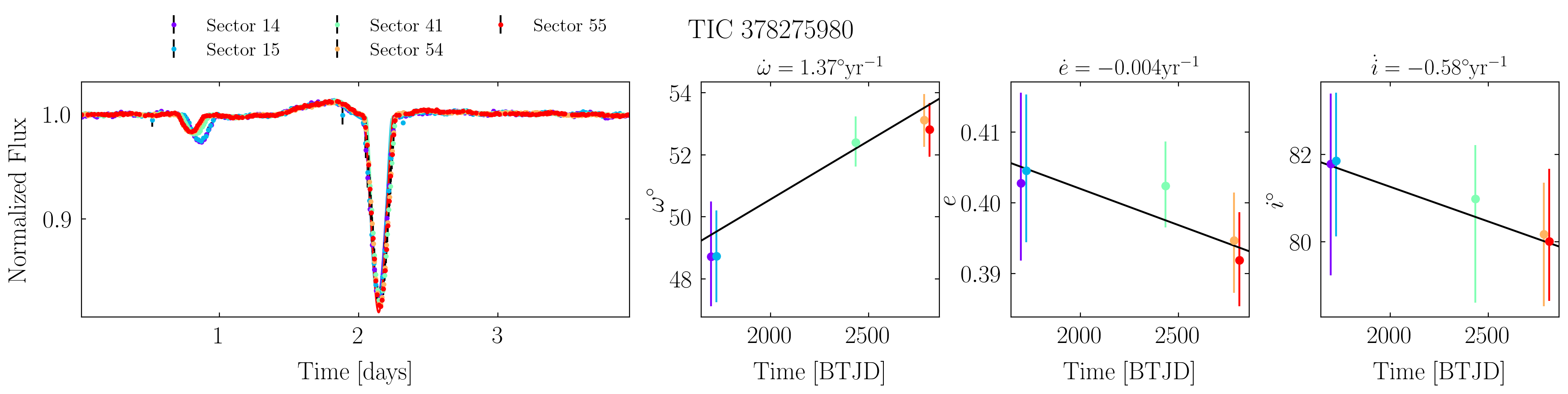}
    \includegraphics[width=0.7\textwidth]{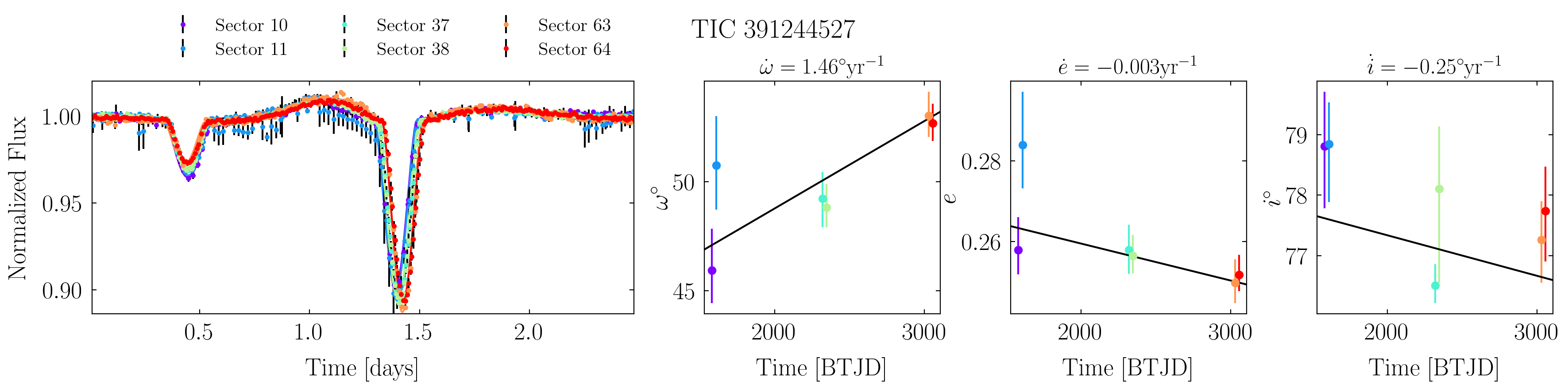}
    \includegraphics[width=0.7\textwidth]{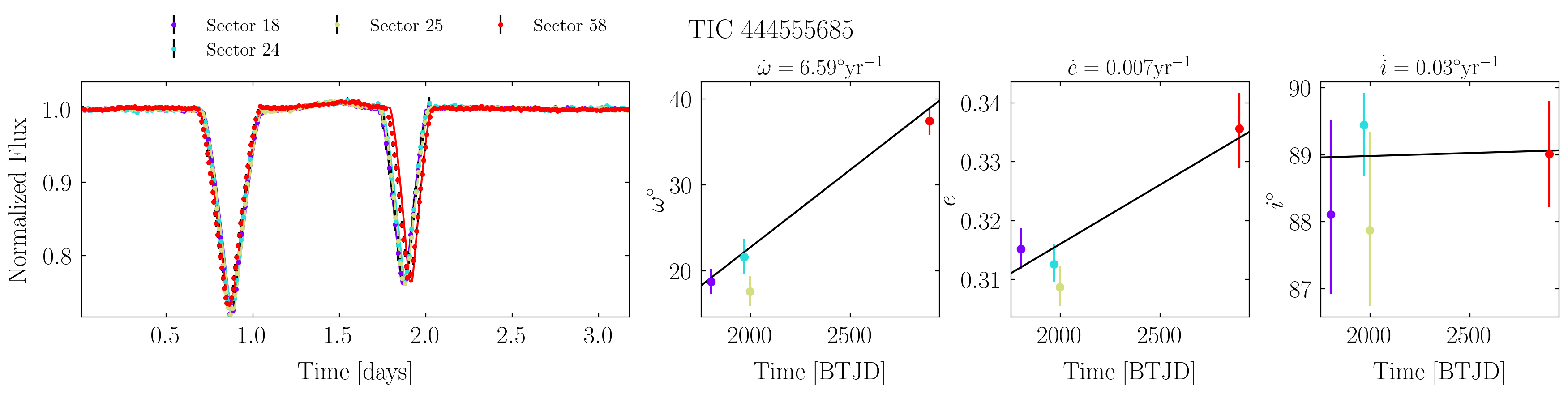}
    \includegraphics[width=0.7\textwidth]{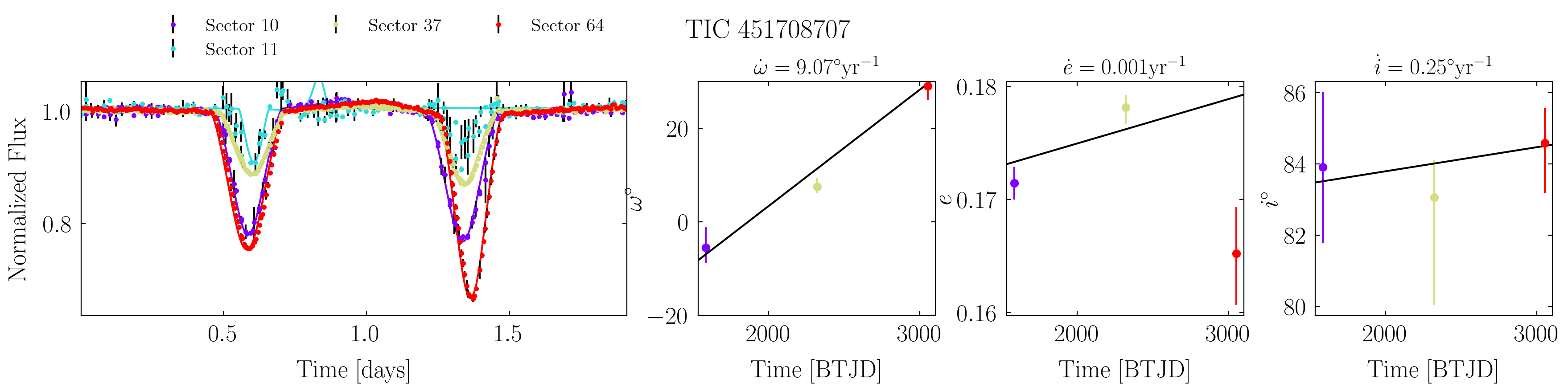}
    \caption{Same as Figure \ref{fig:periastron_advance_2} but for the other $6$ of the $12$ sources. TIC $451708707$ shows the highest periastron advance rate of $9^{\circ} \text{yr}^{-1}$.}
    \label{fig:periastron_advance_2}
\end{figure*}

While this precession should be present in many of the non-eclipsing heartbeats, it is much more difficult to observe in the light curves \citep{Hambleton_2016}. This is because the heartbeat signal is not very sensitive to the value of the argument of periastron at lower inclinations; see Fig. \ref{fig:orbits_ebeer_model}. 

Finally, an attempt was made to measure the radial velocities of TIC~378275980, as well as those of a few other systems in the sample, to further constrain their physical properties. We used a bench-mounted, fiber-fed echelle spectrograph on the 1.5m reflector at the Fred L.\ Whipple Observatory on Mount Hopkins (Arizona, USA), which covers the full optical range. At our resolving power of $R \approx 44,000$, the spectra of TIC 378275980, and also those of TIC~239714064 and TIC~283539216, display a lack of measurable metal lines, showing them to be very hot and rapidly rotating stars (estimated $v \sin i > 200$~km~s$^{-1}$). Unfortunately, this has prevented us from obtaining meaningful radial velocities.

We now discuss our results in the context of close binary formation. 


\section{Discussion} \label{sec:discussion}
The formation mechanisms of close binaries with orbital periods of $P \lesssim 10$ days ($a \lesssim 0.1$ au) are actively being studied \citep{moe_kratter_2018ApJ...854...44M}. Short-period HBs represent a special subset of close binaries due to their high eccentricities. Therefore, an important question to ask is: do such HBs form from the same formation channels as other circularized, short-period binaries? 

The tight relation between the close-binary fraction of $M_{\odot}$ main-sequence and pre-main sequence binaries suggests that the hardening of the binaries takes place in the first few Myr of star formation \citep{kounkel_2019AJ....157..196K}. Simulations of binaries embedded in a circum-nuclear disk (CBD) show that the binary orbits can shrink from interactions with the CBD via viscosity-driven \citep{dittmann_2022MNRAS.513.6158D} or wind-driven \citep{turpin_2024MNRAS.528.7256T} accretion.  \cite{orazio_2021ApJ...914L..21D} find that for an initially eccentric binary, with $e_i > 0.1$, the interactions with the CBD evolve the orbit towards $e \sim 0.4$, while the semi-major axis $a$ continues to shrink. Therefore, HBs with $P \lesssim 10$ days could represent the population of binaries that were initially eccentric and hardened from interactions with the CBD. Disk fragmentation can produce binaries with separations that are already $\lesssim 200$ au, which can then continue to harden from a shared CBD \citep{offner_2023ASPC..534..275O}. 

Alternatively, a combination of Kozai-Lidov oscillations from tertiary companions and tidal dissipation can produce short-period binaries. Observations show that the distribution of short-period binaries is closely linked to the distributions of triple star systems, and the frequency of triple star systems increases with masses of the stars in the inner binary \citep{moe_di_2017ApJS..230...15M,tokovinin_2006A&A...450..681T}. \cite{toonen_2020A&A...640A..16T} find through population synthesis of triple systems that orbits of the inner binary star eccentric all the way up to mass transfer, when one of the stars overflows its Roche Lobe. For A and B spectral-type HB stars, this further requires the presence of close tertiary companions to harden the orbits of these stars on $\lesssim 0.1 $ Gyr timescales. Eccentric short-period binaries with close tertiary companions have been observed in the TESS and the Kepler data. \cite{new_eclipsing_2024A&A...683A.158Z} reported the discovery of $6$ such systems, where one of the systems, ASASSN-V J231028.27+590841.8 had the inner binary period and eccentricity of $2.4$ days and $0.43$ respectively, and a tertiary period of only $4.9$ years. Many HBs in our sample can have such close companions, which can potentially give rise to the observed orbital precession. However, to correctly infer the presence of a tertiary companion, detailed modeling of the eclipse-timing variations is required, such as in \cite{borkovits_2016MNRAS.455.4136B}.

A notable difference in the population of HBs presented in this work is that the period-eccentricity diagram of this population is extremely eccentric even when compared to that of other HBs: a few systems have eccentricities as big as $e \gtrsim 0.3$ with orbital periods of $\sim 2$ days. Moreover, the period-cutoff is slightly lower than inferred from eclipsing binaries in Kepler and TESS data, e.g., \cite{circularizatio_ebs_2023AAS...24143006Z}, although this may be due to a selection effect from HBs being more eccentric. Note that this population of stars is younger and hotter than the ones in the cited works because it primarily comprises A and B-type main-sequence stars, distinct from the systems presented in, e.g., \cite{shporer_2016ApJ...829...34S}. In fact, one of the earliest HBs discovered is composed of two B stars and has a high eccentricity of $e \sim 0.3$ for an orbital period of $3.6$ days \citep{high_ecc_B_2009A&A...508.1375M}. It needs to be understood whether such HBs are more eccentric because their circularization timescales are much larger than their age or whether some other physical mechanism is responsible for their high eccentricity. The former is plausible given the lack of large convective zones in massive stars, where tidal dissipation is efficient, thus reducing the tidal circularization timescales. The underlying formation mechanisms can be distinct for the two populations. A larger sample of HBs with massive stars, for example, in the TESS data, would enable making more quantitative statements about their $P-E$ distributions. More of these systems can be found through currently existing databases of eclipsing binaries that have some of the orbital properties determined, such as in \cite{tess_ebs_2022ApJS..258...16P}.

There are two outliers with even higher eccentricities, TICs 405320687 and 17873953, which can be relatively young systems still actively circularizing. In a sample of $\sim 200$ A and B-type stars with typical main-sequence lifetimes of a few $100$ Myr, it is unsurprising to find $\sim 1-2$ systems with ages $t \lesssim 10$ Myr. These two systems warrant additional stellar modeling and radial velocity follow-ups, which can show whether these systems are young and rapidly circularizing or have high eccentricities that are driven by close and inclined tertiary companions.

Modeling the $12$ systems with light curve changes revealed they all have a prograde apsidal precession. Orbital precession has been observed in other heartbeat systems \citep{Hambleton_2016,new_eclipsing_2024A&A...683A.158Z}, which could be explained by the perturbing gravitational potential from a third body. Apsidal precession can also be produced from tidal bulges on the stars, where the precession scales strongly with the periastron distance. We estimate the contributions to apsidal precession from three body effects and the tidal bulges on the stars. The contribution from the rotational bulge due to the rotation of the stars is much smaller than from the tidal bulge of the star when the stars are pseudo-synchronized  \citep{liu_2015MNRAS.447..747L}. This can be seen very crudely by noting that the ratio of the potentials from the tidal and rotational bulges must scale as the ratio of the star's rotational and orbital kinetic energies. This further scales as $R_{*}^2/d^2 \ll 1$, where $R_{*}$ is its radius and $d$ is the separation of the stars taken at the periastron. Finally, note that the changes in the light curves are unlikely to result from light travel time effects (LTTE) from tertiary companions because these effects are expected to be of the order $\lesssim 1 AU /c \sim 10$ minutes as opposed to a few hours that we see in our phase folded light curves.

To be more quantitative, we specifically consider the case of TIC 444555685, for which we infer a periastron advance rate of $6.59^{\circ}$ per year from the eBEER fits. The fits also give primary and secondary masses of $5.4$ and $4.4 M_{\odot}$ and an eccentricity $e \approx 0.32$. To get a precession rate from three body precession, we need to make assumptions about the mass, eccentricity, and orbital period of the tertiary. The tertiaries typically have smaller masses than the stars in the inner binary \citep{moe_di_2017ApJS..230...15M} and orbital periods $P \gtrsim 1000$ days, and therefore we assume a tertiary mass of $1 M_{\odot}$, and an eccentricity and orbital period of $0.5$ and $1000$ days respectively. The precession rate is given by \citep{moe_kratter_2018ApJ...854...44M}:

\begin{equation}
\dot{\omega}_{\text{TB}} = \frac{8}{15 \pi} \frac{M_1 + M_2 + M_3}{M_3}  \frac{P_{\text{out}}^2}{P_{\text{in}}} \frac{(1 - e_{\text{out}}^2)^{3/2}}{\sqrt{1 - e_{\text{in}^2}}} \, ,
\label{eq:precession_kozai}
\end{equation}
where the subscripts `in' and `out' refer to the inner and outer orbits, respectively. Using the above estimates, we obtain a three-body precession rate of $\dot{\omega} \approx 6 \times 10^{-2}$ degrees per year, a factor of $100$ smaller than what we see. Note that three-body precession happens on a Kozai-Lidov timescale but does not require a relative inclination between the inner and outer orbits. Reducing the tertiary orbital period to $100$ days gives the precession rate to be $\approx 5.9^{\circ}$ per year, much closer to the observed value. While such close tertiaries are rare \citep{tokovinin_2006A&A...450..681T}, they have been observed around other short-period binaries \citep{new_eclipsing_2024A&A...683A.158Z,borkovits_2020MNRAS.496.4624B}. 

Precession from tides may also be crucial for these systems with small periastron distances. The precession rate further depends on the radii and tidal Love numbers for the two stars. The eBEER fits give the stellar radii $2.7$ and $3.1 R_{\odot}$ respectively. Using these radii, we can approximate the tidal Loves numbers for the two stars\citep{apsial_motion_constant_1984MNRAS.207..323J,claret_2020A&A...641A.157C}. The tidal precession rate is given by \cite{liu_2015MNRAS.447..747L}

\begin{widetext}
\begin{equation*}
\dot{\omega}_{\text{tidal}} = \frac{30\pi \left(1 + \frac{3}{2}e^2 + \frac{1}{8}e^2 \right)}{P} \left[ \left(\frac{R_1}{a(1 - e)^2}\right)^5  q k_{2,1}^1 +  \left(\frac{R_2}{a(1 - e)^2}\right)^5 \ (1-q) k_{2,1}^2   \right] \, ,
\label{eq:precession_tidal}
\end{equation*}
\end{widetext}
where $q = M_1 / M_2$ and $k_{2,1}^1 = k_{2,1}^2 = 10^{-2}$ are the tidal Love numbers for stars $1$ and $2$. Using these equations, we get a precession rate of $6.05^{\circ}$ every year, almost exactly equal to what we measure from the eBEER fits. Therefore, the precession from tidal bulges can account for a large part of $\dot{\omega}$ without requiring the presence of very close tertiary companions. 

Gaia can detect binaries with orbital separations of a few AU at distances less than $1$ kpc from Earth \citep{RUWE2Belokurov2020}. Specifically, the presence of tertiaries with orbital periods in the range of $100-1000$ days can be inferred from large GAIA Renormalised Unit Weight Error (RUWE)  values, typically above $1-1.4$, depending on the location of the sky 
\citep{RUWE2Belokurov2020,RUWECastroGinard2024}. We find that all but one of the $12$ systems have RUWE values below $1.4$, with the exception being TIC 283539216 with a RUWE value of 4.76. This system also has very large `astrometric excess noise' and `astrometric excess noise sig' values of 0.564 and 401.51, respectively. Moreover, $6$ of the systems have RUWE values that are below 1. This indicates that the orbital precession is likely driven by tides and not tertiary companions. 

However, tidal forces cannot account for the evolution of the eccentricity and inclination seen 
in some of these sources. For example, a linear fit to the eccentricities estimated from different TESS phase curves of TIC 444555685 yields a non-zero $\dot{e}$ of $7 \times 10^{-3}$ year$^{-1}$. The eccentricity changes are possible through the Kozai-Lidov resonance, although that requires the presence of compact and relatively inclined triple systems, which are rare.

\begin{figure}
    \centering
    \includegraphics[width=0.49\textwidth]{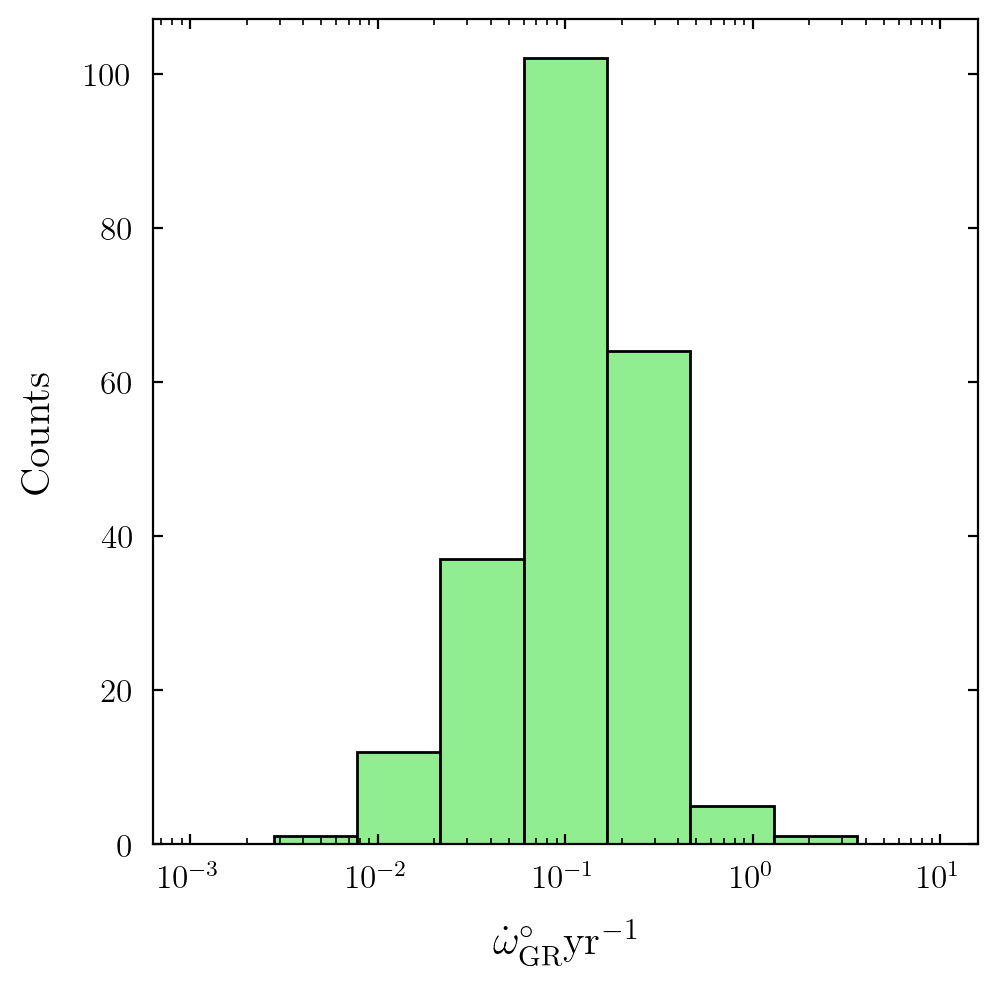}
    \caption{A histogram of the estimated periastron advance rates from general relativistic effects for the population. A precession rate of $\sim 10^{-1}$ degrees per year is typical for most of the HBs; however, this is as big as $1.7^{\circ}$ per year for TIC $405320687$, which was also one of the outliers in the period-eccentricity diagram.}
    \label{fig:gr-precession}
\end{figure}

In the above estimates, we have ignored the contribution to precession from general-relativistic effects, which is expected to be sub-dominant to the tidal precession (see \cite{liu_2015MNRAS.447..747L} for an estimate of the precession rates). However, if the internal structures of the two stars can be accurately modeled, then the precession contribution from tidal effects and rotation can be well constrained. Then, in the absence of other perturbing potentials, the observed precession rates in HBs can be used to test Einstein's theory of general relativity if the expected GR precession rate is larger than measurement uncertainties. A histogram of the GR precession for the HB population is plotted in Figure \ref{fig:gr-precession}, following Eq. (42) of \cite{liu_2015MNRAS.447..747L}. The mean precession rate is around $0.1^{\circ}$ yr$^{-1}$, however, this is as big as $1.7^{\circ}$ per year for TIC $405320687$. The high eccentricity of this system, combined with a large expected GR precession rate, makes this system an interesting candidate for future studies.

We end the discussion by listing some caveats we faced while modeling the systems and considerations that must be taken while interpreting our results. 

A few singly-eclipsing systems with TEOs were challenging to fit because the model confused the oscillations for eclipses. Specifically, these systems are TICs 441626681, 468721314, 42821678, 293950421, and 21325268. The TEO frequencies of these systems contain substantial harmonic power at non-integer multiples of the orbital period and could not be modeled with up to $4$ sinusoids. There were significant residuals when we tried subtracting the best-fit multi-sinusoidal fit from the original light curves. We tried the same procedure on the residuals obtained from subtracting the best-fit eBEER model from the light curve and still could not get a good fit for the TEOs, possibly suggesting that the variability is caused by star spots or other oscillations.

Most of the binaries in our current sample contain primaries with masses $\gtrsim 1.5 M_{\odot}$; see Sec. \ref{subsec:different_hb_systems}. These stars have small convective envelopes, giving rise to larger flux perturbations from tidal forcing than expected from equilibrium tide models \citep{pfahl_2008ApJ...679..783P}. Additionally, the flux perturbations from the photospheres of these stars can have phase lags relative to the tidal forcing, which is also inconsistent with the assumptions used to model the ellipsoidal flux perturbations \citep{pfahl_2008ApJ...679..783P}. This can result in the eBEER model predicting larger stellar masses, radii, and temperatures than their true values. Accurately modeling the light curves of HBs with massive stars would require prior knowledge of the stellar masses and the response functions of these stars to tidal perturbations. Therefore, the estimates of stellar masses, radii, and temperatures presented in Appendix \ref{appendix:list_of_hb_candidates} must not be taken as the ground truth.


\section{Conclusions} \label{sec:conclusion}
We identify $240$ short period ($P \lesssim 10 $ days) systems in the TESS FFI data in sectors 1-67. The light curves of these systems are jointly fit using the Gaia magnitudes and distances, together with the eBEER model of \cite{Engel_2020MNRAS.497.4884E} that accounts for flux variations from Doppler beaming, ellipsoidal variations, reflection from companions, and to which we add a model for eclipses. We evaluate our model by fitting it on previously monitored Kepler sources with well-determined orbits and find excellent agreement between the model and the observed parameters.

After we model the light curves of these systems, we find that $180$ systems have eccentricities over $0.2$, and we define these systems to be HB binaries. Our work extends the number of known HBs in TESS data from  $25$ \citep{tess_heartbeats_2021AA...647A..12K} to over $200$. The HB sample is primarily composed of A and B-spectral type stars with masses $M \gtrsim 1.5 M_{\odot}$, and contains $132$ eclipsing systems, systems with large-amplitude tidally excited oscillations, and $30$ sources that show changes in their phase curves over multi-year timescale.

 The fits on the entire population reveal that the sources in our TESS population are, on average, more eccentric for a given orbital period than the Kepler HBs. Fits to $12$ of the sources show long-term evolution in their light curves, revealing that they all show prograde apsidal precession as high as $\sim 9^{\circ}$ per year. The precession rates can be explained by three-body effects and precession from tidal bulges. In two of these systems, we find additional periodicities in their light curves that are likely internal modes of the stars since they do not occur at the theoretical pseudo-synchronous frequencies of the systems. Finally, we compute the periastron precession rate from general relativistic effects and find that roughly $\sim 0.1^\circ$ yr$^{-1}$ precession is expected for a majority of the systems, and this estimate goes up to $1.7^\circ$ yr$^{-1}$ for TIC $405320687$. 

Radial velocity measurements of these systems, in conjunction with modeling the stars and their eclipse timing variations, can provide tighter constraints on their physical parameters. This can help us understand how these extremely eccentric, short-period systems form. Future work will involve modeling the eclipse timing variations in conjunction to the heartbeat signal to get precise estimates of the orbital parameters and infer the presence of tertiary companions.
 

\begin{acknowledgments}
SS is thankful to Cole Miller, Doug Hamilton, and Omer Blaes for insightful conversations on orbital dynamics and to Sasha Philippov for supporting him through this work. We are grateful to the anonymous referee for their insightful comments. AC and JS were supported through the TESS Cycle-1 GI grant 011100. 
\end{acknowledgments}

\appendix

\section{eBEER Model} \label{Appendix:ebeer_equations}
We follow \cite{Engel_2020MNRAS.497.4884E} for the analytical model for the flux perturbations from eccentric BEaming, Ellipsoidal, and Reflection (eBEER) effects. In addition, we account for eclipses in our model, including a simple one-parameter limb-darkening prescription.

Table \ref{tab:orbital_parameters} lists the free parameters in the model with their corresponding symbols. The subscripts `$i$' and '$s$' for some of the symbols refer to the corresponding parameter for star $i$ or the TESS sector $s$, respectively.

\begin{table}[]
    \centering
    \begin{tabular}{||c|c||}
        \hline
         Parameter Name &  Symbol \\
         \hline \hline
         Stellar Mass & $M_i$\\
         \hline
         Eccentricity & $e$  \\
         \hline
         Inclination & $i$  \\
         \hline
         Argument of Periastron & $\omega$ \\
         \hline
         True Anomaly at $t=0$   & $\phi_0$ \\
         \hline
         Limb-darkening coefficient & $\mu_i$ \\
         \hline
         Gravity-darkening coefficient & $\tau_i$ \\
         \hline
         Radius re-scaling factor & $\beta_{R,i}$ \\
         \hline
         Temperature re-scaling factor & $\beta_{T,i}$ \\
         \hline
         Reflection coefficient & $\alpha_{\text{ref,i}}$ \\
         \hline
         Beaming coefficient & $\alpha_{\text{beam,i}}$  \\
         \hline
         Blending  & $\delta_{s}$\\
         \hline
         Flux normalization & $\Sigma_{s}$ \\
         \hline
         Noise-rescaling & $\sigma_{s}$ \\
         \hline \hline
    \end{tabular}
    \caption{Model parameters and their corresponding symbols that are used in the model for the heartbeat light curve.}
    \label{tab:orbital_parameters}
\end{table}

We compute the orbital separation $d$ and the true anomaly $\nu$ using the orbital parameters by solving Kepler's equation for each phase of the data. Then, the stellar radii and temperatures are determined using equations presented in Appendix \ref{Appendix:Temperature_radius_estimatation}. Finally, we construct the light curves by adding the flux contributions from the Doppler beaming, ellipsoidal variations, reflection from the companion, and eclipses. 

Beaming results in the red-shifting or blue-shifting of fluxes from the two stars based on their line of sight velocities. This effect is generally weaker than the ellipsoidal and reflection effects \citep{faigler_mazeh_2011MNRAS.415.3921F}. The fractional change in flux from the relativistic beaming, $\Delta F_i/F_{\text{beam}}$ , is given by
\begin{equation}
  \frac{\Delta F_i}{F}_{\text{beam}} = \left[-2830 \bar{\alpha}_{\text{beam, i}} \frac{q}{(1+q)^{2/3}} \left(\frac{M_i}{M_{\odot}}\right) \left(\frac{d}{P_{\text{orb}}}\right)\sin i^{\circ} \frac{\cos(\omega + \nu)}{\sqrt{1 - e^2}} \right] \times 10^{-6} 
\end{equation}
where $q = M_2 / M_1$ is the mass ratio and $\bar{\alpha}_{\text{beam, i}}$ is the beaming coefficient which is equal to $\alpha_{\text{beam,i}} f(T) / 4$. $f(T)$ is a piece-wise linear fit to the beaming coefficient, estimated from Figure 5 of \cite{claret_2020A&A...641A.157C} (see Table \ref{tab:beaming_fit}). Note the factor of $4$ comes from different definitions for the beaming coefficient between \cite{claret_2020A&A...641A.157C} and \cite{Engel_2020MNRAS.497.4884E}. We allow further flexibility in the model by allowing the beaming coefficient $\alpha_{\text{beam,i}}$ to vary from source-to-source. 

\begin{table}[]
    \centering
    \begin{tabular}{||c|c||}
        \hline
        $\log T$ & $f(T)$ \\ 
        \hline \hline
        $3.5$ & $6.5$ \\
        \hline
        $3.7$ & $4$ \\
        \hline
        $3.9$ & $2.5$ \\
        \hline
        $4.5$ & $1.2$ \\
        \hline
    \end{tabular}
    \caption{The nodes defining $f(T)$, which is the beaming coefficient as a function of stellar temperature.}
    \label{tab:beaming_fit}
\end{table}

Similarly, the fractional flux contribution from ellipsoidal variations is
\begin{equation}
    \begin{aligned}
       \frac{\Delta F_i}{F}_{\text{ellip}} =  \left[13435 \, 2 \alpha_{e0,1}  (2 - 3\sin^2 i) \left(\frac{M_i}{M_{\odot}} \right)^{-1}
        \left(\frac{P_{\rm orb}}{d}\right)^{-2} \left(1-e \right)^{-3/2} \left(\frac{R_i}{R_{\odot}} \right)^3 \right. \\
         + 13435 \, 3 \alpha_{e0,1}  (2 - 3\sin^2 i) \left(\frac{M_i}{M_{\odot}} \right)^{-1} \frac{q}{1+q}
        \left(\frac{P_{\rm orb}}{d}\right)^{-2} \left(\frac{\beta R_i}{R_{\odot}} \right)^3 \\
        + 759 \, \alpha_{e0b,1} (8 - 40\sin^2 i + 35\sin^{4}i) \left(\frac{M_i}{M_{\odot}}\right)^{-5/3} \frac{q}{(1+q)^{5/3}}
        \left(\frac{P_{\rm orb}}{d}\right)^{-10/3} \left(\frac{\beta R_i}{R_{\odot}} \right)^5 \\
        + 3194 \, \alpha_{e1,1} (4\sin i - 5\sin^3 i) \left(\frac{M_i}{M_{\odot}}\right)^{-4/3} \frac{q}{(1+q)^{4/3}} 
        \left(\frac{P_{\rm orb}}{d}\right)^{-8/3} \left(\frac{\beta R_i}{R_{\odot}} \right)^4 \sin(\omega + \nu) \\
        + 13435 \, \alpha_{e2,1} (\sin^2 i) \left(\frac{M_i}{M_{\odot}}\right)^{-1} \frac{q}{(1+q)}
        \left(\frac{P_{\rm orb}}{d}\right)^{-2} \left(\frac{\beta R_i}{R_{\odot}} \right)^3 \cos(2(\omega + \nu)) \\
        + 759 \, \alpha_{e2b,1} (6\sin^2 i - 7\sin^4 i) \left(\frac{M_i}{M_{\odot}}\right)^{-5/3} \frac{q}{(1+q)^{5/3}} 
        \left(\frac{P_{\rm orb}}{d}\right)^{-10/3} \left(\frac{\beta R_i}{R_{\odot}} \right)^5 \sin(\omega + \nu) \\
        + 3194 \, \alpha_{e3,1} (\sin^3i) \left(\frac{M_i}{M_{\odot}}\right)^{-4/3} \frac{q}{(1+q)^{4/3}} 
        \left(\frac{P_{\rm orb}}{d}\right)^{-8/3} \left(\frac{\beta R_i}{R_{\odot}} \right)^4 \sin(3(\omega + \nu)) \\
        + 759 \, \alpha_{e4,1} (\sin^4i) \left(\frac{M_i}{M_{\odot}}\right)^{-5/3} \frac{q}{(1+q)^{5/3}} 
        \left. \left(\frac{P_{\rm orb}}{d}\right)^{-10/3} \left(\frac{\beta R_i}{R_{\odot}} \right)^5 \cos(4(\omega + \nu))\right] \times 10^{-6}.
    \end{aligned}
\end{equation}

Here $\frac{\Delta F_i}{F}_{\text{ellip}}$ are the fractional flux variations from ellipsoidal variations of star $i$.  $\alpha_i$ are functions of gravity and limb-darkening coefficients, $\mu_i$ and $\tau_i$, which are the model parameters. The $\alpha_i$ are defined in equation (5) of \cite{Engel_2020MNRAS.497.4884E} and $\beta = \frac{1 + e\cos\nu}{1-e^2}$. 

The reflected starlight off the companion also adds to the flux variability. The reflection effects are usually modeled by assuming the companion to be a Lambertian surface \citep{faigler_mazeh_2011MNRAS.415.3921F}. These are given by:
\begin{equation}
\begin{aligned}
    \frac{\Delta F_{\text{ref}, i}}{F} = \left[ 56514 \, \alpha_{\text{ref}, i} (1 + q)^{-2/3} \left(\frac{M_1}{M_{\odot}}\right)^{-2/3} \left(\frac{P_{\text{orb}}}{d}\right)^{-4/3} \left(\frac{\beta R_2}{R_{\odot}}\right)^2 \right. \\ 
    \left. (0.64 - \sin{i}\sin(\omega + \nu) + 0.18\sin^2{i}(1 - cos(2(\omega + \nu))) \right] \times 10^{-6}.
\end{aligned}
\end{equation}

All of our sources have been observed in more than one TESS sector. The light curves from different TESS sectors have different amounts of blending from other sources, and different uncertainties on the data. Furthermore, the light curves once folded on different sectors can have slightly different normalizations. To model different sector data simultaneously, we introduce three additional parameters - blending ($\delta$), noise re-scaling ($\sigma$) and flux normalization $\Sigma$.

The complete set of parameters is $\vec{\theta} = \{M_1, M_2, e, i^{\circ}, \omega, \phi_0, \mu_1, \tau_1, \mu_2, \tau_2, \alpha_{\text{ref, 1}}, \alpha_{\text{ref, 2}}, \gamma_{\text{resc, 1}}, \\ \gamma_{\text{resc, 2}}, \beta_{R,1}, \beta_{R,2}, \beta_{T,1}, \beta_{T,2}, \delta_{s_1}, \sigma_{s_1}, \Sigma_{s_1}, ... \delta_{s_N}, \sigma_{s_N}, \Sigma_{s_N}\}$ The subscripts $1$ and $2$ refer to the two stars and the subscript $s_i$ refers to the $i^{\text{th}}$ sector. 

\section{Temperature and Radius Estimation} \label{Appendix:Temperature_radius_estimatation}
\begin{figure}
    \centering
    \includegraphics[width=0.99\textwidth]{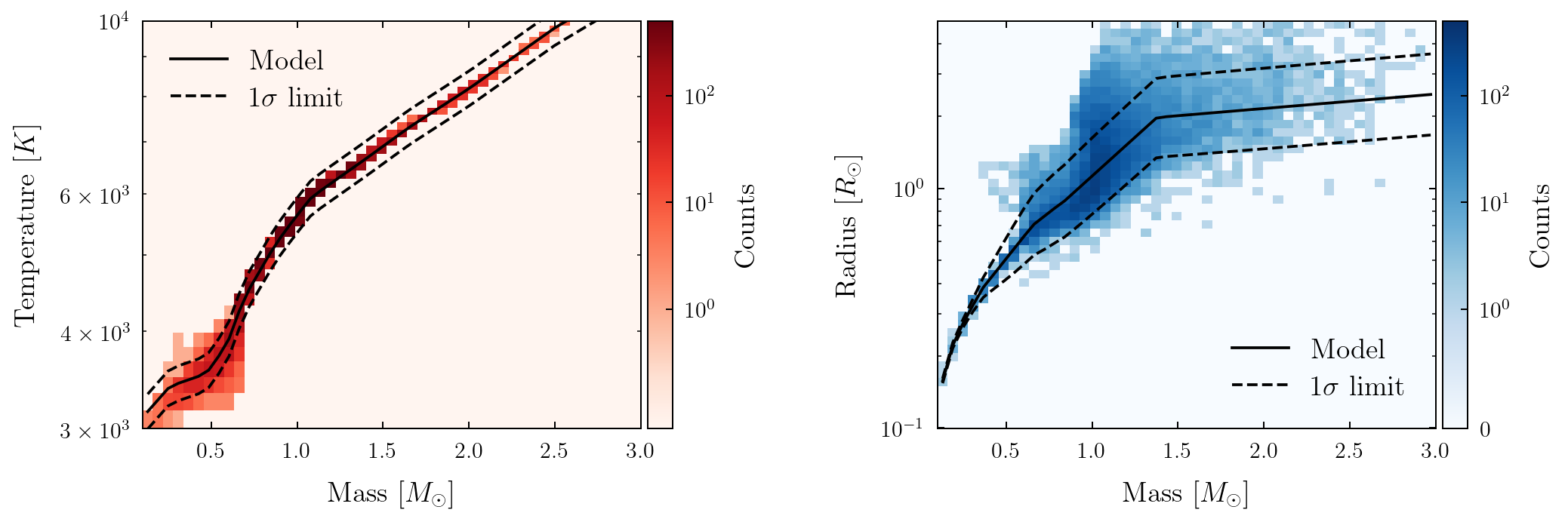}
    \caption{The mass-radius and mass-temperature measurements of $\approx 3 \times 10^5$ sources taken from the TESS Catalog . We fit a piece-wise linear function to the measurements to obtain a scaling relation. We also calculate the $\pm 1 \sigma$ deviations from the mean fit to measure the spread at a given mass.}
    \label{fig:piecewise_fits}
\end{figure}

The model requires a stellar mass-radius and mass-temperature relationship. To do so, we use a piece-wise linear fit to the $3 \times 10^5$ mass-radius and mass-temperature measurements taken from the TESS input catalog. The fits give us mean temperature-mass $T(M)$ and radius-mass $R(M)$ relations and estimates on the 1-$\sigma$ spread in temperatures $\sigma_T(M)$ and radii $\sigma_R(M)$ as a function of mass. These relations are plotted in Figure \ref{fig:piecewise_fits}. The black dashed lines represent the mean relations and the red and the blue lines represent the $\pm \sigma$ estimates for the spread in temperatures and radii respectively. The temperatures of the stars is defined in the model as $T_* = 10^{T(M) + \beta_T \sigma_T(M)}$ and similarly the radii are defined as $R_* = 10^{R(M) + \beta_R \sigma_R(M)}$. From this definition, the model parameters $\beta_T$ and $\beta_R$ capture the deviation from the measured mean temperature and radii distributions in units of $\sigma_{T,R}$. 

Then, each star $i$ in the system is taken to be a blackbody, with temperature $T_i$ and radius $R_i$. In order to compute $T_i$ and $R_i$ from the model parameters in Table \ref{tab:orbital_parameters}, we first use a linear interpolation table to compute the average temperature and radius of the two stars as a function of their mass. These are then re-scaled by the model parameters $\beta_{[T,R],i}$. A normal prior on these parameters accounts for our fit for the observed variations in the mass-radius and mass-temperature relations.
Using the distance to the source from the Gaia DR3, we compute the corresponding G-mag for the system. The difference between the computed and observed G-mags is then used to make a contribution to the likelihood evaluation.

\section{Markov Chain Monte Carlo Method} \label{Appendix:MCMC}
We use a custom MCMC sampler that uses parallel tempering to sample the model parameters. The sampler is written in \textsc{C} and uses \textsc{OPENMP} parallelization, enabling quick computation of the likelihood function. Using $25$ cores takes an average of $4.4 \times 10^{-5}$ seconds per chain update, using a light curve with $500$ data points. The MCMC script is run for at least $10^6$ iterations per chain for each source. 

\subsection{Parallel Tempering}
We use $50$ chains for each of our runs, where the temperature of the $i^{\text{th}}$ chain is $T_i$ and $T_{i+1} / T_i = 1.4$. Likewise, the posterior of the $i^{\text{th}}$ chain is computed as $\pi^i(\vec{\theta}) = \mathcal{L}(\vec{y}, \vec{\theta})^{1/T_i} \Pi(\vec{\theta})$, where $\vec{\theta}$ and $\vec{y}$ are the sampled parameters and light curve array respectively, and $\pi^i$ is the posterior for chain $i$, $\mathcal{L}$ is the likelihood and $\Pi$ is the prior. Note that the posterior approaches the prior for the chains with the larger temperatures, thus enabling an effective scan of the parameter space.
A pair of chains $i$ and $j$ are randomly selected after every iteration, and their temperatures are swapped if $\exp(-\frac{\ln \mathcal{L}_j - \ln \mathcal{L}_i}{1/T_j - 1/T_i}) > \beta$, where $\beta$ is uniformly sampled from $(0,1]$. 

\subsection{Proposals}
The new parameters are sampled using a combination of Gaussian and differential evolution proposals. In the Gaussian proposal, each parameter's displacement vector is sampled independently from a Gaussian distribution with a zero mean and unit variance. The result is then scaled by a parameter-specific prefactor and the square root of the chain's temperature. This way, hotter chains take larger steps in the parameter space than the colder ones.
After the first $500$ iterations, the parameters are also sampled using the differential evolution method. In the latter, each chain's parameter history is stored from the previous $500$ iterations, and a displacement vector connecting two randomly chosen samples from the history is drawn. Each element of the displacement is scaled by a number drawn from a Gaussian with zero mean and a unit norm.
Each of the above two methods has an equal probability of being called to generate a proposal after the first $500$ iterations. 

\subsection{Priors}
We use a combination of uniform and Gaussian priors depending on the model parameter. Table \ref{tab:uniform_priors} contains the list of parameters with uniform priors along with their upper and lower limits. Table \ref{tab:gaussian_priors} lists the parameters with Gaussian priors with their upper and lower limits and also the means and standard deviation for the priors.

\begin{table}[]
    \centering
    \begin{tabular}{||c|c|c||}
    \hline
    Parameter & Lower Limit & Upper Limit \\
    \hline
    $\log M_i$ & $-1.5$ & $2$ \\
    \hline
    $e$ & $0.$ & $1$ \\
    \hline
    $\cos i$ & $0 $ & $1$ \\
    \hline
    $\omega$ & $-\pi$ & $\pi$ \\
    \hline
    $t_0$ & $0$ & $P$ \\
    \hline
    $\delta_s$ & $0$ & $1$ \\
    \hline
    $\Sigma_s$ & $0.99$ & $1.01$ \\
    \hline
    \end{tabular}
    \caption{Model parameters with uniform priors along with their lower and upper limits while making proposals in the MCMC.}
    \label{tab:uniform_priors}
\end{table}

\begin{table}[]
    \centering
    \begin{tabular}{||c|c|c|c|c||}
    \hline
    Parameter & Lower Limit & Upper Limit & Mean & Standard Deviation \\
    \hline
        $\beta_{R,i}$ & $-5$ & $5$ & $0$ & $1$ \\
        \hline
        $\beta_{T,i}$ & $-5 $& $5$ & $0$ & $1$ \\
        \hline
        $\mu_i$  & $0$ & $1$ & $0.2$ & $0.2$ \\
        \hline
        $\tau_i$  & $0.15$ & $0.55$ & $0.4$ & $0.2$ \\
        \hline
        $\alpha_{\text{ref},i}$  & $0$ & $2$ & $1$ & $0.2$ \\
        \hline
        $\alpha_{\text{beam},i}$  & $-0.3$ & $0.3$ & $0$ & $0.1$ \\
        \hline
        $\sigma_s$  & $-0.1$ & $0.1$ & $0$ & $0.03$ \\
        \hline
    \end{tabular}
    \caption{Same as Table \ref{tab:uniform_priors} but for parameters that use Gaussian priors. Additionally, we also mention the mean and standard deviation of these priors.}
    \label{tab:gaussian_priors}
\end{table}

\section{List of Heartbeat Candidates} \label{appendix:list_of_hb_candidates}
We list the TICs, absolute Gaia magnitudes, periods and some of the system parameters estimated from the eBEER fits. The upper and lower limits on uncertainties in the model parameters bracket the $16^{\text{th}}$ and $84^{\text{th}}$ percentile values from the posteriors, respectively. The current list only contains systems with $e \geq 0.2$ that we could fit with the eBEER model.

\begin{longtable}{||c|c|c|c|c|c|c|c|c||}
\caption{The Gaia magnitudes, periods, and other inferred parameters from the MCMC fits to the phase curves of the $180$ heartbeat candidates. The errorbars for $M_1, M_2, e, i$ and $\omega$ contain $16^{\text{th}}-84^{\text{th}}$ percentile values from the posteriors.} \label{tab:HB_params} \\
 \hline
 \textbf{TIC} & \textbf{$M_G$} & \textbf{Period [days]} & \textbf{$M_{1} [M_{\odot}]$} & \textbf{$M_{2} [M_{\odot}]$} & \textbf{$e$} & \textbf{$i^{\circ}$} & \textbf{$\omega^{\circ}$} & \textbf{blending} \\
 \hline
 \endfirsthead

\multicolumn{9}{c}%
{\tablename\ \thetable\ -- \textit{Continued from previous page}} \\ \hline
 \textbf{TIC} & \textbf{$M_G$} & \textbf{Period [days]} & \textbf{$M_{1} [M_{\odot}]$} & \textbf{$M_{2} [M_{\odot}]$} & \textbf{$e$} & \textbf{$i^{\circ}$} & \textbf{$\omega^{\circ}$} & \textbf{blending} \\ \hline
\endhead
$8772474$ &$-1.85 \pm 1.15$ &$3.13178 \pm 0.00006$ &${3.3^{+2.5}_{-0.8}}$ &${0.8^{+0.4}_{-0.2}}$ &${0.372^{+0.003}_{-0.003}}$ &${48.21^{+1.03}_{-0.63}}$ &${-49.92^{+0.57}_{-0.68}}$ &${0.24^{+0.08}_{-0.08}}$\\ 
$13951387$ &$-3.52 \pm 0.83$ &$3.68465 \pm 0.00009$ &${11.2^{+5.3}_{-3.6}}$ &${3.5^{+2.5}_{-1.7}}$ &${0.327^{+0.008}_{-0.009}}$ &${43.98^{+5.61}_{-2.62}}$ &${-23.17^{+3.05}_{-2.51}}$ &${0.24^{+0.15}_{-0.11}}$\\ 
$19971063$ &$1.21 \pm 0.23$ &$13.99151 \pm 0.00152$ &${1.5^{+0.2}_{-0.1}}$ &${1.3^{+0.2}_{-0.1}}$ &${0.606^{+0.001}_{-0.001}}$ &${89.27^{+0.45}_{-0.73}}$ &${-1.00^{+0.60}_{-0.64}}$ &${0.07^{+0.08}_{-0.05}}$\\ 
$28431075$ &$0.38 \pm 0.43$ &$2.96264 \pm 0.00005$ &${2.2^{+0.3}_{-0.3}}$ &${0.3^{+0.1}_{-0.0}}$ &${0.297^{+0.006}_{-0.006}}$ &${42.19^{+0.58}_{-0.64}}$ &${-5.77^{+1.11}_{-1.21}}$ &${0.24^{+0.10}_{-0.13}}$\\ 
$37710507$ &$2.30 \pm 0.16$ &$5.32352 \pm 0.00022$ &${1.6^{+0.1}_{-0.1}}$ &${1.2^{+0.1}_{-0.1}}$ &${0.342^{+0.003}_{-0.002}}$ &${84.70^{+0.22}_{-0.19}}$ &${-21.26^{+1.17}_{-1.24}}$ &${0.29^{+0.06}_{-0.05}}$\\ 
$49924598$ &$-0.71 \pm 0.26$ &$4.49943 \pm 0.00020$ &${3.2^{+0.3}_{-0.2}}$ &${2.2^{+0.2}_{-0.3}}$ &${0.371^{+0.003}_{-0.003}}$ &${78.08^{+0.47}_{-0.43}}$ &${-66.88^{+0.33}_{-0.31}}$ &${0.45^{+0.07}_{-0.09}}$\\ 
$50280622$ &$-0.53 \pm 0.44$ &$3.85060 \pm 0.00006$ &${2.6^{+0.3}_{-0.3}}$ &${1.0^{+0.9}_{-0.3}}$ &${0.297^{+0.002}_{-0.003}}$ &${47.61^{+0.78}_{-1.10}}$ &${-17.37^{+0.63}_{-0.64}}$ &${0.17^{+0.12}_{-0.07}}$\\ 
$51243999$ &$-0.30 \pm 0.22$ &$5.99401 \pm 0.00029$ &${3.8^{+0.2}_{-0.2}}$ &${2.6^{+0.3}_{-0.2}}$ &${0.480^{+0.002}_{-0.002}}$ &${86.93^{+0.38}_{-0.40}}$ &${19.66^{+0.57}_{-0.66}}$ &${0.15^{+0.06}_{-0.05}}$\\ 
$53194798$ &$-0.85 \pm 0.69$ &$3.45109 \pm 0.00005$ &${5.0^{+1.0}_{-1.6}}$ &${3.6^{+0.8}_{-1.1}}$ &${0.299^{+0.003}_{-0.009}}$ &${84.21^{+0.97}_{-1.09}}$ &${49.16^{+0.53}_{-1.85}}$ &${0.17^{+0.19}_{-0.16}}$\\ 
$53824793$ &$0.63 \pm 0.25$ &$3.41680 \pm 0.00010$ &${2.4^{+0.2}_{-0.2}}$ &${0.5^{+0.1}_{-0.1}}$ &${0.441^{+0.004}_{-0.004}}$ &${65.08^{+3.18}_{-5.64}}$ &${70.27^{+0.98}_{-0.96}}$ &${0.25^{+0.09}_{-0.16}}$\\ 
$59090149$ &$-1.09 \pm 0.31$ &$4.64187 \pm 0.00012$ &${3.3^{+0.3}_{-0.2}}$ &${3.0^{+0.2}_{-0.3}}$ &${0.321^{+0.001}_{-0.001}}$ &${79.97^{+0.21}_{-0.38}}$ &${8.79^{+0.41}_{-0.48}}$ &${0.34^{+0.04}_{-0.03}}$\\ 
$64290127$ &$-1.11 \pm 0.59$ &$8.55666 \pm 0.00959$ &${5.9^{+3.0}_{-1.7}}$ &${0.8^{+0.7}_{-0.2}}$ &${0.726^{+0.007}_{-0.009}}$ &${78.48^{+4.97}_{-10.12}}$ &${76.20^{+2.33}_{-3.55}}$ &${0.25^{+0.15}_{-0.15}}$\\ 
$65201080$ &$1.47 \pm 0.22$ &$4.22062 \pm 0.00011$ &${1.5^{+0.2}_{-0.2}}$ &${0.5^{+0.1}_{-0.1}}$ &${0.219^{+0.002}_{-0.002}}$ &${76.96^{+0.67}_{-0.67}}$ &${-18.18^{+0.85}_{-0.99}}$ &${0.15^{+0.10}_{-0.07}}$\\ 
$72834545$ &$-1.16 \pm 0.42$ &$4.59276 \pm 0.00808$ &${3.2^{+3.6}_{-0.9}}$ &${0.5^{+0.5}_{-0.2}}$ &${0.524^{+0.241}_{-0.012}}$ &${40.31^{+34.52}_{-2.00}}$ &${-5.92^{+74.51}_{-5.61}}$ &${0.39^{+0.19}_{-0.22}}$\\ 
$78379043$ &$-0.11 \pm 0.17$ &$3.09693 \pm 0.00010$ &${3.6^{+0.3}_{-0.3}}$ &${1.0^{+0.5}_{-0.2}}$ &${0.462^{+0.004}_{-0.004}}$ &${67.86^{+1.87}_{-1.87}}$ &${-27.55^{+1.36}_{-1.36}}$ &${0.17^{+0.11}_{-0.09}}$\\ 
$78379048$ &$1.66 \pm 0.15$ &$3.12414 \pm 0.00011$ &${1.9^{+0.2}_{-0.2}}$ &${0.5^{+0.1}_{-0.0}}$ &${0.447^{+0.006}_{-0.006}}$ &${70.86^{+1.23}_{-1.67}}$ &${-22.11^{+2.30}_{-2.31}}$ &${0.21^{+0.12}_{-0.13}}$\\ 
$79686189$ &$-0.12 \pm 0.53$ &$5.94037 \pm 0.00031$ &${6.9^{+2.7}_{-3.9}}$ &${4.1^{+1.5}_{-2.5}}$ &${0.685^{+0.008}_{-0.098}}$ &${66.94^{+8.81}_{-33.20}}$ &${-20.36^{+51.07}_{-5.86}}$ &${0.36^{+0.14}_{-0.20}}$\\ 
$86128744$ &$2.46 \pm 0.45$ &$1.99160 \pm 0.00100$ &${0.8^{+0.3}_{-0.1}}$ &${0.7^{+0.1}_{-0.3}}$ &${0.283^{+0.004}_{-0.004}}$ &${73.71^{+1.51}_{-0.79}}$ &${20.79^{+4.81}_{-1.90}}$ &${0.70^{+0.03}_{-0.07}}$\\ 
$89522181$ &$1.63 \pm 0.09$ &$2.34698 \pm 0.00003$ &${2.4^{+0.2}_{-0.1}}$ &${1.5^{+0.1}_{-0.1}}$ &${0.323^{+0.001}_{-0.001}}$ &${86.67^{+0.26}_{-0.23}}$ &${9.38^{+0.41}_{-0.62}}$ &${0.19^{+0.02}_{-0.01}}$\\ 
$90547242$ &$-0.56 \pm 0.33$ &$6.32273 \pm 0.00018$ &${3.4^{+0.4}_{-0.4}}$ &${3.0^{+0.4}_{-0.6}}$ &${0.305^{+0.003}_{-0.003}}$ &${80.55^{+0.70}_{-0.86}}$ &${38.51^{+1.35}_{-1.15}}$ &${0.11^{+0.14}_{-0.06}}$\\ 
$102289966$ &$0.48 \pm 0.21$ &$6.25702 \pm 0.00016$ &${2.4^{+0.1}_{-0.1}}$ &${1.4^{+0.1}_{-0.2}}$ &${0.312^{+0.003}_{-0.003}}$ &${83.44^{+0.45}_{-0.83}}$ &${36.01^{+0.75}_{-0.74}}$ &${0.15^{+0.12}_{-0.11}}$\\ 
$106886673$ &$0.44 \pm 0.25$ &$3.62355 \pm 0.00005$ &${2.5^{+0.2}_{-0.3}}$ &${2.1^{+0.2}_{-0.2}}$ &${0.314^{+0.001}_{-0.001}}$ &${80.54^{+0.24}_{-0.24}}$ &${11.34^{+0.42}_{-0.60}}$ &${0.19^{+0.06}_{-0.10}}$\\ 
$110106107$ &$-0.69 \pm 0.53$ &$4.24699 \pm 0.00008$ &${3.5^{+0.4}_{-0.4}}$ &${1.9^{+0.3}_{-0.2}}$ &${0.291^{+0.001}_{-0.001}}$ &${78.52^{+0.66}_{-0.66}}$ &${-15.96^{+0.57}_{-0.66}}$ &${0.16^{+0.10}_{-0.07}}$\\ 
$110602878$ &$-0.04 \pm 0.28$ &$5.35246 \pm 0.00013$ &${2.8^{+0.3}_{-0.3}}$ &${1.0^{+0.1}_{-0.1}}$ &${0.356^{+0.001}_{-0.001}}$ &${89.09^{+0.50}_{-0.61}}$ &${-19.54^{+0.50}_{-0.53}}$ &${0.08^{+0.08}_{-0.04}}$\\ 
$115394297$ &$-1.37 \pm 0.74$ &$5.06355 \pm 0.00016$ &${5.4^{+0.9}_{-1.1}}$ &${4.7^{+0.7}_{-1.3}}$ &${0.299^{+0.001}_{-0.001}}$ &${76.05^{+0.80}_{-0.36}}$ &${-31.25^{+0.43}_{-0.30}}$ &${0.10^{+0.14}_{-0.05}}$\\ 
$115866969$ &$-0.61 \pm 7.77$ &$2.75258 \pm 0.00004$ &${6.2^{+1.5}_{-1.2}}$ &${5.3^{+1.6}_{-0.7}}$ &${0.253^{+0.001}_{-0.001}}$ &${79.34^{+0.41}_{-0.41}}$ &${-10.45^{+0.61}_{-0.49}}$ &${0.37^{+0.04}_{-0.04}}$\\ 
$116066534$ &$-1.45 \pm 1.06$ &$2.48426 \pm 0.00003$ &${3.0^{+3.4}_{-0.6}}$ &${2.6^{+2.7}_{-0.6}}$ &${0.375^{+0.002}_{-0.001}}$ &${83.08^{+1.32}_{-0.95}}$ &${-4.23^{+0.43}_{-5.26}}$ &${0.38^{+0.12}_{-0.19}}$\\ 
$118305806$ &$0.52 \pm 0.24$ &$3.67851 \pm 0.00008$ &${2.0^{+0.3}_{-0.2}}$ &${1.8^{+0.2}_{-0.3}}$ &${0.352^{+0.002}_{-0.002}}$ &${74.54^{+2.48}_{-0.90}}$ &${46.26^{+2.31}_{-0.91}}$ &${0.26^{+0.20}_{-0.12}}$\\ 
$118798174$ &$1.64 \pm 0.15$ &$7.56722 \pm 0.00019$ &${5.6^{+1.0}_{-1.4}}$ &${1.3^{+0.2}_{-0.2}}$ &${0.565^{+0.013}_{-0.014}}$ &${40.70^{+3.01}_{-2.25}}$ &${-28.37^{+1.57}_{-1.79}}$ &${0.30^{+0.12}_{-0.21}}$\\ 
$120684604$ &$1.15 \pm 0.22$ &$2.18911 \pm 0.00002$ &${2.3^{+0.3}_{-0.2}}$ &${2.0^{+0.2}_{-0.1}}$ &${0.282^{+0.001}_{-0.001}}$ &${88.55^{+0.93}_{-1.00}}$ &${-15.86^{+0.59}_{-0.67}}$ &${0.10^{+0.03}_{-0.04}}$\\ 
$124412957$ &$-2.33 \pm 1.02$ &$3.41066 \pm 0.00011$ &${4.1^{+0.9}_{-0.6}}$ &${0.1^{+0.0}_{-0.0}}$ &${0.203^{+0.006}_{-0.006}}$ &${63.15^{+8.00}_{-2.21}}$ &${-5.56^{+3.63}_{-2.79}}$ &${0.16^{+0.13}_{-0.10}}$\\ 
$127079833$ &$-0.47 \pm 0.62$ &$3.13653 \pm 0.00006$ &${2.3^{+0.3}_{-0.2}}$ &${0.5^{+0.3}_{-0.1}}$ &${0.210^{+0.003}_{-0.003}}$ &${45.42^{+0.85}_{-0.64}}$ &${-38.74^{+0.73}_{-0.84}}$ &${0.16^{+0.08}_{-0.09}}$\\ 
$137810570$ &$1.78 \pm 0.06$ &$3.84173 \pm 0.00009$ &${1.3^{+0.2}_{-0.2}}$ &${0.5^{+0.4}_{-0.3}}$ &${0.297^{+0.003}_{-0.003}}$ &${43.82^{+6.03}_{-3.69}}$ &${9.25^{+0.85}_{-0.86}}$ &${0.35^{+0.07}_{-0.11}}$\\ 
$145974291$ &$1.09 \pm 0.16$ &$5.21445 \pm 0.00012$ &${2.6^{+0.1}_{-0.2}}$ &${2.2^{+0.2}_{-0.2}}$ &${0.549^{+0.003}_{-0.003}}$ &${87.20^{+0.15}_{-0.18}}$ &${-49.98^{+0.42}_{-0.33}}$ &${0.26^{+0.05}_{-0.07}}$\\ 
$147960368$ &$0.73 \pm 0.43$ &$4.05025 \pm 0.00015$ &${2.2^{+0.3}_{-0.2}}$ &${0.7^{+0.2}_{-0.2}}$ &${0.421^{+0.007}_{-0.007}}$ &${48.30^{+1.59}_{-1.09}}$ &${-5.55^{+1.95}_{-1.90}}$ &${0.22^{+0.16}_{-0.13}}$\\ 
$149411271$ &$-3.94 \pm 0.94$ &$6.20414 \pm 0.00017$ &${91.9^{+5.6}_{-11.3}}$ &${21.8^{+4.7}_{-3.7}}$ &${0.316^{+0.002}_{-0.002}}$ &${88.88^{+0.69}_{-0.87}}$ &${11.35^{+1.95}_{-3.81}}$ &${0.12^{+0.09}_{-0.08}}$\\ 
$150284425$ &$-0.28 \pm 0.11$ &$3.00443 \pm 0.00005$ &${2.9^{+0.3}_{-0.2}}$ &${2.4^{+0.2}_{-0.2}}$ &${0.286^{+0.003}_{-0.002}}$ &${78.84^{+0.39}_{-0.22}}$ &${51.08^{+0.53}_{-0.38}}$ &${0.23^{+0.35}_{-0.21}}$\\ 
$151467181$ &$-2.79 \pm 0.85$ &$5.07779 \pm 0.00018$ &${8.4^{+16.4}_{-1.9}}$ &${5.4^{+10.4}_{-1.0}}$ &${0.263^{+0.003}_{-0.004}}$ &${79.81^{+0.20}_{-0.15}}$ &${50.04^{+0.63}_{-0.68}}$ &${0.04^{+0.06}_{-0.03}}$\\ 
$153043302$ &$-0.35 \pm 0.47$ &$3.62329 \pm 0.00009$ &${2.4^{+0.3}_{-0.2}}$ &${0.3^{+0.1}_{-0.0}}$ &${0.411^{+0.006}_{-0.006}}$ &${37.74^{+0.59}_{-0.60}}$ &${-29.39^{+1.20}_{-1.12}}$ &${0.13^{+0.17}_{-0.08}}$\\ 
$158824564$ &$1.45 \pm 0.52$ &$5.72476 \pm 0.00033$ &${1.8^{+0.3}_{-0.2}}$ &${1.4^{+0.3}_{-0.2}}$ &${0.231^{+0.001}_{-0.001}}$ &${83.91^{+0.63}_{-0.23}}$ &${11.19^{+0.57}_{-0.71}}$ &${0.10^{+0.14}_{-0.07}}$\\ 
$169398679$ &$1.56 \pm 0.23$ &$4.85237 \pm 0.00021$ &${2.5^{+0.3}_{-0.4}}$ &${0.3^{+0.1}_{-0.1}}$ &${0.672^{+0.020}_{-0.070}}$ &${72.33^{+8.73}_{-15.29}}$ &${-72.84^{+20.51}_{-6.43}}$ &${0.31^{+0.16}_{-0.17}}$\\ 
$172985206$ &$1.23 \pm 0.17$ &$5.70475 \pm 0.00016$ &${1.5^{+0.2}_{-0.3}}$ &${0.4^{+0.3}_{-0.2}}$ &${0.375^{+0.004}_{-0.004}}$ &${45.37^{+2.59}_{-1.98}}$ &${-7.40^{+0.93}_{-0.92}}$ &${0.18^{+0.13}_{-0.09}}$\\ 
$174094640$ &$1.29 \pm 0.13$ &$6.19336 \pm 0.00023$ &${3.5^{+0.9}_{-1.7}}$ &${2.6^{+0.7}_{-1.8}}$ &${0.694^{+0.017}_{-0.075}}$ &${50.43^{+12.71}_{-10.96}}$ &${-53.73^{+29.51}_{-10.42}}$ &${0.22^{+0.12}_{-0.13}}$\\ 
$178008663$ &$0.70 \pm 0.38$ &$2.75457 \pm 0.00006$ &${2.8^{+0.2}_{-0.2}}$ &${2.4^{+0.2}_{-0.2}}$ &${0.362^{+0.002}_{-0.002}}$ &${83.41^{+0.51}_{-0.84}}$ &${17.16^{+0.86}_{-0.75}}$ &${0.11^{+0.29}_{-0.06}}$\\ 
$178436848$ &$1.53 \pm 0.23$ &$3.13724 \pm 0.00008$ &${1.9^{+0.1}_{-0.2}}$ &${0.7^{+0.0}_{-0.0}}$ &${0.202^{+0.002}_{-0.002}}$ &${81.22^{+0.33}_{-0.37}}$ &${-27.06^{+0.98}_{-1.29}}$ &${0.13^{+0.07}_{-0.07}}$\\ 
$178739533$ &$-0.39 \pm 0.96$ &$1.47550 \pm 0.00002$ &${4.2^{+1.3}_{-0.4}}$ &${3.1^{+0.3}_{-0.3}}$ &${0.294^{+0.005}_{-0.005}}$ &${73.15^{+2.07}_{-0.76}}$ &${29.24^{+0.71}_{-1.37}}$ &${0.22^{+0.09}_{-0.16}}$\\ 
$186260283$ &$0.31 \pm 0.11$ &$3.60432 \pm 0.00008$ &${1.9^{+0.2}_{-0.2}}$ &${0.3^{+0.1}_{-0.0}}$ &${0.298^{+0.004}_{-0.005}}$ &${36.78^{+0.46}_{-0.40}}$ &${8.27^{+0.96}_{-1.09}}$ &${0.19^{+0.17}_{-0.11}}$\\ 
$187920824$ &$-0.25 \pm 0.31$ &$4.50997 \pm 0.00018$ &${3.3^{+0.3}_{-0.3}}$ &${2.3^{+0.3}_{-0.3}}$ &${0.438^{+0.004}_{-0.004}}$ &${89.29^{+0.42}_{-0.50}}$ &${44.81^{+0.63}_{-0.65}}$ &${0.15^{+0.11}_{-0.10}}$\\ 
$187971301$ &$-0.61 \pm 0.30$ &$7.27253 \pm 0.00045$ &${4.3^{+0.9}_{-2.1}}$ &${3.8^{+1.0}_{-2.5}}$ &${0.619^{+0.003}_{-0.005}}$ &${88.84^{+0.69}_{-0.65}}$ &${22.10^{+0.98}_{-2.11}}$ &${0.32^{+0.17}_{-0.16}}$\\ 
$189333345$ &$-2.91 \pm 0.63$ &$6.42285 \pm 0.00022$ &${8.0^{+1.6}_{-2.3}}$ &${6.1^{+2.3}_{-2.1}}$ &${0.264^{+0.001}_{-0.001}}$ &${83.25^{+0.37}_{-0.47}}$ &${17.29^{+0.54}_{-0.52}}$ &${0.27^{+0.05}_{-0.06}}$\\ 
$190708336$ &$-2.28 \pm 0.58$ &$7.53380 \pm 0.00044$ &${5.3^{+1.0}_{-0.8}}$ &${4.6^{+0.6}_{-0.8}}$ &${0.267^{+0.001}_{-0.001}}$ &${77.91^{+0.33}_{-0.35}}$ &${14.71^{+0.69}_{-0.85}}$ &${0.37^{+0.09}_{-0.11}}$\\ 
$191457397$ &$-0.13 \pm 0.33$ &$4.23948 \pm 0.00011$ &${3.2^{+0.4}_{-0.4}}$ &${2.7^{+0.5}_{-0.5}}$ &${0.440^{+0.001}_{-0.001}}$ &${85.38^{+0.66}_{-0.36}}$ &${1.38^{+1.98}_{-1.33}}$ &${0.20^{+0.12}_{-0.06}}$\\ 
$192876875$ &$-1.50 \pm 0.81$ &$2.93658 \pm 0.00006$ &${3.9^{+1.6}_{-0.6}}$ &${0.9^{+1.9}_{-0.4}}$ &${0.275^{+0.005}_{-0.005}}$ &${32.51^{+0.52}_{-0.51}}$ &${-27.46^{+0.99}_{-0.92}}$ &${0.15^{+0.12}_{-0.08}}$\\ 
$194633998$ &$-1.43 \pm 0.32$ &$5.21941 \pm 0.00014$ &${4.4^{+0.4}_{-0.5}}$ &${3.3^{+0.4}_{-0.3}}$ &${0.323^{+0.003}_{-0.001}}$ &${80.54^{+0.95}_{-0.50}}$ &${19.98^{+1.08}_{-0.40}}$ &${0.06^{+0.28}_{-0.04}}$\\ 
$206704992$ &$-0.21 \pm 0.53$ &$6.64486 \pm 0.00044$ &${2.4^{+0.5}_{-0.4}}$ &${1.1^{+0.4}_{-0.4}}$ &${0.475^{+0.006}_{-0.005}}$ &${57.22^{+9.87}_{-8.32}}$ &${64.91^{+1.43}_{-1.97}}$ &${0.22^{+0.14}_{-0.14}}$\\ 
$209558524$ &$-1.44 \pm 0.29$ &$3.81656 \pm 0.00011$ &${3.8^{+0.6}_{-0.4}}$ &${2.5^{+0.3}_{-0.4}}$ &${0.215^{+0.002}_{-0.002}}$ &${71.79^{+0.36}_{-0.55}}$ &${39.76^{+0.50}_{-0.68}}$ &${0.26^{+0.06}_{-0.07}}$\\ 
$213223576$ &$-3.23 \pm 1.88$ &$2.77724 \pm 0.00005$ &${5.5^{+1.2}_{-2.6}}$ &${4.7^{+0.9}_{-2.6}}$ &${0.318^{+0.006}_{-0.008}}$ &${76.91^{+0.46}_{-0.44}}$ &${-42.18^{+1.72}_{-1.29}}$ &${0.84^{+0.06}_{-0.07}}$\\ 
$219707463$ &$1.78 \pm 0.05$ &$4.50725 \pm 0.00010$ &${2.0^{+0.2}_{-0.2}}$ &${1.8^{+0.2}_{-0.2}}$ &${0.530^{+0.003}_{-0.003}}$ &${81.92^{+0.69}_{-0.98}}$ &${49.11^{+0.90}_{-0.88}}$ &${0.22^{+0.14}_{-0.12}}$\\ 
$230752185$ &$1.24 \pm 0.26$ &$4.81467 \pm 0.00013$ &${1.8^{+0.2}_{-0.1}}$ &${1.7^{+0.1}_{-0.1}}$ &${0.471^{+0.001}_{-0.001}}$ &${80.16^{+0.66}_{-0.46}}$ &${22.87^{+0.54}_{-0.38}}$ &${0.07^{+0.09}_{-0.05}}$\\ 
$232446839$ &$-1.45 \pm 0.80$ &$4.72508 \pm 0.00024$ &${3.2^{+0.3}_{-0.3}}$ &${2.4^{+0.2}_{-0.3}}$ &${0.357^{+0.003}_{-0.003}}$ &${77.28^{+0.33}_{-0.56}}$ &${29.23^{+1.09}_{-1.28}}$ &${0.63^{+0.03}_{-0.05}}$\\ 
$233841767$ &$-1.25 \pm 0.33$ &$12.76491 \pm 0.00077$ &${7.3^{+1.2}_{-1.4}}$ &${4.8^{+0.6}_{-0.4}}$ &${0.685^{+0.002}_{-0.002}}$ &${85.97^{+0.13}_{-0.17}}$ &${27.28^{+0.66}_{-0.66}}$ &${0.33^{+0.15}_{-0.16}}$\\ 
$234964382$ &$1.21 \pm 0.20$ &$3.62726 \pm 0.00500$ &${1.7^{+0.2}_{-0.2}}$ &${0.8^{+0.3}_{-0.3}}$ &${0.307^{+0.007}_{-0.007}}$ &${42.73^{+6.83}_{-3.04}}$ &${-28.12^{+1.57}_{-3.08}}$ &${0.13^{+0.13}_{-0.05}}$\\ 
$236881602$ &$0.35 \pm 0.27$ &$3.48762 \pm 0.00007$ &${1.9^{+0.2}_{-0.2}}$ &${0.2^{+0.1}_{-0.0}}$ &${0.283^{+0.003}_{-0.003}}$ &${44.28^{+1.19}_{-0.93}}$ &${58.34^{+0.48}_{-0.47}}$ &${0.14^{+0.22}_{-0.11}}$\\ 
$237702040$ &$-0.08 \pm 0.40$ &$4.87674 \pm 0.00022$ &${3.0^{+0.4}_{-0.3}}$ &${1.7^{+0.2}_{-0.2}}$ &${0.472^{+0.003}_{-0.003}}$ &${74.52^{+1.63}_{-3.66}}$ &${-11.91^{+0.84}_{-0.98}}$ &${0.16^{+0.08}_{-0.08}}$\\ 
$237957506$ &$1.78 \pm 0.13$ &$2.55866 \pm 0.00003$ &${2.0^{+0.1}_{-0.1}}$ &${1.1^{+0.0}_{-0.1}}$ &${0.316^{+0.002}_{-0.001}}$ &${78.74^{+0.16}_{-0.24}}$ &${11.24^{+1.44}_{-0.64}}$ &${0.23^{+0.07}_{-0.04}}$\\ 
$240918551$ &$-0.19 \pm 0.61$ &$3.81634 \pm 0.00009$ &${5.5^{+0.6}_{-0.6}}$ &${2.6^{+0.2}_{-0.2}}$ &${0.507^{+0.002}_{-0.002}}$ &${89.60^{+0.23}_{-0.27}}$ &${34.49^{+0.48}_{-0.49}}$ &${0.10^{+0.20}_{-0.08}}$\\ 
$251972126$ &$1.08 \pm 0.46$ &$3.05964 \pm 0.00006$ &${1.7^{+0.2}_{-0.2}}$ &${0.6^{+0.2}_{-0.1}}$ &${0.246^{+0.002}_{-0.002}}$ &${52.22^{+2.45}_{-1.40}}$ &${16.67^{+0.66}_{-0.60}}$ &${0.10^{+0.10}_{-0.06}}$\\ 
$252588526$ &$1.36 \pm 0.12$ &$4.22984 \pm 0.00011$ &${1.7^{+0.2}_{-0.2}}$ &${0.3^{+0.1}_{-0.0}}$ &${0.525^{+0.004}_{-0.005}}$ &${40.65^{+0.67}_{-0.47}}$ &${59.54^{+0.86}_{-0.84}}$ &${0.17^{+0.13}_{-0.08}}$\\ 
$255876795$ &$0.90 \pm 0.33$ &$2.80977 \pm 0.00005$ &${2.0^{+0.6}_{-0.4}}$ &${1.7^{+0.5}_{-0.4}}$ &${0.256^{+0.002}_{-0.002}}$ &${71.61^{+0.97}_{-0.55}}$ &${-53.42^{+0.98}_{-1.52}}$ &${0.10^{+0.15}_{-0.07}}$\\ 
$261620164$ &$0.03 \pm 0.31$ &$2.36782 \pm 0.00003$ &${3.1^{+0.2}_{-0.3}}$ &${1.9^{+0.2}_{-0.2}}$ &${0.262^{+0.003}_{-0.003}}$ &${75.59^{+0.26}_{-0.30}}$ &${54.79^{+0.73}_{-0.67}}$ &${0.18^{+0.12}_{-0.12}}$\\ 
$265473090$ &$-1.09 \pm 7.77$ &$4.36129 \pm 0.00011$ &${11.9^{+4.8}_{-2.9}}$ &${0.5^{+0.2}_{-0.1}}$ &${0.325^{+0.007}_{-0.007}}$ &${59.60^{+1.96}_{-1.41}}$ &${-22.56^{+2.14}_{-2.05}}$ &${0.22^{+0.15}_{-0.14}}$\\ 
$269511526$ &$-0.32 \pm 0.49$ &$2.74726 \pm 0.00004$ &${2.7^{+0.6}_{-0.5}}$ &${1.7^{+0.8}_{-0.7}}$ &${0.221^{+0.002}_{-0.003}}$ &${43.28^{+3.19}_{-3.95}}$ &${15.97^{+0.73}_{-2.31}}$ &${0.20^{+0.15}_{-0.13}}$\\ 
$269692669$ &$0.55 \pm 0.42$ &$4.95600 \pm 0.00020$ &${1.7^{+0.2}_{-0.2}}$ &${0.4^{+0.1}_{-0.1}}$ &${0.308^{+0.003}_{-0.003}}$ &${55.51^{+1.45}_{-1.18}}$ &${-74.80^{+0.66}_{-0.63}}$ &${0.22^{+0.12}_{-0.10}}$\\ 
$270859445$ &$1.07 \pm 0.13$ &$5.58375 \pm 0.00014$ &${1.7^{+0.4}_{-0.1}}$ &${1.5^{+0.2}_{-0.4}}$ &${0.306^{+0.001}_{-0.001}}$ &${87.91^{+1.04}_{-2.50}}$ &${-9.25^{+0.77}_{-1.64}}$ &${0.35^{+0.05}_{-0.26}}$\\ 
$271554988$ &$1.56 \pm 0.10$ &$4.37970 \pm 0.00009$ &${1.5^{+0.1}_{-0.1}}$ &${0.4^{+0.2}_{-0.0}}$ &${0.299^{+0.003}_{-0.003}}$ &${50.25^{+2.19}_{-0.86}}$ &${-2.64^{+0.85}_{-0.84}}$ &${0.36^{+0.08}_{-0.09}}$\\ 
$272822324$ &$3.86 \pm 0.12$ &$4.50132 \pm 0.00012$ &${0.7^{+0.1}_{-0.0}}$ &${0.7^{+0.0}_{-0.1}}$ &${0.351^{+0.001}_{-0.002}}$ &${81.58^{+0.70}_{-0.40}}$ &${12.12^{+0.60}_{-2.35}}$ &${0.40^{+0.08}_{-0.20}}$\\ 
$272822330$ &$0.87 \pm 0.21$ &$4.50134 \pm 0.00011$ &${1.8^{+0.2}_{-0.3}}$ &${1.4^{+0.2}_{-0.2}}$ &${0.352^{+0.001}_{-0.001}}$ &${82.56^{+0.58}_{-0.88}}$ &${10.42^{+0.96}_{-0.63}}$ &${0.34^{+0.08}_{-0.14}}$\\ 
$274686141$ &$-0.42 \pm 0.36$ &$2.76759 \pm 0.00003$ &${6.1^{+1.4}_{-1.0}}$ &${0.5^{+0.2}_{-0.1}}$ &${0.410^{+0.019}_{-0.020}}$ &${40.80^{+13.22}_{-9.73}}$ &${66.55^{+3.89}_{-4.90}}$ &${0.31^{+0.04}_{-0.04}}$\\ 
$275586333$ &$0.40 \pm 0.46$ &$2.83295 \pm 0.00006$ &${1.9^{+0.4}_{-0.3}}$ &${0.2^{+0.3}_{-0.1}}$ &${0.269^{+0.004}_{-0.004}}$ &${40.21^{+0.76}_{-0.96}}$ &${-48.12^{+0.87}_{-0.73}}$ &${0.21^{+0.17}_{-0.11}}$\\ 
$277236190$ &$0.30 \pm 0.30$ &$3.58116 \pm 0.00008$ &${2.0^{+0.3}_{-0.3}}$ &${1.1^{+0.2}_{-0.1}}$ &${0.315^{+0.001}_{-0.002}}$ &${75.96^{+0.31}_{-0.33}}$ &${-17.54^{+0.66}_{-0.77}}$ &${0.39^{+0.22}_{-0.11}}$\\ 
$282118355$ &$-0.30 \pm 0.67$ &$3.96133 \pm 0.00013$ &${3.9^{+2.4}_{-2.4}}$ &${2.6^{+1.6}_{-1.6}}$ &${0.284^{+0.005}_{-0.002}}$ &${75.61^{+0.39}_{-0.44}}$ &${-20.51^{+0.70}_{-2.46}}$ &${0.37^{+0.07}_{-0.06}}$\\ 
$282876586$ &$-1.45 \pm 0.63$ &$2.38762 \pm 0.00004$ &${6.6^{+1.1}_{-1.1}}$ &${4.8^{+0.7}_{-0.5}}$ &${0.292^{+0.003}_{-0.003}}$ &${76.15^{+1.23}_{-1.35}}$ &${61.53^{+0.54}_{-0.55}}$ &${0.30^{+0.12}_{-0.19}}$\\ 
$283539216$ &$-0.82 \pm 0.85$ &$2.66169 \pm 0.00004$ &${3.8^{+1.5}_{-1.1}}$ &${3.3^{+1.6}_{-1.0}}$ &${0.261^{+0.001}_{-0.002}}$ &${82.72^{+0.32}_{-0.46}}$ &${21.22^{+0.68}_{-1.22}}$ &${0.41^{+0.08}_{-0.08}}$\\ 
$284144129$ &$0.82 \pm 0.32$ &$3.76651 \pm 0.00010$ &${2.2^{+0.5}_{-0.4}}$ &${1.4^{+0.4}_{-0.4}}$ &${0.540^{+0.009}_{-0.004}}$ &${72.85^{+2.78}_{-2.34}}$ &${-24.57^{+1.38}_{-1.22}}$ &${0.42^{+0.21}_{-0.27}}$\\ 
$285128834$ &$-6.54 \pm 1.64$ &$5.14849 \pm 0.00016$ &${6.8^{+19.7}_{-2.1}}$ &${5.3^{+12.8}_{-2.0}}$ &${0.441^{+0.001}_{-0.001}}$ &${82.93^{+0.85}_{-0.58}}$ &${-1.63^{+0.74}_{-0.43}}$ &${0.34^{+0.11}_{-0.08}}$\\ 
$293228326$ &$-0.62 \pm 0.12$ &$6.82633 \pm 0.00041$ &${2.5^{+0.4}_{-0.3}}$ &${1.6^{+0.7}_{-0.8}}$ &${0.355^{+0.003}_{-0.003}}$ &${67.16^{+7.96}_{-6.20}}$ &${-3.24^{+0.77}_{-0.81}}$ &${0.23^{+0.15}_{-0.14}}$\\ 
$293618358$ &$2.32 \pm 0.09$ &$10.48601 \pm 0.00029$ &${1.3^{+0.3}_{-0.2}}$ &${0.3^{+0.0}_{-0.0}}$ &${0.654^{+0.001}_{-0.001}}$ &${88.06^{+1.02}_{-1.04}}$ &${-2.05^{+0.61}_{-0.63}}$ &${0.12^{+0.18}_{-0.09}}$\\ 
$296183275$ &$1.96 \pm 0.15$ &$4.37213 \pm 0.00019$ &${1.5^{+1.0}_{-0.3}}$ &${0.5^{+0.2}_{-0.2}}$ &${0.557^{+0.134}_{-0.009}}$ &${45.79^{+25.67}_{-2.80}}$ &${-8.08^{+21.31}_{-2.56}}$ &${0.24^{+0.26}_{-0.17}}$\\ 
$302828770$ &$1.93 \pm 0.34$ &$2.95162 \pm 0.00006$ &${1.5^{+0.3}_{-0.3}}$ &${0.3^{+0.1}_{-0.1}}$ &${0.421^{+0.005}_{-0.005}}$ &${72.22^{+5.74}_{-14.86}}$ &${45.05^{+1.34}_{-1.58}}$ &${0.28^{+0.21}_{-0.17}}$\\ 
$303577635$ &$-7.25 \pm 2.27$ &$8.66950 \pm 0.00045$ &${8.8^{+3.6}_{-3.2}}$ &${4.9^{+2.3}_{-1.5}}$ &${0.372^{+0.001}_{-0.001}}$ &${75.74^{+0.49}_{-0.60}}$ &${-23.57^{+0.47}_{-0.43}}$ &${0.29^{+0.11}_{-0.13}}$\\ 
$305252841$ &$0.73 \pm 0.34$ &$6.11764 \pm 0.00020$ &${1.5^{+0.1}_{-0.3}}$ &${1.1^{+0.3}_{-0.1}}$ &${0.257^{+0.002}_{-0.002}}$ &${78.11^{+0.89}_{-0.28}}$ &${-32.08^{+2.13}_{-0.70}}$ &${0.29^{+0.29}_{-0.09}}$\\ 
$305454334$ &$-2.54 \pm 0.56$ &$4.81686 \pm 0.00010$ &${11.3^{+1.7}_{-1.4}}$ &${7.0^{+1.8}_{-1.2}}$ &${0.466^{+0.002}_{-0.001}}$ &${78.50^{+0.44}_{-0.19}}$ &${-19.58^{+0.51}_{-0.48}}$ &${0.17^{+0.09}_{-0.10}}$\\ 
$306107122$ &$1.85 \pm 0.09$ &$7.75985 \pm 0.00022$ &${1.2^{+0.3}_{-0.1}}$ &${1.1^{+0.1}_{-0.1}}$ &${0.252^{+0.002}_{-0.002}}$ &${89.39^{+0.40}_{-0.59}}$ &${-17.22^{+1.56}_{-1.68}}$ &${0.15^{+0.08}_{-0.06}}$\\ 
$307813020$ &$0.07 \pm 0.37$ &$3.55543 \pm 0.00008$ &${3.1^{+0.3}_{-0.4}}$ &${2.6^{+0.3}_{-0.2}}$ &${0.294^{+0.001}_{-0.001}}$ &${82.22^{+0.77}_{-1.10}}$ &${-2.85^{+1.14}_{-1.77}}$ &${0.32^{+0.05}_{-0.04}}$\\ 
$309682332$ &$1.55 \pm 0.20$ &$7.18878 \pm 0.00041$ &${1.9^{+0.1}_{-0.1}}$ &${0.8^{+0.1}_{-0.1}}$ &${0.253^{+0.002}_{-0.002}}$ &${89.50^{+0.33}_{-0.54}}$ &${-8.49^{+6.42}_{-1.86}}$ &${0.16^{+0.10}_{-0.10}}$\\ 
$312344965$ &$-0.27 \pm 0.30$ &$5.81836 \pm 0.00025$ &${2.5^{+0.2}_{-0.2}}$ &${0.7^{+0.2}_{-0.2}}$ &${0.257^{+0.006}_{-0.006}}$ &${63.93^{+6.58}_{-4.36}}$ &${-4.27^{+2.04}_{-2.07}}$ &${0.27^{+0.14}_{-0.16}}$\\ 
$312344969$ &$1.38 \pm 0.33$ &$5.82055 \pm 0.00023$ &${1.6^{+0.1}_{-0.1}}$ &${0.5^{+0.1}_{-0.1}}$ &${0.351^{+0.006}_{-0.006}}$ &${70.20^{+3.41}_{-2.86}}$ &${-13.68^{+1.54}_{-1.72}}$ &${0.19^{+0.16}_{-0.10}}$\\ 
$313176361$ &$0.55 \pm 0.24$ &$7.86603 \pm 0.00025$ &${1.8^{+0.1}_{-0.1}}$ &${1.4^{+0.1}_{-0.3}}$ &${0.324^{+0.003}_{-0.002}}$ &${79.05^{+0.22}_{-0.09}}$ &${-38.34^{+0.52}_{-0.56}}$ &${0.07^{+0.08}_{-0.06}}$\\ 
$316119373$ &$1.07 \pm 0.27$ &$7.90980 \pm 0.00031$ &${2.5^{+0.2}_{-0.3}}$ &${0.3^{+0.0}_{-0.0}}$ &${0.708^{+0.003}_{-0.003}}$ &${84.10^{+0.31}_{-0.27}}$ &${-23.87^{+0.84}_{-0.85}}$ &${0.13^{+0.10}_{-0.08}}$\\ 
$316621853$ &$1.46 \pm 0.44$ &$3.10096 \pm 0.00007$ &${1.5^{+0.2}_{-0.2}}$ &${0.2^{+0.0}_{-0.0}}$ &${0.289^{+0.006}_{-0.006}}$ &${47.07^{+1.17}_{-1.04}}$ &${-6.33^{+1.63}_{-1.59}}$ &${0.26^{+0.09}_{-0.17}}$\\ 
$316668953$ &$-3.25 \pm 1.01$ &$6.95107 \pm 0.00030$ &${7.7^{+2.0}_{-1.5}}$ &${6.3^{+1.5}_{-1.5}}$ &${0.431^{+0.002}_{-0.002}}$ &${79.06^{+0.45}_{-0.27}}$ &${-32.97^{+0.55}_{-0.39}}$ &${0.29^{+0.10}_{-0.12}}$\\ 
$321952939$ &$1.24 \pm 0.29$ &$6.63015 \pm 0.00024$ &${1.7^{+0.2}_{-0.2}}$ &${1.0^{+0.2}_{-0.2}}$ &${0.279^{+0.015}_{-0.009}}$ &${78.19^{+1.14}_{-1.26}}$ &${-63.06^{+1.73}_{-1.69}}$ &${0.11^{+0.22}_{-0.08}}$\\ 
$326374705$ &$-1.30 \pm 0.27$ &$5.40604 \pm 0.00016$ &${3.4^{+0.3}_{-0.3}}$ &${3.0^{+0.3}_{-0.4}}$ &${0.281^{+0.002}_{-0.002}}$ &${84.20^{+1.20}_{-0.29}}$ &${65.29^{+0.25}_{-0.25}}$ &${0.09^{+0.17}_{-0.07}}$\\ 
$326484902$ &$1.89 \pm 0.18$ &$2.58910 \pm 0.00004$ &${1.6^{+0.2}_{-0.2}}$ &${0.3^{+0.0}_{-0.0}}$ &${0.260^{+0.005}_{-0.005}}$ &${81.56^{+0.37}_{-0.36}}$ &${43.21^{+0.87}_{-0.95}}$ &${0.09^{+0.09}_{-0.05}}$\\ 
$327725463$ &$1.97 \pm 0.07$ &$4.75026 \pm 0.00006$ &${1.3^{+0.2}_{-0.3}}$ &${0.7^{+0.1}_{-0.1}}$ &${0.235^{+0.005}_{-0.007}}$ &${73.67^{+1.22}_{-1.24}}$ &${50.13^{+1.21}_{-2.09}}$ &${0.38^{+0.17}_{-0.24}}$\\ 
$330605074$ &$4.88 \pm 7.77$ &$5.34610 \pm 0.00017$ &${2.6^{+2.4}_{-0.8}}$ &${2.0^{+1.6}_{-0.7}}$ &${0.226^{+0.003}_{-0.003}}$ &${75.02^{+0.41}_{-1.24}}$ &${48.07^{+1.07}_{-0.76}}$ &${0.43^{+0.08}_{-0.10}}$\\ 
$336091799$ &$1.84 \pm 0.21$ &$2.16482 \pm 0.00013$ &${1.3^{+0.3}_{-0.2}}$ &${0.3^{+0.4}_{-0.1}}$ &${0.210^{+0.005}_{-0.006}}$ &${38.65^{+6.08}_{-1.05}}$ &${45.39^{+1.41}_{-1.59}}$ &${0.22^{+0.14}_{-0.10}}$\\ 
$336538437$ &$0.52 \pm 0.47$ &$3.89355 \pm 0.00009$ &${1.9^{+0.6}_{-0.7}}$ &${1.7^{+0.6}_{-0.7}}$ &${0.437^{+0.001}_{-0.001}}$ &${81.33^{+0.34}_{-0.88}}$ &${4.69^{+0.45}_{-0.37}}$ &${0.38^{+0.07}_{-0.09}}$\\ 
$336823975$ &$0.13 \pm 7.77$ &$4.66422 \pm 0.00016$ &${1.9^{+0.9}_{-0.5}}$ &${0.6^{+0.3}_{-0.2}}$ &${0.405^{+0.003}_{-0.004}}$ &${77.55^{+0.87}_{-0.75}}$ &${-21.39^{+1.32}_{-2.02}}$ &${0.51^{+0.07}_{-0.09}}$\\ 
$337376644$ &$-1.47 \pm 0.42$ &$5.28619 \pm 0.00018$ &${14.8^{+12.0}_{-6.7}}$ &${5.2^{+2.1}_{-2.1}}$ &${0.619^{+0.016}_{-0.014}}$ &${26.18^{+9.81}_{-12.31}}$ &${-62.25^{+13.54}_{-11.30}}$ &${0.39^{+0.09}_{-0.16}}$\\ 
$338282749$ &$-2.89 \pm 0.89$ &$9.92119 \pm 0.00056$ &${7.4^{+3.0}_{-1.8}}$ &${5.3^{+2.9}_{-1.7}}$ &${0.609^{+0.002}_{-0.003}}$ &${84.20^{+0.94}_{-0.86}}$ &${-16.44^{+0.72}_{-0.57}}$ &${0.37^{+0.13}_{-0.19}}$\\ 
$342520115$ &$1.13 \pm 0.22$ &$8.66907 \pm 0.00046$ &${1.7^{+0.2}_{-0.2}}$ &${0.7^{+0.1}_{-0.1}}$ &${0.431^{+0.002}_{-0.002}}$ &${85.84^{+0.88}_{-0.65}}$ &${-14.95^{+0.65}_{-0.66}}$ &${0.06^{+0.14}_{-0.04}}$\\ 
$343173397$ &$1.46 \pm 0.13$ &$2.87477 \pm 0.00005$ &${1.6^{+0.1}_{-0.3}}$ &${1.4^{+0.1}_{-0.2}}$ &${0.429^{+0.001}_{-0.001}}$ &${75.63^{+0.45}_{-0.15}}$ &${-7.25^{+0.67}_{-0.33}}$ &${0.34^{+0.07}_{-0.04}}$\\ 
$343626774$ &$1.42 \pm 0.36$ &$3.66314 \pm 0.00008$ &${1.7^{+0.4}_{-0.4}}$ &${1.5^{+0.3}_{-0.4}}$ &${0.256^{+0.002}_{-0.002}}$ &${80.64^{+0.67}_{-0.72}}$ &${-16.25^{+1.38}_{-1.18}}$ &${0.25^{+0.07}_{-0.11}}$\\ 
$343878759$ &$-2.49 \pm 0.82$ &$2.53680 \pm 0.00093$ &${6.5^{+2.2}_{-1.5}}$ &${0.9^{+0.9}_{-0.2}}$ &${0.211^{+0.005}_{-0.005}}$ &${43.45^{+1.64}_{-0.83}}$ &${-36.91^{+1.81}_{-1.59}}$ &${0.23^{+0.14}_{-0.11}}$\\ 
$344586348$ &$1.08 \pm 0.11$ &$7.73929 \pm 0.00021$ &${1.4^{+0.1}_{-0.2}}$ &${0.8^{+0.1}_{-0.1}}$ &${0.337^{+0.001}_{-0.001}}$ &${87.28^{+0.53}_{-0.36}}$ &${26.15^{+0.36}_{-0.49}}$ &${0.09^{+0.09}_{-0.06}}$\\ 
$347268123$ &$1.07 \pm 0.28$ &$2.31622 \pm 0.00002$ &${1.7^{+0.2}_{-0.2}}$ &${0.2^{+0.1}_{-0.0}}$ &${0.285^{+0.003}_{-0.003}}$ &${42.83^{+0.68}_{-0.60}}$ &${-46.30^{+0.72}_{-0.68}}$ &${0.22^{+0.09}_{-0.12}}$\\ 
$348508349$ &$-2.86 \pm 0.81$ &$4.15821 \pm 0.00011$ &${8.3^{+1.0}_{-1.1}}$ &${4.5^{+0.6}_{-0.5}}$ &${0.294^{+0.001}_{-0.001}}$ &${72.79^{+0.62}_{-0.59}}$ &${15.85^{+0.62}_{-0.46}}$ &${0.27^{+0.11}_{-0.12}}$\\ 
$352770261$ &$-1.41 \pm 0.50$ &$6.52954 \pm 0.00035$ &${2.7^{+0.6}_{-0.5}}$ &${0.4^{+0.8}_{-0.2}}$ &${0.414^{+0.010}_{-0.004}}$ &${45.62^{+25.41}_{-0.66}}$ &${66.75^{+0.40}_{-0.40}}$ &${0.18^{+0.20}_{-0.14}}$\\ 
$352835929$ &$-0.13 \pm 0.30$ &$3.32735 \pm 0.00006$ &${8.5^{+1.3}_{-1.1}}$ &${3.4^{+0.5}_{-0.4}}$ &${0.377^{+0.002}_{-0.002}}$ &${80.41^{+0.54}_{-1.79}}$ &${-8.12^{+0.91}_{-0.85}}$ &${0.12^{+0.10}_{-0.07}}$\\ 
$356169556$ &$0.61 \pm 0.23$ &$17.64534 \pm 0.00120$ &${2.6^{+0.1}_{-0.5}}$ &${0.9^{+0.1}_{-0.1}}$ &${0.635^{+0.001}_{-0.002}}$ &${87.10^{+0.74}_{-4.22}}$ &${11.74^{+0.56}_{-0.64}}$ &${0.32^{+0.12}_{-0.30}}$\\ 
$358370373$ &$-2.30 \pm 0.85$ &$6.22525 \pm 0.00016$ &${3.3^{+0.9}_{-0.7}}$ &${0.4^{+0.7}_{-0.1}}$ &${0.334^{+0.005}_{-0.004}}$ &${39.26^{+0.51}_{-0.49}}$ &${-8.24^{+0.90}_{-0.77}}$ &${0.27^{+0.19}_{-0.14}}$\\ 
$362081271$ &$1.45 \pm 0.07$ &$11.63190 \pm 0.00071$ &${1.1^{+0.1}_{-0.1}}$ &${0.2^{+0.5}_{-0.1}}$ &${0.423^{+0.007}_{-0.008}}$ &${40.53^{+5.76}_{-1.07}}$ &${-6.07^{+1.14}_{-1.42}}$ &${0.41^{+0.05}_{-0.25}}$\\ 
$363674500$ &$0.84 \pm 0.33$ &$3.02983 \pm 0.00005$ &${2.3^{+0.3}_{-0.2}}$ &${0.4^{+0.1}_{-0.1}}$ &${0.325^{+0.005}_{-0.005}}$ &${67.95^{+3.24}_{-2.59}}$ &${11.36^{+1.61}_{-1.57}}$ &${0.19^{+0.16}_{-0.12}}$\\ 
$363679519$ &$-0.03 \pm 0.23$ &$3.13259 \pm 0.00006$ &${2.5^{+0.2}_{-0.2}}$ &${2.1^{+0.1}_{-0.2}}$ &${0.312^{+0.001}_{-0.001}}$ &${80.30^{+0.44}_{-0.42}}$ &${-1.94^{+0.45}_{-0.53}}$ &${0.29^{+0.07}_{-0.05}}$\\ 
$365831185$ &$0.11 \pm 0.11$ &$5.55091 \pm 0.00025$ &${6.8^{+4.3}_{-3.2}}$ &${2.8^{+0.5}_{-0.6}}$ &${0.420^{+0.086}_{-0.019}}$ &${73.10^{+5.77}_{-8.53}}$ &${-79.19^{+4.51}_{-4.38}}$ &${0.39^{+0.14}_{-0.08}}$\\ 
$367944808$ &$0.87 \pm 0.26$ &$3.61145 \pm 0.00008$ &${2.6^{+0.4}_{-0.4}}$ &${2.2^{+0.3}_{-0.3}}$ &${0.421^{+0.004}_{-0.004}}$ &${76.09^{+1.57}_{-2.05}}$ &${-50.29^{+1.87}_{-2.19}}$ &${0.27^{+0.17}_{-0.21}}$\\ 
$369033532$ &$0.84 \pm 0.46$ &$6.04914 \pm 0.00021$ &${1.9^{+0.4}_{-0.4}}$ &${1.5^{+0.3}_{-0.4}}$ &${0.463^{+0.002}_{-0.003}}$ &${78.97^{+0.49}_{-0.55}}$ &${22.93^{+0.70}_{-1.03}}$ &${0.35^{+0.12}_{-0.15}}$\\ 
$369999283$ &$1.24 \pm 0.32$ &$3.55119 \pm 0.00009$ &${3.1^{+1.0}_{-0.7}}$ &${2.1^{+0.5}_{-0.6}}$ &${0.450^{+0.022}_{-0.026}}$ &${61.64^{+11.81}_{-15.90}}$ &${-73.97^{+8.07}_{-5.36}}$ &${0.39^{+0.16}_{-0.17}}$\\ 
$370209445$ &$0.94 \pm 0.28$ &$3.75863 \pm 0.00009$ &${1.7^{+0.2}_{-0.2}}$ &${0.4^{+0.1}_{-0.1}}$ &${0.202^{+0.002}_{-0.002}}$ &${65.71^{+1.73}_{-1.55}}$ &${-12.89^{+0.78}_{-0.70}}$ &${0.22^{+0.08}_{-0.08}}$\\ 
$370269453$ &$0.26 \pm 0.26$ &$5.92922 \pm 0.00021$ &${1.8^{+0.2}_{-0.2}}$ &${0.4^{+0.1}_{-0.0}}$ &${0.321^{+0.006}_{-0.006}}$ &${42.34^{+0.61}_{-0.56}}$ &${-36.69^{+1.36}_{-1.13}}$ &${0.31^{+0.18}_{-0.18}}$\\ 
$371584261$ &$-0.97 \pm 0.27$ &$4.11336 \pm 0.00011$ &${2.4^{+0.3}_{-0.2}}$ &${0.5^{+0.3}_{-0.1}}$ &${0.256^{+0.002}_{-0.002}}$ &${52.14^{+0.50}_{-0.60}}$ &${-31.89^{+0.71}_{-0.68}}$ &${0.16^{+0.13}_{-0.12}}$\\ 
$371966579$ &$2.57 \pm 0.08$ &$3.67718 \pm 0.00008$ &${1.5^{+0.2}_{-0.2}}$ &${1.2^{+0.1}_{-0.1}}$ &${0.327^{+0.004}_{-0.002}}$ &${83.93^{+0.49}_{-0.27}}$ &${-8.27^{+0.22}_{-5.35}}$ &${0.37^{+0.06}_{-0.07}}$\\ 
$372116736$ &$1.40 \pm 0.33$ &$2.71166 \pm 0.00004$ &${1.9^{+0.3}_{-0.2}}$ &${0.6^{+0.1}_{-0.1}}$ &${0.349^{+0.006}_{-0.005}}$ &${56.46^{+4.04}_{-2.32}}$ &${-41.68^{+1.66}_{-1.37}}$ &${0.25^{+0.13}_{-0.13}}$\\ 
$372863796$ &$0.35 \pm 0.41$ &$4.91977 \pm 0.00017$ &${7.8^{+2.9}_{-3.2}}$ &${1.6^{+1.3}_{-0.8}}$ &${0.475^{+0.026}_{-0.017}}$ &${66.12^{+9.46}_{-15.48}}$ &${66.35^{+10.63}_{-8.85}}$ &${0.36^{+0.14}_{-0.16}}$\\ 
$374130970$ &$-1.94 \pm 0.67$ &$5.26465 \pm 0.00019$ &${3.3^{+0.8}_{-0.5}}$ &${0.5^{+0.5}_{-0.1}}$ &${0.327^{+0.004}_{-0.004}}$ &${51.48^{+0.89}_{-0.79}}$ &${4.59^{+0.70}_{-0.85}}$ &${0.21^{+0.15}_{-0.13}}$\\ 
$374211609$ &$-0.51 \pm 0.36$ &$4.69545 \pm 0.00016$ &${3.4^{+0.4}_{-0.4}}$ &${0.6^{+0.2}_{-0.1}}$ &${0.449^{+0.009}_{-0.008}}$ &${49.54^{+1.89}_{-1.24}}$ &${-15.58^{+3.07}_{-2.44}}$ &${0.14^{+0.17}_{-0.07}}$\\ 
$376499580$ &$0.85 \pm 7.77$ &$2.87476 \pm 0.00004$ &${9.6^{+87.7}_{-2.1}}$ &${6.3^{+17.3}_{-2.0}}$ &${0.275^{+0.000}_{-0.000}}$ &${88.35^{+0.44}_{-0.65}}$ &${-0.01^{+0.56}_{-0.45}}$ &${0.26^{+0.05}_{-0.03}}$\\ 
$378275980$ &$-2.93 \pm 0.84$ &$15.79718 \pm 0.00106$ &${9.4^{+75.6}_{-2.2}}$ &${6.0^{+32.7}_{-1.7}}$ &${0.396^{+0.002}_{-0.002}}$ &${80.86^{+0.79}_{-0.78}}$ &${52.26^{+0.27}_{-0.29}}$ &${0.17^{+0.11}_{-0.14}}$\\ 
$383518759$ &$-0.14 \pm 0.41$ &$4.09045 \pm 0.00009$ &${2.2^{+0.2}_{-0.2}}$ &${0.2^{+0.7}_{-0.0}}$ &${0.384^{+0.004}_{-0.003}}$ &${42.49^{+21.46}_{-0.45}}$ &${39.79^{+0.80}_{-2.40}}$ &${0.13^{+0.38}_{-0.08}}$\\ 
$386210336$ &$0.06 \pm 0.33$ &$2.28326 \pm 0.00004$ &${2.7^{+0.7}_{-0.3}}$ &${2.4^{+0.2}_{-0.3}}$ &${0.259^{+0.002}_{-0.002}}$ &${76.33^{+0.29}_{-0.73}}$ &${-14.98^{+1.95}_{-1.03}}$ &${0.40^{+0.10}_{-0.19}}$\\ 
$386262459$ &$1.08 \pm 0.15$ &$2.67712 \pm 0.00005$ &${2.6^{+0.2}_{-0.2}}$ &${0.6^{+0.1}_{-0.1}}$ &${0.229^{+0.003}_{-0.003}}$ &${82.01^{+0.32}_{-0.43}}$ &${2.25^{+1.28}_{-1.47}}$ &${0.06^{+0.05}_{-0.04}}$\\ 
$389829007$ &$0.28 \pm 1.74$ &$7.91653 \pm 0.00049$ &${4.9^{+1.1}_{-1.1}}$ &${4.4^{+0.3}_{-1.2}}$ &${0.521^{+0.001}_{-0.001}}$ &${85.92^{+0.38}_{-0.41}}$ &${4.04^{+0.60}_{-0.52}}$ &${0.15^{+0.11}_{-0.10}}$\\ 
$390334189$ &$0.99 \pm 7.77$ &$2.35419 \pm 0.00003$ &${2.1^{+0.9}_{-0.5}}$ &${0.6^{+0.2}_{-0.1}}$ &${0.404^{+0.005}_{-0.005}}$ &${57.24^{+3.04}_{-2.51}}$ &${14.42^{+1.49}_{-1.61}}$ &${0.15^{+0.15}_{-0.10}}$\\ 
$390337472$ &$1.27 \pm 0.81$ &$3.34906 \pm 0.00006$ &${1.1^{+0.4}_{-0.2}}$ &${0.9^{+0.3}_{-0.1}}$ &${0.353^{+0.002}_{-0.002}}$ &${83.88^{+1.38}_{-1.81}}$ &${15.10^{+1.40}_{-1.32}}$ &${0.27^{+0.41}_{-0.12}}$\\ 
$391085159$ &$-0.86 \pm 1.16$ &$4.84919 \pm 0.00014$ &${2.6^{+0.3}_{-0.3}}$ &${0.6^{+0.2}_{-0.1}}$ &${0.276^{+0.003}_{-0.003}}$ &${55.82^{+1.98}_{-1.55}}$ &${31.04^{+0.62}_{-0.79}}$ &${0.18^{+0.11}_{-0.08}}$\\ 
$391244527$ &$-0.79 \pm 0.41$ &$2.47872 \pm 0.00002$ &${6.3^{+1.3}_{-3.1}}$ &${3.0^{+0.9}_{-0.9}}$ &${0.265^{+0.009}_{-0.004}}$ &${79.94^{+0.28}_{-0.44}}$ &${49.19^{+1.86}_{-0.70}}$ &${0.07^{+0.26}_{-0.06}}$\\ 
$395413286$ &$-0.24 \pm 0.24$ &$3.10833 \pm 0.00006$ &${2.8^{+0.6}_{-0.4}}$ &${2.0^{+0.4}_{-0.9}}$ &${0.295^{+0.004}_{-0.005}}$ &${76.14^{+2.50}_{-7.05}}$ &${48.57^{+1.07}_{-2.78}}$ &${0.44^{+0.18}_{-0.27}}$\\ 
$396201191$ &$-2.13 \pm 0.73$ &$4.23807 \pm 0.00012$ &${7.3^{+2.2}_{-2.8}}$ &${4.6^{+1.3}_{-1.2}}$ &${0.410^{+0.003}_{-0.003}}$ &${80.26^{+1.43}_{-0.69}}$ &${32.96^{+0.79}_{-0.77}}$ &${0.32^{+0.13}_{-0.11}}$\\ 
$405320687$ &$0.38 \pm 0.29$ &$2.89828 \pm 0.00005$ &${5.3^{+1.2}_{-1.0}}$ &${0.4^{+0.1}_{-0.1}}$ &${0.745^{+0.007}_{-0.006}}$ &${81.92^{+0.58}_{-6.34}}$ &${58.08^{+1.26}_{-7.58}}$ &${0.17^{+0.17}_{-0.10}}$\\ 
$408618201$ &$0.37 \pm 7.77$ &$4.13535 \pm 0.00012$ &${3.2^{+4.4}_{-2.1}}$ &${2.8^{+2.1}_{-1.9}}$ &${0.483^{+0.001}_{-0.001}}$ &${83.16^{+0.71}_{-0.55}}$ &${-6.43^{+0.49}_{-0.40}}$ &${0.52^{+0.07}_{-0.14}}$\\ 
$412581945$ &$-4.17 \pm 2.73$ &$3.23882 \pm 0.00304$ &${11.1^{+12.5}_{-3.5}}$ &${6.5^{+9.4}_{-1.9}}$ &${0.345^{+0.004}_{-0.004}}$ &${76.68^{+0.79}_{-0.98}}$ &${36.56^{+2.86}_{-1.62}}$ &${0.66^{+0.04}_{-0.05}}$\\ 
$418183908$ &$0.98 \pm 0.50$ &$8.89975 \pm 0.00045$ &${2.7^{+1.2}_{-0.5}}$ &${0.4^{+0.2}_{-0.1}}$ &${0.719^{+0.028}_{-0.029}}$ &${66.16^{+19.53}_{-25.44}}$ &${-60.35^{+4.98}_{-6.18}}$ &${0.33^{+0.15}_{-0.12}}$\\ 
$420388688$ &$-0.30 \pm 0.28$ &$4.23554 \pm 0.00015$ &${2.7^{+0.3}_{-0.2}}$ &${0.5^{+0.2}_{-0.1}}$ &${0.336^{+0.004}_{-0.004}}$ &${46.65^{+0.68}_{-0.65}}$ &${-2.61^{+1.19}_{-1.17}}$ &${0.13^{+0.13}_{-0.07}}$\\ 
$421924260$ &$-0.86 \pm 0.77$ &$5.75772 \pm 0.00020$ &${2.5^{+0.4}_{-0.3}}$ &${0.5^{+0.2}_{-0.2}}$ &${0.276^{+0.005}_{-0.005}}$ &${47.26^{+1.08}_{-0.86}}$ &${-19.85^{+1.15}_{-1.22}}$ &${0.20^{+0.14}_{-0.12}}$\\ 
$422065233$ &$0.91 \pm 0.32$ &$6.75514 \pm 0.00018$ &${2.3^{+0.1}_{-0.1}}$ &${1.5^{+0.2}_{-0.1}}$ &${0.523^{+0.002}_{-0.002}}$ &${86.78^{+0.29}_{-0.25}}$ &${33.76^{+0.39}_{-0.42}}$ &${0.19^{+0.08}_{-0.04}}$\\ 
$426256249$ &$-6.54 \pm 2.08$ &$7.84702 \pm 0.00029$ &${8.2^{+2.2}_{-1.6}}$ &${5.7^{+1.7}_{-1.1}}$ &${0.391^{+0.001}_{-0.001}}$ &${89.36^{+0.41}_{-0.50}}$ &${-1.47^{+0.28}_{-0.29}}$ &${0.04^{+0.04}_{-0.03}}$\\ 
$427377458$ &$4.49 \pm 0.12$ &$2.53771 \pm 0.00006$ &${0.6^{+0.1}_{-0.1}}$ &${0.4^{+0.1}_{-0.0}}$ &${0.365^{+0.008}_{-0.007}}$ &${77.01^{+0.88}_{-1.11}}$ &${-47.52^{+2.59}_{-2.84}}$ &${0.79^{+0.03}_{-0.04}}$\\ 
$427377463$ &$0.56 \pm 0.61$ &$2.53766 \pm 0.00007$ &${2.3^{+0.3}_{-0.3}}$ &${0.7^{+0.1}_{-0.2}}$ &${0.358^{+0.005}_{-0.005}}$ &${77.14^{+1.13}_{-2.79}}$ &${-35.40^{+2.10}_{-2.33}}$ &${0.61^{+0.09}_{-0.15}}$\\ 
$428117606$ &$0.36 \pm 0.22$ &$2.14945 \pm 0.00003$ &${2.3^{+0.2}_{-0.2}}$ &${1.5^{+0.4}_{-0.2}}$ &${0.230^{+0.003}_{-0.003}}$ &${72.32^{+1.02}_{-1.51}}$ &${42.99^{+1.00}_{-1.00}}$ &${0.43^{+0.05}_{-0.04}}$\\ 
$431952504$ &$1.84 \pm 0.17$ &$5.98106 \pm 0.00023$ &${2.5^{+0.8}_{-0.5}}$ &${0.5^{+0.2}_{-0.1}}$ &${0.568^{+0.026}_{-0.023}}$ &${56.05^{+15.69}_{-18.80}}$ &${61.71^{+7.03}_{-7.08}}$ &${0.44^{+0.14}_{-0.27}}$\\ 
$434376215$ &$-1.59 \pm 0.47$ &$2.73142 \pm 0.00005$ &${6.4^{+0.7}_{-1.3}}$ &${5.1^{+0.6}_{-0.9}}$ &${0.330^{+0.002}_{-0.001}}$ &${82.99^{+2.09}_{-1.22}}$ &${-8.54^{+0.43}_{-0.42}}$ &${0.12^{+0.14}_{-0.10}}$\\ 
$444446481$ &$0.41 \pm 0.41$ &$3.40154 \pm 0.00007$ &${2.0^{+0.3}_{-0.3}}$ &${0.2^{+0.1}_{-0.0}}$ &${0.305^{+0.006}_{-0.006}}$ &${52.43^{+2.64}_{-1.77}}$ &${18.04^{+1.82}_{-2.01}}$ &${0.21^{+0.16}_{-0.12}}$\\ 
$444555685$ &$-2.75 \pm 0.82$ &$3.18181 \pm 0.00006$ &${5.4^{+1.5}_{-1.5}}$ &${4.4^{+1.9}_{-1.3}}$ &${0.319^{+0.002}_{-0.002}}$ &${89.03^{+0.57}_{-0.55}}$ &${26.92^{+0.72}_{-0.64}}$ &${0.39^{+0.01}_{-0.01}}$\\ 
$448860246$ &$0.43 \pm 0.13$ &$5.40089 \pm 0.00030$ &${3.0^{+0.3}_{-0.3}}$ &${2.5^{+0.3}_{-0.2}}$ &${0.290^{+0.004}_{-0.005}}$ &${85.29^{+0.59}_{-0.32}}$ &${-15.02^{+5.27}_{-2.39}}$ &${0.19^{+0.13}_{-0.09}}$\\ 
$452366964$ &$-1.80 \pm 0.36$ &$6.39238 \pm 0.00022$ &${5.7^{+1.2}_{-1.3}}$ &${3.3^{+0.8}_{-0.4}}$ &${0.439^{+0.003}_{-0.003}}$ &${77.24^{+0.81}_{-0.86}}$ &${51.14^{+0.77}_{-0.58}}$ &${0.37^{+0.11}_{-0.09}}$\\ 
$453278567$ &$1.00 \pm 0.19$ &$3.00898 \pm 0.00006$ &${1.8^{+0.4}_{-0.3}}$ &${0.9^{+1.0}_{-0.5}}$ &${0.249^{+0.007}_{-0.009}}$ &${29.54^{+18.41}_{-2.80}}$ &${43.30^{+3.35}_{-7.96}}$ &${0.17^{+0.16}_{-0.10}}$\\ 
$456042860$ &$-0.44 \pm 0.39$ &$2.78286 \pm 0.00005$ &${2.6^{+0.7}_{-0.5}}$ &${0.4^{+0.7}_{-0.2}}$ &${0.267^{+0.005}_{-0.006}}$ &${27.85^{+1.26}_{-0.79}}$ &${33.37^{+0.94}_{-1.06}}$ &${0.16^{+0.21}_{-0.10}}$\\ 
$457264281$ &$0.58 \pm 0.76$ &$3.22971 \pm 0.00005$ &${1.7^{+0.2}_{-0.2}}$ &${0.2^{+0.2}_{-0.0}}$ &${0.200^{+0.004}_{-0.003}}$ &${42.70^{+1.07}_{-0.74}}$ &${-15.43^{+0.88}_{-1.22}}$ &${0.15^{+0.13}_{-0.10}}$\\ 
$461541766$ &$-0.14 \pm 0.17$ &$8.94488 \pm 0.00051$ &${5.1^{+16.0}_{-2.0}}$ &${0.6^{+1.5}_{-0.2}}$ &${0.551^{+0.027}_{-0.027}}$ &${74.62^{+6.21}_{-8.34}}$ &${44.08^{+6.52}_{-9.09}}$ &${0.26^{+0.21}_{-0.16}}$\\ 
$462292185$ &$-5.12 \pm 1.45$ &$2.52092 \pm 0.00003$ &${12.3^{+12.0}_{-5.0}}$ &${2.9^{+2.6}_{-1.3}}$ &${0.233^{+0.006}_{-0.005}}$ &${38.82^{+1.12}_{-0.65}}$ &${16.15^{+1.94}_{-1.91}}$ &${0.26^{+0.15}_{-0.14}}$\\ 
$462940910$ &$0.21 \pm 7.77$ &$4.88476 \pm 0.00009$ &${3.4^{+3.3}_{-1.7}}$ &${2.7^{+2.2}_{-1.4}}$ &${0.438^{+0.004}_{-0.003}}$ &${78.85^{+1.39}_{-1.15}}$ &${68.73^{+1.01}_{-1.15}}$ &${0.39^{+0.23}_{-0.10}}$\\ 
$469289800$ &$0.43 \pm 0.37$ &$3.55041 \pm 0.00015$ &${5.4^{+1.5}_{-1.5}}$ &${3.0^{+1.0}_{-1.3}}$ &${0.496^{+0.027}_{-0.024}}$ &${20.35^{+8.53}_{-6.02}}$ &${46.26^{+19.90}_{-36.45}}$ &${0.36^{+0.18}_{-0.22}}$\\ 
$470847250$ &$-1.74 \pm 0.62$ &$4.33039 \pm 0.00012$ &${4.3^{+0.6}_{-0.4}}$ &${0.5^{+0.1}_{-0.0}}$ &${0.281^{+0.003}_{-0.003}}$ &${59.42^{+0.58}_{-0.59}}$ &${29.35^{+0.71}_{-0.72}}$ &${0.12^{+0.07}_{-0.08}}$\\ 
$670290948$ &$2.46 \pm 0.45$ &$1.99155 \pm 0.00002$ &${0.9^{+0.7}_{-0.2}}$ &${0.4^{+0.6}_{-0.1}}$ &${0.294^{+0.003}_{-0.003}}$ &${75.35^{+1.24}_{-1.53}}$ &${34.14^{+1.23}_{-1.75}}$ &${0.36^{+0.17}_{-0.10}}$\\ 
$745751393$ &$-4.17 \pm 2.73$ &$3.23937 \pm 0.00005$ &${9.4^{+9.6}_{-2.5}}$ &${6.1^{+4.0}_{-1.5}}$ &${0.353^{+0.005}_{-0.004}}$ &${76.94^{+0.86}_{-1.34}}$ &${41.23^{+4.72}_{-3.82}}$ &${0.64^{+0.06}_{-0.10}}$\\ 
$746425948$ &$1.73 \pm 0.24$ &$17.77823 \pm 0.00231$ &${1.7^{+0.2}_{-0.1}}$ &${1.0^{+0.2}_{-0.2}}$ &${0.638^{+0.001}_{-0.048}}$ &${87.26^{+0.68}_{-6.43}}$ &${10.10^{+0.56}_{-14.95}}$ &${0.18^{+0.16}_{-0.04}}$\\ 
$2021686529$ &$1.84 \pm 0.21$ &$2.16482 \pm 0.00003$ &${1.3^{+0.2}_{-0.2}}$ &${0.3^{+0.2}_{-0.1}}$ &${0.201^{+0.007}_{-0.005}}$ &${38.09^{+4.33}_{-0.94}}$ &${46.71^{+1.60}_{-1.26}}$ &${0.27^{+0.13}_{-0.12}}$\\ 
$2021686530$ &$2.55 \pm 0.21$ &$2.16482 \pm 0.00003$ &${1.1^{+0.2}_{-0.1}}$ &${0.5^{+0.2}_{-0.2}}$ &${0.205^{+0.003}_{-0.006}}$ &${43.89^{+5.07}_{-5.72}}$ &${47.58^{+0.80}_{-1.52}}$ &${0.28^{+0.12}_{-0.18}}$\\ 
$2046878952$ &$-2.49 \pm 0.82$ &$2.53347 \pm 0.00004$ &${6.5^{+1.9}_{-1.4}}$ &${0.9^{+0.3}_{-0.2}}$ &${0.207^{+0.005}_{-0.005}}$ &${40.80^{+1.04}_{-0.73}}$ &${-37.07^{+1.68}_{-1.65}}$ &${0.23^{+0.14}_{-0.14}}$\\
\hline
\end{longtable}

\clearpage
\bibliography{main}{}

\begin{thebibliography}{}
\expandafter\ifx\csname natexlab\endcsname\relax\def\natexlab#1{#1}\fi
\providecommand{\url}[1]{\href{#1}{#1}}
\providecommand{\dodoi}[1]{doi:~\href{http://doi.org/#1}{\nolinkurl{#1}}}
\providecommand{\doeprint}[1]{\href{http://ascl.net/#1}{\nolinkurl{http://ascl.net/#1}}}
\providecommand{\doarXiv}[1]{\href{https://arxiv.org/abs/#1}{\nolinkurl{https://arxiv.org/abs/#1}}}

\bibitem[{{Astropy Collaboration} {et~al.}(2022){Astropy Collaboration}, {Price-Whelan}, {Lim}, {Earl}, {Starkman}, {Bradley}, {Shupe}, {Patil}, {Corrales}, {Brasseur}, {N{"o}the}, {Donath}, {Tollerud}, {Morris}, {Ginsburg}, {Vaher}, {Weaver}, {Tocknell}, {Jamieson}, {van Kerkwijk}, {Robitaille}, {Merry}, {Bachetti}, {G{"u}nther}, {Aldcroft}, {Alvarado-Montes}, {Archibald}, {B{'o}di}, {Bapat}, {Barentsen}, {Baz{'a}n}, {Biswas}, {Boquien}, {Burke}, {Cara}, {Cara}, {Conroy}, {Conseil}, {Craig}, {Cross}, {Cruz}, {D'Eugenio}, {Dencheva}, {Devillepoix}, {Dietrich}, {Eigenbrot}, {Erben}, {Ferreira}, {Foreman-Mackey}, {Fox}, {Freij}, {Garg}, {Geda}, {Glattly}, {Gondhalekar}, {Gordon}, {Grant}, {Greenfield}, {Groener}, {Guest}, {Gurovich}, {Handberg}, {Hart}, {Hatfield-Dodds}, {Homeier}, {Hosseinzadeh}, {Jenness}, {Jones}, {Joseph}, {Kalmbach}, {Karamehmetoglu}, {Ka{l}uszy{'n}ski}, {Kelley}, {Kern}, {Kerzendorf}, {Koch}, {Kulumani}, {Lee}, {Ly}, {Ma}, {MacBride}, {Maljaars}, {Muna}, {Murphy}, {Norman}, {O'Steen},
  {Oman}, {Pacifici}, {Pascual}, {Pascual-Granado}, {Patil}, {Perren}, {Pickering}, {Rastogi}, {Roulston}, {Ryan}, {Rykoff}, {Sabater}, {Sakurikar}, {Salgado}, {Sanghi}, {Saunders}, {Savchenko}, {Schwardt}, {Seifert-Eckert}, {Shih}, {Jain}, {Shukla}, {Sick}, {Simpson}, {Singanamalla}, {Singer}, {Singhal}, {Sinha}, {Sip{H{o}}cz}, {Spitler}, {Stansby}, {Streicher}, {{{S}}umak}, {Swinbank}, {Taranu}, {Tewary}, {Tremblay}, {Val-Borro}, {Van Kooten}, {Vasovi{'c}}, {Verma}, {de Miranda Cardoso}, {Williams}, {Wilson}, {Winkel}, {Wood-Vasey}, {Xue}, {Yoachim}, {Zhang}, {Zonca}, \& {Astropy Project Contributors}}]{astropy:2022}
{Astropy Collaboration}, {Price-Whelan}, A.~M., {Lim}, P.~L., {et~al.} 2022, \apj, 935, 167, \dodoi{10.3847/1538-4357/ac7c74}

\bibitem[{{Balona}(2017)}]{star_spots_2017MNRAS.467.1830B}
{Balona}, L.~A. 2017, \mnras, 467, 1830, \dodoi{10.1093/mnras/stx265}

\bibitem[{{Beck} {et~al.}(2014){Beck}, {Hambleton}, {Vos}, {Kallinger}, {Bloemen}, {Tkachenko}, {Garc{\'\i}a}, {{\O}stensen}, {Aerts}, {Kurtz}, {De Ridder}, {Hekker}, {Pavlovski}, {Mathur}, {De Smedt}, {Derekas}, {Corsaro}, {Mosser}, {Van Winckel}, {Huber}, {Degroote}, {Davies}, {Pr{\v{s}}a}, {Debosscher}, {Elsworth}, {Nemeth}, {Siess}, {Schmid}, {P{\'a}pics}, {de Vries}, {van Marle}, {Marcos-Arenal}, \& {Lobel}}]{beck_2014AA...564A..36B}
{Beck}, P.~G., {Hambleton}, K., {Vos}, J., {et~al.} 2014, \aap, 564, A36, \dodoi{10.1051/0004-6361/201322477}

\bibitem[{Belokurov {et~al.}(2020)Belokurov, Penoyre, Oh, Iorio, Hodgkin, Evans, Everall, Koposov, Tout, Izzard, Clarke, \& Brown}]{RUWE2Belokurov2020}
Belokurov, V., Penoyre, Z., Oh, S., {et~al.} 2020, Monthly Notices of the Royal Astronomical Society, 496, 1922–1940, \dodoi{10.1093/mnras/staa1522}

\bibitem[{{Borkovits} {et~al.}(2016){Borkovits}, {Hajdu}, {Sztakovics}, {Rappaport}, {Levine}, {B{\'\i}r{\'o}}, \& {Klagyivik}}]{borkovits_2016MNRAS.455.4136B}
{Borkovits}, T., {Hajdu}, T., {Sztakovics}, J., {et~al.} 2016, \mnras, 455, 4136, \dodoi{10.1093/mnras/stv2530}

\bibitem[{{Borkovits} {et~al.}(2020){Borkovits}, {Rappaport}, {Tan}, {Gagliano}, {Jacobs}, {Huang}, {Mitnyan}, {Hambsch}, {Kaye}, {Maxted}, {P{\'a}l}, \& {Schmitt}}]{borkovits_2020MNRAS.496.4624B}
{Borkovits}, T., {Rappaport}, S.~A., {Tan}, T.~G., {et~al.} 2020, \mnras, 496, 4624, \dodoi{10.1093/mnras/staa1817}

\bibitem[{{Burkart} {et~al.}(2012){Burkart}, {Quataert}, {Arras}, \& {Weinberg}}]{burkart_elliot_2012MNRAS.421..983B}
{Burkart}, J., {Quataert}, E., {Arras}, P., \& {Weinberg}, N.~N. 2012, \mnras, 421, 983, \dodoi{10.1111/j.1365-2966.2011.20344.x}

\bibitem[{Castro-Ginard {et~al.}(2024)Castro-Ginard, Penoyre, Casey, Brown, Belokurov, Cantat-Gaudin, Drimmel, Fouesneau, Khanna, Kurbatov, Price-Whelan, Rix, \& Smart}]{RUWECastroGinard2024}
Castro-Ginard, A., Penoyre, Z., Casey, A.~R., {et~al.} 2024, Astronomy \&amp; Astrophysics, 688, A1, \dodoi{10.1051/0004-6361/202450172}

\bibitem[{{Cheng} {et~al.}(2020){Cheng}, {Fuller}, {Guo}, {Lehman}, \& {Hambleton}}]{shelley_cheng_2020ApJ...903..122C}
{Cheng}, S.~J., {Fuller}, J., {Guo}, Z., {Lehman}, H., \& {Hambleton}, K. 2020, \apj, 903, 122, \dodoi{10.3847/1538-4357/abb46d}

\bibitem[{{Claret} {et~al.}(2020){Claret}, {Cukanovaite}, {Burdge}, {Tremblay}, {Parsons}, \& {Marsh}}]{claret_2020A&A...641A.157C}
{Claret}, A., {Cukanovaite}, E., {Burdge}, K., {et~al.} 2020, \aap, 641, A157, \dodoi{10.1051/0004-6361/202038436}

\bibitem[{{Dittmann} \& {Ryan}(2022)}]{dittmann_2022MNRAS.513.6158D}
{Dittmann}, A.~J., \& {Ryan}, G. 2022, \mnras, 513, 6158, \dodoi{10.1093/mnras/stac935}

\bibitem[{{D'Orazio} \& {Duffell}(2021)}]{orazio_2021ApJ...914L..21D}
{D'Orazio}, D.~J., \& {Duffell}, P.~C. 2021, \apjl, 914, L21, \dodoi{10.3847/2041-8213/ac0621}

\bibitem[{{Engel} {et~al.}(2020){Engel}, {Faigler}, {Shahaf}, \& {Mazeh}}]{Engel_2020MNRAS.497.4884E}
{Engel}, M., {Faigler}, S., {Shahaf}, S., \& {Mazeh}, T. 2020, \mnras, 497, 4884, \dodoi{10.1093/mnras/staa2182}

\bibitem[{{Faigler}(2016)}]{lambert_2016arXiv161208846F}
{Faigler}, S. 2016, arXiv e-prints, arXiv:1612.08846, \dodoi{10.48550/arXiv.1612.08846}

\bibitem[{{Faigler} \& {Mazeh}(2011)}]{faigler_mazeh_2011MNRAS.415.3921F}
{Faigler}, S., \& {Mazeh}, T. 2011, \mnras, 415, 3921, \dodoi{10.1111/j.1365-2966.2011.19011.x}

\bibitem[{{Feinstein} {et~al.}(2019){Feinstein}, {Montet}, {Foreman-Mackey}, {Bedell}, {Saunders}, {Bean}, {Christiansen}, {Hedges}, {Luger}, {Scolnic}, \& {Cardoso}}]{eleanor_2019PASP..131i4502F}
{Feinstein}, A.~D., {Montet}, B.~T., {Foreman-Mackey}, D., {et~al.} 2019, \pasp, 131, 094502, \dodoi{10.1088/1538-3873/ab291c}

\bibitem[{{Fuller}(2017)}]{fuller_2017MNRAS.472.1538F}
{Fuller}, J. 2017, \mnras, 472, 1538, \dodoi{10.1093/mnras/stx2135}

\bibitem[{{Fuller} \& {Lai}(2012)}]{fuller_lai_2012MNRAS.420.3126F}
{Fuller}, J., \& {Lai}, D. 2012, \mnras, 420, 3126, \dodoi{10.1111/j.1365-2966.2011.20237.x}

\bibitem[{{Guo} {et~al.}(2020){Guo}, {Shporer}, {Hambleton}, \& {Isaacson}}]{teos_2020ApJ...888...95G}
{Guo}, Z., {Shporer}, A., {Hambleton}, K., \& {Isaacson}, H. 2020, \apj, 888, 95, \dodoi{10.3847/1538-4357/ab58c2}

\bibitem[{Hambleton {et~al.}(2016)Hambleton, Kurtz, Prša, Quinn, Fuller, Murphy, Thompson, Latham, \& Shporer}]{Hambleton_2016}
Hambleton, K., Kurtz, D.~W., Prša, A., {et~al.} 2016, Monthly Notices of the Royal Astronomical Society, 463, 1199–1212, \dodoi{10.1093/mnras/stw1970}

\bibitem[{{Hambleton} {et~al.}(2013){Hambleton}, {Kurtz}, {Pr{\v{s}}a}, {Guzik}, {Pavlovski}, {Bloemen}, {Southworth}, {Conroy}, {Littlefair}, \& {Fuller}}]{hambleton_2013MNRAS.434..925H}
{Hambleton}, K.~M., {Kurtz}, D.~W., {Pr{\v{s}}a}, A., {et~al.} 2013, \mnras, 434, 925, \dodoi{10.1093/mnras/stt886}

\bibitem[{{Hut}(1981)}]{hut_1981A&A....99..126H}
{Hut}, P. 1981, \aap, 99, 126

\bibitem[{{Jayasinghe} {et~al.}(2021){Jayasinghe}, {Kochanek}, {Strader}, {Stanek}, {Vallely}, {Thompson}, {Hinkle}, {Shappee}, {Dupree}, {Auchettl}, {Chomiuk}, {Aydi}, {Dage}, {Hughes}, {Shishkovsky}, {Sokolovsky}, {Swihart}, {Voggel}, \& {Thompson}}]{jayasinghe_2021MNRAS.506.4083J}
{Jayasinghe}, T., {Kochanek}, C.~S., {Strader}, J., {et~al.} 2021, \mnras, 506, 4083, \dodoi{10.1093/mnras/stab1920}

\bibitem[{{Jeffery}(1984)}]{apsial_motion_constant_1984MNRAS.207..323J}
{Jeffery}, C.~S. 1984, \mnras, 207, 323, \dodoi{10.1093/mnras/207.2.323}

\bibitem[{{Kirk} {et~al.}(2016){Kirk}, {Conroy}, {Pr{\v{s}}a}, {Abdul-Masih}, {Kochoska}, {Matijevi{\v{c}}}, {Hambleton}, {Barclay}, {Bloemen}, {Boyajian}, {Doyle}, {Fulton}, {Hoekstra}, {Jek}, {Kane}, {Kostov}, {Latham}, {Mazeh}, {Orosz}, {Pepper}, {Quarles}, {Ragozzine}, {Shporer}, {Southworth}, {Stassun}, {Thompson}, {Welsh}, {Agol}, {Derekas}, {Devor}, {Fischer}, {Green}, {Gropp}, {Jacobs}, {Johnston}, {LaCourse}, {Saetre}, {Schwengeler}, {Toczyski}, {Werner}, {Garrett}, {Gore}, {Martinez}, {Spitzer}, {Stevick}, {Thomadis}, {Vrijmoet}, {Yenawine}, {Batalha}, \& {Borucki}}]{kirk_2016AJ....151...68K}
{Kirk}, B., {Conroy}, K., {Pr{\v{s}}a}, A., {et~al.} 2016, \aj, 151, 68, \dodoi{10.3847/0004-6256/151/3/68}

\bibitem[{{Ko{\l}aczek-Szyma{\'n}ski} {et~al.}(2021){Ko{\l}aczek-Szyma{\'n}ski}, {Pigulski}, {Michalska}, {Mo{\'z}dzierski}, \& {R{\'o}{\.z}a{\'n}ski}}]{tess_heartbeats_2021AA...647A..12K}
{Ko{\l}aczek-Szyma{\'n}ski}, P.~A., {Pigulski}, A., {Michalska}, G., {Mo{\'z}dzierski}, D., \& {R{\'o}{\.z}a{\'n}ski}, T. 2021, \aap, 647, A12, \dodoi{10.1051/0004-6361/202039553}

\bibitem[{{Kostov} {et~al.}(2022){Kostov}, {Powell}, {Rappaport}, {Borkovits}, {Gagliano}, {Jacobs}, {Kristiansen}, {LaCourse}, {Omohundro}, {Orosz}, {Schmitt}, {Schwengeler}, {Terentev}, {Torres}, {Barclay}, {Friedman}, {Kruse}, {Olmschenk}, {Vanderburg}, \& {Welsh}}]{2022ApJS..259...66K}
{Kostov}, V.~B., {Powell}, B.~P., {Rappaport}, S.~A., {et~al.} 2022, \apjs, 259, 66, \dodoi{10.3847/1538-4365/ac5458}

\bibitem[{{Kostov} {et~al.}(2024){Kostov}, {Powell}, {Rappaport}, {Borkovits}, {Gagliano}, {Jacobsy}, {Jayaraman}, {Kristiansen}, {LaCourse}, {Mitnyan}, {Omohundro}, {Orosz}, {P{\'a}l}, {Schmitt}, {Schwengeler}, {Terentev}, {Torres}, {Barclay}, {Vanderburg}, \& {Welsh}}]{2024MNRAS.527.3995K}
---. 2024, \mnras, 527, 3995, \dodoi{10.1093/mnras/stad2947}

\bibitem[{{Kounkel} {et~al.}(2019){Kounkel}, {Covey}, {Moe}, {Kratter}, {Su{\'a}rez}, {Stassun}, {Rom{\'a}n-Z{\'u}{\~n}iga}, {Hernandez}, {Kim}, {Pe{\~n}a Ram{\'\i}rez}, {Roman-Lopes}, {Stringfellow}, {Jaehnig}, {Borissova}, {Tofflemire}, {Krolikowski}, {Rizzuto}, {Kraus}, {Badenes}, {Longa-Pe{\~n}a}, {G{\'o}mez Maqueo Chew}, {Barba}, {Nidever}, {Brown}, {De Lee}, {Pan}, {Bizyaev}, {Oravetz}, \& {Oravetz}}]{kounkel_2019AJ....157..196K}
{Kounkel}, M., {Covey}, K., {Moe}, M., {et~al.} 2019, \aj, 157, 196, \dodoi{10.3847/1538-3881/ab13b1}

\bibitem[{{Kumar} {et~al.}(1995){Kumar}, {Ao}, \& {Quataert}}]{Kumar_1995ApJ...449..294K}
{Kumar}, P., {Ao}, C.~O., \& {Quataert}, E.~J. 1995, \apj, 449, 294, \dodoi{10.1086/176055}

\bibitem[{{Liu} {et~al.}(2015){Liu}, {Mu{\~n}oz}, \& {Lai}}]{liu_2015MNRAS.447..747L}
{Liu}, B., {Mu{\~n}oz}, D.~J., \& {Lai}, D. 2015, \mnras, 447, 747, \dodoi{10.1093/mnras/stu2396}

\bibitem[{{Maceroni} {et~al.}(2009){Maceroni}, {Montalb{\'a}n}, {Michel}, {Harmanec}, {Prsa}, {Briquet}, {Niemczura}, {Morel}, {Ladjal}, {Auvergne}, {Baglin}, {Baudin}, {Catala}, {Samadi}, \& {Aerts}}]{high_ecc_B_2009A&A...508.1375M}
{Maceroni}, C., {Montalb{\'a}n}, J., {Michel}, E., {et~al.} 2009, \aap, 508, 1375, \dodoi{10.1051/0004-6361/200913311}

\bibitem[{{Moe} \& {Di Stefano}(2017)}]{moe_di_2017ApJS..230...15M}
{Moe}, M., \& {Di Stefano}, R. 2017, \apjs, 230, 15, \dodoi{10.3847/1538-4365/aa6fb6}

\bibitem[{{Moe} \& {Kratter}(2018)}]{moe_kratter_2018ApJ...854...44M}
{Moe}, M., \& {Kratter}, K.~M. 2018, \apj, 854, 44, \dodoi{10.3847/1538-4357/aaa6d2}

\bibitem[{{Morris}(1985)}]{morris_1985ApJ...295..143M}
{Morris}, S.~L. 1985, \apj, 295, 143, \dodoi{10.1086/163359}

\bibitem[{{Offner} {et~al.}(2023){Offner}, {Moe}, {Kratter}, {Sadavoy}, {Jensen}, \& {Tobin}}]{offner_2023ASPC..534..275O}
{Offner}, S.~S.~R., {Moe}, M., {Kratter}, K.~M., {et~al.} 2023, in Astronomical Society of the Pacific Conference Series, Vol. 534, Protostars and Planets VII, ed. S.~{Inutsuka}, Y.~{Aikawa}, T.~{Muto}, K.~{Tomida}, \& M.~{Tamura}, 275, \dodoi{10.48550/arXiv.2203.10066}

\bibitem[{Olmschenk {et~al.}(2021)Olmschenk, Silva, Rau, Barry, Kruse, Cacciapuoti, Kostov, Powell, Wyrwas, Schnittman, \& Barclay}]{olmschenk2021transit}
Olmschenk, G., Silva, S.~I., Rau, G., {et~al.} 2021, The Astronomical Journal, 161, 273, \dodoi{10.3847/1538-3881/abf4c6}

\bibitem[{{Pecaut} \& {Mamajek}(2013)}]{pre-ms-colors_2013ApJS..208....9P}
{Pecaut}, M.~J., \& {Mamajek}, E.~E. 2013, \apjs, 208, 9, \dodoi{10.1088/0067-0049/208/1/9}

\bibitem[{{Pfahl} {et~al.}(2008){Pfahl}, {Arras}, \& {Paxton}}]{pfahl_2008ApJ...679..783P}
{Pfahl}, E., {Arras}, P., \& {Paxton}, B. 2008, \apj, 679, 783, \dodoi{10.1086/586878}

\bibitem[{{Powell} {et~al.}(2021){Powell}, {Kostov}, {Rappaport}, {Borkovits}, {Zasche}, {Tokovinin}, {Kruse}, {Latham}, {Montet}, {Jensen}, {Jayaraman}, {Collins}, {Ma{\v{s}}ek}, {Hellier}, {Evans}, {Tan}, {Schlieder}, {Torres}, {Smale}, {Friedman}, {Barclay}, {Gagliano}, {Quintana}, {Jacobs}, {Gilbert}, {Kristiansen}, {Col{\'o}n}, {LaCourse}, {Olmschenk}, {Omohundro}, {Schnittman}, {Schwengeler}, {Barry}, {Terentev}, {Boyd}, {Schmitt}, {Quinn}, {Vanderburg}, {Palle}, {Armstrong}, {Ricker}, {Vanderspek}, {Seager}, {Winn}, {Jenkins}, {Caldwell}, {Wohler}, {Shiao}, {Burke}, {Daylan}, \& {Villase{\~n}or}}]{2021AJ....161..162P}
{Powell}, B.~P., {Kostov}, V.~B., {Rappaport}, S.~A., {et~al.} 2021, \aj, 161, 162, \dodoi{10.3847/1538-3881/abddb5}

\bibitem[{{Pr{\v{s}}a} {et~al.}(2022){Pr{\v{s}}a}, {Kochoska}, {Conroy}, {Eisner}, {Hey}, {IJspeert}, {Kruse}, {Fleming}, {Johnston}, {Kristiansen}, {LaCourse}, {Mortensen}, {Pepper}, {Stassun}, {Torres}, {Abdul-Masih}, {Chakraborty}, {Gagliano}, {Guo}, {Hambleton}, {Hong}, {Jacobs}, {Jones}, {Kostov}, {Lee}, {Omohundro}, {Orosz}, {Page}, {Powell}, {Rappaport}, {Reed}, {Schnittman}, {Schwengeler}, {Shporer}, {Terentev}, {Vanderburg}, {Welsh}, {Caldwell}, {Doty}, {Jenkins}, {Latham}, {Ricker}, {Seager}, {Schlieder}, {Shiao}, {Vanderspek}, \& {Winn}}]{tess_ebs_2022ApJS..258...16P}
{Pr{\v{s}}a}, A., {Kochoska}, A., {Conroy}, K.~E., {et~al.} 2022, \apjs, 258, 16, \dodoi{10.3847/1538-4365/ac324a}

\bibitem[{{Ricker} {et~al.}(2015){Ricker}, {Winn}, {Vanderspek}, {Latham}, {Bakos}, {Bean}, {Berta-Thompson}, {Brown}, {Buchhave}, {Butler}, {Butler}, {Chaplin}, {Charbonneau}, {Christensen-Dalsgaard}, {Clampin}, {Deming}, {Doty}, {De Lee}, {Dressing}, {Dunham}, {Endl}, {Fressin}, {Ge}, {Henning}, {Holman}, {Howard}, {Ida}, {Jenkins}, {Jernigan}, {Johnson}, {Kaltenegger}, {Kawai}, {Kjeldsen}, {Laughlin}, {Levine}, {Lin}, {Lissauer}, {MacQueen}, {Marcy}, {McCullough}, {Morton}, {Narita}, {Paegert}, {Palle}, {Pepe}, {Pepper}, {Quirrenbach}, {Rinehart}, {Sasselov}, {Sato}, {Seager}, {Sozzetti}, {Stassun}, {Sullivan}, {Szentgyorgyi}, {Torres}, {Udry}, \& {Villasenor}}]{tess_2015JATIS...1a4003R}
{Ricker}, G.~R., {Winn}, J.~N., {Vanderspek}, R., {et~al.} 2015, Journal of Astronomical Telescopes, Instruments, and Systems, 1, 014003, \dodoi{10.1117/1.JATIS.1.1.014003}

\bibitem[{{Shporer} {et~al.}(2016){Shporer}, {Fuller}, {Isaacson}, {Hambleton}, {Thompson}, {Pr{\v{s}}a}, {Kurtz}, {Howard}, \& {O'Leary}}]{shporer_2016ApJ...829...34S}
{Shporer}, A., {Fuller}, J., {Isaacson}, H., {et~al.} 2016, \apj, 829, 34, \dodoi{10.3847/0004-637X/829/1/34}

\bibitem[{{Stassun} {et~al.}(2019){Stassun}, {Oelkers}, {Paegert}, {Torres}, {Pepper}, {De Lee}, {Collins}, {Latham}, {Muirhead}, {Chittidi}, {Rojas-Ayala}, {Fleming}, {Rose}, {Tenenbaum}, {Ting}, {Kane}, {Barclay}, {Bean}, {Brassuer}, {Charbonneau}, {Ge}, {Lissauer}, {Mann}, {McLean}, {Mullally}, {Narita}, {Plavchan}, {Ricker}, {Sasselov}, {Seager}, {Sharma}, {Shiao}, {Sozzetti}, {Stello}, {Vanderspek}, {Wallace}, \& {Winn}}]{tic_catalog_2019AJ....158..138S}
{Stassun}, K.~G., {Oelkers}, R.~J., {Paegert}, M., {et~al.} 2019, \aj, 158, 138, \dodoi{10.3847/1538-3881/ab3467}

\bibitem[{{Thompson} {et~al.}(2012){Thompson}, {Everett}, {Mullally}, {Barclay}, {Howell}, {Still}, {Rowe}, {Christiansen}, {Kurtz}, {Hambleton}, {Twicken}, {Ibrahim}, \& {Clarke}}]{susan_thompson_2012ApJ...753...86T}
{Thompson}, S.~E., {Everett}, M., {Mullally}, F., {et~al.} 2012, \apj, 753, 86, \dodoi{10.1088/0004-637X/753/1/86}

\bibitem[{{Tokovinin} {et~al.}(2006){Tokovinin}, {Thomas}, {Sterzik}, \& {Udry}}]{tokovinin_2006A&A...450..681T}
{Tokovinin}, A., {Thomas}, S., {Sterzik}, M., \& {Udry}, S. 2006, \aap, 450, 681, \dodoi{10.1051/0004-6361:20054427}

\bibitem[{{Toonen} {et~al.}(2020){Toonen}, {Portegies Zwart}, {Hamers}, \& {Bandopadhyay}}]{toonen_2020A&A...640A..16T}
{Toonen}, S., {Portegies Zwart}, S., {Hamers}, A.~S., \& {Bandopadhyay}, D. 2020, \aap, 640, A16, \dodoi{10.1051/0004-6361/201936835}

\bibitem[{{Turpin} \& {Nelson}(2024)}]{turpin_2024MNRAS.528.7256T}
{Turpin}, G.~A., \& {Nelson}, R.~P. 2024, \mnras, 528, 7256, \dodoi{10.1093/mnras/stae109}

\bibitem[{{Welsh} {et~al.}(2011){Welsh}, {Orosz}, {Aerts}, {Brown}, {Brugamyer}, {Cochran}, {Gilliland}, {Guzik}, {Kurtz}, {Latham}, {Marcy}, {Quinn}, {Zima}, {Allen}, {Batalha}, {Bryson}, {Buchhave}, {Caldwell}, {Gautier}, {Howell}, {Kinemuchi}, {Ibrahim}, {Isaacson}, {Jenkins}, {Prsa}, {Still}, {Street}, {Wohler}, {Koch}, \& {Borucki}}]{welsh_koi_2011ApJS..197....4W}
{Welsh}, W.~F., {Orosz}, J.~A., {Aerts}, C., {et~al.} 2011, \apjs, 197, 4, \dodoi{10.1088/0067-0049/197/1/4}

\bibitem[{{Wrona} {et~al.}(2022{\natexlab{a}}){Wrona}, {Ko{\l}aczek-Szyma{\'n}ski}, {Ratajczak}, \& {Koz{\l}owski}}]{ogle_2_2022ApJ...928..135W}
{Wrona}, M., {Ko{\l}aczek-Szyma{\'n}ski}, P.~A., {Ratajczak}, M., \& {Koz{\l}owski}, S. 2022{\natexlab{a}}, \apj, 928, 135, \dodoi{10.3847/1538-4357/ac56e6}

\bibitem[{{Wrona} {et~al.}(2022{\natexlab{b}}){Wrona}, {Ratajczak}, {Ko{\l}aczek-Szyma{\'n}ski}, {Koz{\l}owski}, {Soszy{\'n}ski}, {Iwanek}, {Udalski}, {Szyma{\'n}ski}, {Pietrukowicz}, {Skowron}, {Skowron}, {Mr{\'o}z}, {Poleski}, {Gromadzki}, {Ulaczyk}, \& {Rybicki}}]{ogle_2022ApJS..259...16W}
{Wrona}, M., {Ratajczak}, M., {Ko{\l}aczek-Szyma{\'n}ski}, P.~A., {et~al.} 2022{\natexlab{b}}, \apjs, 259, 16, \dodoi{10.3847/1538-4365/ac4018}

\bibitem[{{Zanazzi}(2023)}]{circularizatio_ebs_2023AAS...24143006Z}
{Zanazzi}, J. 2023, in American Astronomical Society Meeting Abstracts, Vol. 241, American Astronomical Society Meeting Abstracts, 430.06

\bibitem[{{Zasche} {et~al.}(2024){Zasche}, {Henzl}, \& {Wolf}}]{new_eclipsing_2024A&A...683A.158Z}
{Zasche}, P., {Henzl}, Z., \& {Wolf}, M. 2024, \aap, 683, A158, \dodoi{10.1051/0004-6361/202348501}

\end{thebibliography}
\bibliographystyle{aasjournal}

\end{document}